\documentclass[12pt]{article}
\pdfoutput=1
\usepackage{graphicx,psfrag,epsf,color}
\usepackage[dvipsnames]{xcolor}
\usepackage{amsmath,amssymb,amsfonts, booktabs}
\usepackage{appendix}
\usepackage{multirow}
\usepackage{geometry}
\usepackage{float}
\usepackage{mathrsfs}  
\usepackage{authblk}
\usepackage{array}
\usepackage{cite}
\usepackage{slashed}
\usepackage{graphicx}
\usepackage{siunitx}
\usepackage{rotating}
\usepackage{slashed, cancel}
\usepackage[caption=false]{subfig}
\usepackage{comment}
\usepackage{hyperref}
\usepackage{tabularx}
\newcolumntype{C}{>{\centering\arraybackslash}X}
\def\be{\begin{equation}}
\def\ee{\end{equation}}
\def\beq{\begin{eqnarray}}
\def\eeq{\end{eqnarray}}
\newcommand{\bea}{\begin{eqnarray}}
\newcommand{\eea}{\end{eqnarray}}
\newcommand{\beas}{\begin{eqnarray*}}
\newcommand{\eeas}{\end{eqnarray*}}

\newcommand{\g}{\lambda}
\newcommand{\mb}{m_Q}
\newcommand{\mB}{m_H}
\newcommand{\B}{H}
\newcommand{\pB}{{p_H}}
\newcommand{\plus}[1]{n_-#1}
\newcommand{\minus}[1]{n_+#1}
\newcommand{\lam}{b}
\newcommand{\Q}{Q}
\newcommand{\aCFopi}{\frac{\alpha_s C_F}{4\pi}}
\newcommand{\as}{\alpha_s}
\newcommand{\LamQCD}{\Lambda_{\rm QCD}}
\newcommand{\LamUV}{{\Lambda_{\rm UV}}}

\newcounter{MBQ}

\newcounter{GFQ}

\begin{document}
\allowdisplaybreaks

\begin{titlepage}
\begin{flushright}
{\small
TUM-HEP-1455/23\\
Nikhef-2023-003\\
May 10, 2023
}
\end{flushright}

\vskip0.8cm
\begin{center}
{\Large \bf\boldmath QCD Light-Cone Distribution Amplitudes of 
Heavy\\[0.15cm] 
Mesons from boosted HQET}
\end{center}

\vspace{0.4cm}
\begin{center}
{\sc Martin~Beneke,$^a$ 
Gael Finauri,$^{a}$ K. Keri Vos,$^{b,c}$ Yanbing Wei$^{a,d}$} \\[6mm]
{\it $^a$Physik Department T31,\\
James-Franck-Stra\ss{}e~1, 
Technische Universit\"at M\"unchen,\\
D--85748 Garching, Germany}\\[0.3cm]

{\it $^b$Gravitational 
Waves and Fundamental Physics (GWFP),\\ 
Maastricht University, Duboisdomein 30,\\ 
NL-6229 GT Maastricht, the
Netherlands}\\[0.3cm]

{\it $^c$Nikhef, Science Park 105,\\ 
NL-1098 XG Amsterdam, the Netherlands}\\[0.3cm]

{\it $^d$Faculty of Science,\\
Beijing University of Technology,\\
Beijing 100124, China}
\end{center}

\vspace{0.6cm}
\begin{abstract}
\vskip0.2cm\noindent
Light-cone distribution amplitudes (LCDAs) frequently arise in factorization theorems involving light and heavy mesons.
The QCD LCDA for heavy mesons includes short-distance physics at energy scales of the heavy-quark mass. In this paper we achieve the separation of this perturbative scale from the purely hadronic effects by matching the QCD LCDA to the convolution of a perturbative function with the universal, quark-mass independent LCDA defined in heavy-quark effective theory. This factorization allows to resum potentially large logarithms between $\LamQCD$ and $m_Q$  as well as between $m_Q$ and the scale $Q$ of the hard process in the production of boosted heavy mesons at colliders. As an application we derive new theoretical predictions for the branching ratio of the 
decay $W^\pm \to B^\pm \gamma$.
Furthermore, we provide phenomenological models for the QCD LCDAs of both the $\bar{B}$ and $D$ mesons expressed as expansions in Gegenbauer polynomials.

\end{abstract}
\end{titlepage}

\pagenumbering{roman}
\tableofcontents
\newpage



\section{Introduction}

\pagenumbering{arabic}
For hard, exclusive processes, which are dominated by light-like distances, the hadronic physics is described by light-cone distribution amplitudes (LCDAs), defined as hadron-to-vacuum matrix elements of non-local operators spread along the light-cone. The leading-twist QCD LCDA for light mesons has been widely investigated and constrained by non-perturbative techniques such as lattice QCD and sum rules as well as by experimental data.

In the present paper, we envisage the production of a heavy pseudoscalar meson $\B$ from a generic hard process with characteristic scale $\Q \gg \mb$, with $\mb$ denoting the heavy-quark mass. The relevant leading-twist QCD LCDA $\phi(u;\mu)$ is defined as~\cite{Braun:1988qv}
\begin{equation}
\label{eq:QCDLCDA}
\langle H(\pB)| \bar{Q}(0) \slashed{n}_+ \gamma^5[0,tn_+]q(tn_+) |0\rangle = 
-i f_H n_+\cdot p_H \int_0^1 du\, e^{i u t n_+ \cdot \pB} \phi(u;\mu) \,,
\end{equation}
where $n_+$ is a light-like vector, and $[0,tn_+]$ denotes a finite-distance Wilson line. The variable $u$ can be interpreted as the light-cone momentum fraction of the light anti-quark in the meson. The above definition is identical to the one for the LCDA of a light meson. However, for a 
heavy meson, the LCDA describes the hadronic physics at two distinct 
scales, $\mb$ and $\LamQCD$, of which the former should be amenable 
by perturbative methods. The LCDA does not depend on the hard scale 
$\Q$ due to boost invariance.

The LCDAs for heavy mesons are commonly defined in the framework of heavy quark effective theory (HQET), that is, in the infinite-quark mass 
limit \cite{Grozin:1996pq,Beneke:2000wa}. This LCDA is directly 
relevant to exclusive $B$ decays into energetic particles 
at leading power in the expansion $\LamQCD/\mb$, when $\mb$ is 
the only short-distance scale. However, due to the prior limit 
$m_Q\to\infty$, it does not describe 
the physics between the scales $Q$ and $m_Q$, when the heavy meson 
is produced in a hard process with $Q\gg m_Q$.

Our goal is therefore to establish a factorization formula for the QCD LCDA \eqref{eq:QCDLCDA} of a heavy meson in order to separate the perturbative physics at energies of order $m_Q$ from the non-perturbative hadronic effects at the QCD scale. Once the scale $m_Q$ is integrated out, 
the infinite-quark mass limit can be taken. We then expect the 
long-distance physics to be described by the quark-mass independent, 
universal, leading-twist HQET LCDA
 $\varphi_+(\omega;\mu)$, which definition reads
\begin{equation}
\label{eq:phiplusdef}
    \langle H_v| \bar{h}_v(0) \slashed{n}_+\gamma^5[0,tn_+]q_s(tn_+) |0\rangle = -i F_{\rm stat}(\mu) \,n_+\cdot v \int_0^\infty d\omega\, e^{i \omega t n_+\cdot v} \varphi_+(\omega;\mu) \,,
\end{equation} 
where $h_v$ denotes the 
effective heavy quark field with four-velocity $v$, 
$|H_v\rangle$ is the $\mb$-independent heavy meson state and 
$F_{\rm stat}(\mu)$ the static HQET decay constant.

Because the meson is highly boosted, we will work within soft-collinear effective theory (SCET), which will be then matched to boosted HQET (bHQET) \cite{Fleming:2007qr,Fleming:2007xt} by integrating out the heavy-quark mass scale. Intuitively, one expects the LCDA at the matching 
scale of order $m_Q$ to be highly asymmetric with the light anti-quark 
carrying typical light-cone momentum fraction $u\sim \LamQCD/\mb$.
It is therefore necessary to consider separately the matching of the QCD LCDA $\phi(u)$ for the case $u\sim \LamQCD/\mb$, corresponding to the parametric location of the ``peak'' of the LCDA at the 
matching scale of order $m_Q$, and the case $u\sim 1$, which we 
refer to as the ``tail''. The factorization formula will take the 
form
\begin{equation}
\label{eq:ffpreview}
\phi(u) =
\begin{cases}
&\mathcal{J}_p(u,\omega)\otimes \varphi_+(\omega)\,, \qquad u \sim \LamQCD/\mb\,, \\
&\mathcal{J}_{\rm tail}(u)\,, \qquad \qquad \qquad \;\, u \sim 1\,,
\end{cases}
\end{equation}
to leading order in $\LamQCD/\mb$, 
where $\otimes$ denotes a convolution in the variable $\omega$, the light-cone component of the momentum of the light spectator anti-quark. The full LCDA is 
obtained by merging the two regions in the intermediate 
region $\LamQCD/\mb\ll u\ll 1$, where both expressions hold. 
All dependence on the mass of the heavy quark is contained in the 
perturbative functions $\mathcal{J}_p$, $\mathcal{J}_{\rm tail}$,
while the non-perturbative information is encoded in the HQET LCDA $\varphi_+(\omega)$. This implies that the QCD LCDAs of any heavy pseudoscalar mesons can be expressed in terms of a single LCDA, the 
universal HQET LCDA, making manifest the heavy-quark flavour 
symmetry of low-energy QCD. Eq.~\eqref{eq:ffpreview} is to be 
viewed as the initial condition at $\mu\sim m_Q$ for the evolution 
of the QCD LCDA \eqref{eq:QCDLCDA} to $\mu\sim Q$, which is 
computed as usual in terms of the Efremov-Radyushkin-Brodsky-Lepage 
(ERBL) renormalization kernel~\cite{Lepage:1979zb,Lepage:1980fj,Efremov:1979qk}.

The paper is organized as follows:
we proceed by reviewing the construction of bHQET in Section~\ref{sec:bHQET}. 
The matching of the LCDA in the two regions mentioned above and their merging is 
performed in Sections~\ref{sec:Matching},~\ref{sec:tail} and~\ref{sec:merging}. 
In Section~\ref{sec:lcdaprop} we verify that the obtained LCDA 
satisfies the required properties of the QCD LCDA and compare our 
result to a previous, significantly different one \cite{Ishaq:2019dst}. 
Section~\ref{sec:numerics} contains a numerical study of the 
QCD LCDA of the $\bar{B}$ and $D$ mesons, which is then employed  
in Section~\ref{sec:WtoBgamma} to obtain 
new, fully resummed predictions for the $W^\pm \to B^\pm\gamma$  rate. We conclude in Section~\ref{sec:conclusion}.
Several appendices collect technical details and supplementary results.

\section{Theoretical Framework}
\label{sec:bHQET}

We consider a heavy meson produced in a hard process with large energy of order $\Q \gg \mb$ in the center-of-mass frame of the hard collision. Since a heavy quark is involved in the process, HQET would naturally be the appropriate effective theory to separate the perturbative scale $\mb$ from the non-perturbative hadronic physics in an expansion in the parameter
\begin{equation}
\g = \frac{\LamQCD}{\mb} \ll 1\,. 
\end{equation}
However, the large boost of the meson suggests a treatment in SCET 
\cite{Bauer:2000yr,Bauer:2001yt,Beneke:2002ph,Beneke:2002ni}, governed by the small expansion parameter
\begin{equation}
\label{eq:lam}
\lam = \frac{\mB}{\Q} \ll 1 \,,
\end{equation}
where  $\mB \sim \mathcal{O}(\mb)$ is the heavy meson mass.
The merging of these two frameworks is bHQET~\cite{Fleming:2007qr, Fleming:2007xt}. To set up bHQET, we consider a generic pseudoscalar heavy meson 
$\B=(Q\bar{q})$  
containing the heavy quark $Q$ and massless anti-quark $\bar{q}$ 
in a boosted frame $S$, such that its 
four-momentum 
\begin{equation}
p_H^\mu = \mB v^\mu = \Q\frac{n^\mu_-}{2} + \frac{\mB^2}{\Q}\frac{n^\mu_+}{2}\,,
\end{equation}
has large $n_+ p_H=Q$ while $n_- p_H\ll n_+ p_H$. Here $n_-$ and $n_+$ are two 
light-like reference vectors, satisfying 
$n_- \cdot \,n_+ = 2$ and $n_-^2 = n_+^2 = 0$. We employ the convention 
$p = (\minus{p}, p_\perp, \plus{p})$ for the components of momenta. 
The direction of $n_-^\mu$ is called the  ``collinear'' direction. 

The rest frame $S^*$ of the heavy meson is related to the frame $S$ by a boost 
into the collinear direction with parameter 
\begin{equation}
\lam = \sqrt{\frac{\plus{\pB}}{\minus{\pB}}} = \frac{\mB}{\Q}\,.
\end{equation}
The components of a generic vector $V^\mu$ transform as
\begin{equation}
\label{eq:boost}
\plus{V}^* = \frac{1}{\lam}\plus{V} \,,\qquad {V_\perp^*}^\mu = V_\perp^\mu\,,\qquad \minus{V}^* =\lam \,\minus{V}\,.
\end{equation}
The scaling of the momenta of the heavy and light quark 
constituents of $H$ in $S^*$ is
\begin{equation}
p_Q^* \sim (\mb, \LamQCD, \mb) \sim (\lam, \g \lam,\lam)\Q \,,\qquad p_q^* \sim (\LamQCD,\LamQCD,\LamQCD) \sim (\lam,\lam,\lam)\g \Q\,.
\end{equation}
The presence of $\mb$ in $p_Q^*$ introduces virtual $\mb$-hard modes
\begin{equation}
p_h^* \sim (\mb, \mb, \mb) \sim (\lam,\lam,\lam)\Q \,,
\end{equation}
which scale in the boosted frame $S$ as
\begin{equation}
\label{eq:finscal}
p_h \sim (1,\lam,\lam^2)\Q \quad \text{``hard-collinear"}  \,,\qquad p_q \sim (1,\lam,\lam^2)\g \Q \quad \text{``soft-collinear"} \,. 
\end{equation}
These two modes with virtuality $m_Q^2$ and 
$\LamQCD^2$, respectively, will be the relevant modes for the subsequent 
analysis.

\subsection{Construction of bHQET}
\label{sec:ConstbHQET}
In HQET, the heavy quark is close to its mass-shell, consequently it only interacts with soft gluons. When changing the reference frame the soft modes will acquire a preferred direction, giving rise to soft-collinear modes, as seen for the spectator quark in~\eqref{eq:finscal}. In the boosted frame, the heavy quark momentum can be parametrized as
\begin{equation}
p_Q^\mu = \mb v^\mu + k^\mu\,,
\end{equation}
where $v\sim (1/\lam,1,\lam)$ and $k\sim (1/\lam,1,\lam)\LamQCD$ is soft-collinear. This scaling differs from HQET which is usually defined in a frame ``close" to the heavy-meson rest frame, implicitly assuming $v \sim (1,1,1)$ and $\lam \sim 1$.

We derive the bHQET Lagrangian starting from the HQET one. In Appendix~\ref{sec:appSCET} following \cite{Dai:2021mxb}, we give the equivalent derivation starting from massive SCET. The leading-power (in $\g$) HQET Lagrangian~\cite{Georgi:1990um,Grinstein:1990mj} reads
\begin{equation}
\label{eq:HQETLagrangian}
\mathcal{L}_{\rm HQET} = \bar{h}_v(x) i v \cdot D h_v(x)\,, 
\end{equation}
with field $h_v$ defined as
\begin{equation}\label{eq:hvdef}
    h_v(x) = e^{i\mb v \cdot x} \frac{1+\slashed{v}}{2}Q(x)\,,
\end{equation}
while $Q(x)$ is the heavy-quark field defined in full QCD.
We define the bHQET field $h_n$ as \cite{Dai:2021mxb}
\begin{equation}
\label{eq:hndefinition}
h_n(x)\equiv \sqrt{\frac{2}{\minus{v}}}e^{i \mb v \cdot x}\frac{\slashed{n}_-\slashed{n}_+}{4}Q(x)\,,
\end{equation}
which projects $Q(x)$ onto its large component $\xi(x) = \frac{\slashed{n}_- \slashed{n}_+}{4}Q(x)$, defined as in SCET, and then subtracts the rapid phase variations from the large momentum piece $m_Q v$  through the exponential factor as in HQET, leaving the residual momentum $k$. The normalization factor in \eqref{eq:hndefinition} is chosen such that the $h_n$ field will have a scaling independent of the boost as shown below.

Starting from the definition of $h_v$ in \eqref{eq:hvdef}, employing the definition of the bHQET field $h_n$ in \eqref{eq:hndefinition}, as well as the known relations between SCET spinors (given in Appendix~\ref{sec:appSCET}), we find
\begin{equation}
\label{eq:hvhnExact}
h_v(x)=\sqrt{\frac{\minus{v}}{2}}\frac{1+\slashed{v}}{2}\Bigl(1-\frac{\slashed{n}_+}{2}\frac{i\slashed{D}_\perp + \mb \slashed{v}_\perp -\mb}{i\minus{D} + \mb \minus{v}}\Bigr)h_n(x)\,,
\end{equation}
which is an exact relation. We note that the covariant derivative $D^\mu$ scales as $k^\mu$, i.e. like a soft-collinear momentum.

Making explicit use of the scaling of the velocity, it is instructive to expand this result in powers of $\lam$ and $\g$:
\begin{align}
\label{eq:hvhn}
h_v(x)&= \sqrt{\frac{\minus{v}}{2}}\biggl[1 + \biggl(\frac{1+\slashed{v}_\perp}{\minus{v}}\frac{\slashed{n}_+}{2} - \frac{i \slashed{D}_\perp}{2\mb} -\frac{1-\slashed{v}_\perp}{2}\frac{i \minus{D}}{\mb \minus{v}} \biggr) + \mathcal{O}(\g^2,\g \lam) \biggr]h_n(x) \nonumber \\ 
&=\sqrt{\frac{\minus{v}}{2}}\biggl[1 + \sum_{k=1}^{\infty}\mathcal{O}(\g^k)+ \frac{\slashed{n}_+}{2}\mathcal{O}(\lam)\sum_{k=0}^\infty\mathcal{O}(\g^k)\biggr]h_n(x)\,.
\end{align}
This relation between the HQET field and the bHQET field, both in the boosted frame $S$, shows that both are equal at leading power in $\lam$ and $\g$ up to a rescaling factor $\sqrt{\minus{v}/2}$.
We observe that, to all orders in the expansion, only power corrections  linear in $\lam$ arise. This is due to the fact that the relative scaling between $i\minus{D}$ and $\mb \minus{v}$ in the denominator of~\eqref{eq:hvhnExact} is of order $\g$, therefore no corrections of order $\lam$ other than those present from the numerator are generated. The same holds for the relation between the QCD spinor and the collinear SCET spinor.

To obtain the bHQET Lagrangian from the HQET Lagrangian requires the subleading power terms in \eqref{eq:hvhn} as well, because the leading order terms vanish due to the projection properties of $h_n$, i.e.
\begin{equation}
\frac{\slashed{n}_-\slashed{n}_+}{4}h_n = h_n\,, \qquad \slashed{n}_- h_n = 0 \,.
\end{equation}
Employing \eqref{eq:hvhn}, we find 
\begin{eqnarray}
\mathcal{L}_{\rm HQET} &=& \bar{h}_v(x) i v \cdot D h_v(x) = \frac{\minus{v}}{2}\bar{h}_n(x)\Bigl(1+\frac{1-\slashed{v}_\perp}{\minus{v}}\frac{\slashed{n}_+}{2} \Bigr) i v \cdot D \Bigl(1+\frac{1+\slashed{v}_\perp}{\minus{v}}\frac{\slashed{n}_+}{2} \Bigr) h_n(x)
\nonumber \\
&=& \bar{h}_n(x)i v \cdot D \frac{\slashed{n}_+}{2} h_n(x)\Bigl(1 + \mathcal{O}(\g)\Bigr) = \mathcal{L}_{\rm bHQET}\Bigl(1 + \mathcal{O}(\g)\Bigr)\,,
\label{eq:bHQETLfromHQET}
\end{eqnarray}
in agreement with~\cite{Dai:2021mxb}.

We remark that the Lagrangian has no corrections of $\mathcal{O}(\lam)$, in analogy with the derivation of the SCET collinear Lagrangian from the QCD one. 
Furthermore, we note that the corrections of order $\mathcal{O}(\g)$ arise only because of the definition of the bHQET field $h_n$ adopted in~\eqref{eq:hndefinition}. This definition is practical because it is directly connected to SCET, making matching computations simpler. However, because there is no projector $(1+\slashed{v})/2$ acting on $Q(x)$, it fails to completely eliminate the small component of the QCD spinor, giving rise to the series of corrections in $\g$ in the relation~\eqref{eq:hvhn}. 

Alternatively, one could choose to define $h_n$ as
\begin{equation}
\label{eq:newhn}
    h_n^{\rm new}(x) \equiv \sqrt{\frac{2}{\minus{v}}}\frac{\slashed{n}_-\slashed{n}_+}{4}h_v(x) = \sqrt{\frac{2}{\minus{v}}}e^{i\mb v\cdot x}\frac{\slashed{n}_-\slashed{n}_+}{4}\frac{1+\slashed{v}}{2}Q(x)\,,
\end{equation}
instead of \eqref{eq:hndefinition}, which would lead to 
\begin{equation}
    \frac{\slashed{n}_+\slashed{n}_-}{4}h_v(x) = \sqrt{\frac{\minus{v}}{2}}\frac{1+\slashed{v}_\perp}{\minus{v}}\frac{\slashed{n}_+}{2}h_n^{\rm new}(x)\,.
\end{equation}
Then one obtains only the linear corrections of $\mathcal{O}(\lam)$ in the analogue of~\eqref{eq:hvhn},
\begin{equation}
  h_v(x) = \biggl(\frac{\slashed{n}_-\slashed{n}_+}{4}+\frac{\slashed{n}_+\slashed{n}_-}{4}\biggr)h_v(x) = \sqrt{\frac{\minus{v}}{2}}\biggl[1+\frac{1+\slashed{v}_\perp}{\minus{v}}\frac{\slashed{n}_+}{2} \biggr]h_n^{\rm new}(x)\,,  
\end{equation}
and the Lagrangians are identical, $\mathcal{L}_{\rm HQET} = \mathcal{L}_{\rm bHQET}$. The difference in the definition of the boosted HQET field would not affect the leading-power theory, but would result in a reshuffle of the power corrections.

It is then clear that, depending on the definition chosen for $h_n$, the relation between the leading-power Lagrangians~\eqref{eq:bHQETLfromHQET} could suffer from power corrections, but that including all the power corrections in~\eqref{eq:HQETLagrangian}, and correspondingly in \eqref{eq:bHQETLfromHQET}, would render the two theories equivalent, as required by boost invariance.
In fact, one could avoid building the bHQET Lagrangian with the new field $h_n$, by simply using standard HQET with its Feynman rules \cite{Fleming:2007qr}\cite{Fleming:2007xt}. This would lead to the correct result, but at the price of having an inhomogeneous power counting in the parameter $\lam$, because the field $h_v$ contains both, the large and the small SCET spinors.

\subsection{Systematics of the bHQET Expansion}
\label{sec:systematicsbHQET}

The scaling of the bHQET field \eqref{eq:hndefinition} can be easily derived from 
requiring that the kinetic term in the action~\eqref{eq:bHQETLfromHQET}
\begin{equation}
   \int d^4x\; \bar{h}_n(x) iv \cdot \partial \frac{\slashed{n}_+}{2} h_n(x) \sim 1\,.
\end{equation}
Since $h_n$ has soft-collinear space-time variations, the derivative 
scales like a soft-collinear momentum, hence $i v \cdot D \sim \LamQCD$.  
Together with $d^4x \sim \LamQCD^{-4}$ we find
\begin{equation}
h_n \sim \LamQCD^{3/2}\,,
\end{equation}
which is independent on the boost $\lam$. 
Some care needs to be taken in applying the same reasoning to the HQET Lagrangian in the boosted frame. Due the projection properties of $h_v$ we need to take into account that the boost suppresses the combination $\slashed{n}_- h_v$,
\begin{equation}
    \frac{\slashed{n}_-}{2}h_v = \frac{1}{\minus{v}}\Bigl(1-\slashed{v}_\perp + \mathcal{O}(\lam)\Bigr)h_v \sim \mathcal{O}(\lam)h_v\,,
\end{equation}
resulting in the Lagrangian
\begin{equation}
\bar{h}_v(x) iv \cdot \partial h_v(x) = \bar{h}_v(x) \biggl(\frac{\slashed{n}_-\slashed{n}_+}{4} + \frac{\slashed{n}_+\slashed{n}_-}{4}\biggr)iv \cdot \partial h_v(x) =\frac{2}{\minus{v}}  \bar{h}_v(x) iv \cdot \partial \frac{\slashed{n}_+}{2} h_v(x)\,,
\end{equation}
which gives the following scaling for the field
\begin{equation}
h_v \sim \sqrt{\minus{v}}\LamQCD^{3/2} \,.
\end{equation}
The scaling of the bHQET field agrees with the scaling of the HQET field in the 
rest frame, but differs from the scaling of the HQET field in a general frame, consistent with~\eqref{eq:hvhn}.
In the usual HQET frame, $\minus{v}\sim 1$, such that the heavy-quark field has the known soft scaling.

To describe the light anti-quark in the boosted frame we have to introduce the soft-collinear field $\xi_{sc}$, with the same projection properties as the standard collinear spinors in SCET. It is hence natural to split the QCD field $q(x)$ into large and small components $\xi_{sc}$ and $\eta_{sc}$, 
\begin{equation}
    \xi_{sc}(x) = \frac{\slashed{n}_-\slashed{n}_+}{4}q(x)\,, \qquad \eta_{sc}(x) = \frac{\slashed{n}_+\slashed{n}_-}{4}q(x) = \frac{i \slashed{D}_\perp}{i\minus{D}}\frac{\slashed{n}_+}{2}\xi_{sc}(x)\,,
\end{equation}
and integrate out the $\eta_{sc}$ field from the Lagrangian in order to have an homogeneous theory in $\lam$. The result is the usual collinear SCET Lagrangian, but with the soft-collinear spinor $\xi_{sc}$.
The scaling of the soft-collinear field can also be obtained from any of the kinetic terms in the action by requiring, for example, 
\begin{equation}
    \int d^4x\; \bar{\xi}_{sc}(x)i\plus{D} \frac{\slashed{n}_+}{2}\xi_{sc}(x) \sim 1\,.
\end{equation}
With $i\plus{D}\sim \lam\LamQCD$ from the scaling of the soft-collinear momentum and $d^4 x\sim \LamQCD^{-4}$, this results in 
\begin{equation}
    \xi_{sc}(x) \sim \sqrt{\frac{1}{\lam}}\LamQCD^{3/2}\,, \qquad \eta_{sc}(x) \sim \sqrt{\lam}\LamQCD^{3/2}\,.
\end{equation}

Before concluding this section we want to briefly emphasize the fact that, due to the presence of the reference vectors $n_\pm$, bHQET is reparametrization invariant (RPI) under the same set of transformations as SCET~\cite{Manohar:2002fd}. 
This is crucial in building a complete set of operators at a given power in $\lam$, because bHQET contains the large parameter $\minus{v}$ and one must control the way it can appear in operators to all orders.
In the following we want to systematically derive a basis for two-particle operators at order $\LamQCD^3$.

Transformations of type-III are the most relevant ones here since they control how $n_+$ and $n_-$ can enter the operators. The type-III transformations 
\begin{equation}
    n_- \to \alpha\, n_-\,, \qquad n_+ \to \frac{1}{\alpha}\,n_+ 
\quad \mbox{($\alpha$ real)}\,, 
\end{equation}
 correspond to boosts which leave the perpendicular components of Lorentz vectors as~\eqref{eq:boost} unchanged, so that we can safely set $v_\perp = 0$ from now on.  
It follows from  $v^2=1$ that $\plus{v}=1/\minus{v}$.
The $\xi$ fields are invariant under this transformation, whereas, due to the prefactor in~\eqref{eq:hndefinition}, the bHQET field transforms as
\begin{equation}
    h_n \to \sqrt{\alpha}\,h_n\,.
\end{equation}
It is therefore advantageous to employ the type-III reparametrization-invariant building blocks
\begin{equation}
    \sqrt{\frac{\minus{v}}{2}}h_n \sim b^{-1/2} \LamQCD^{3/2}\,, \qquad \xi_{sc} \sim b^{-1/2}\LamQCD^{3/2}\,,
\end{equation}
to build the most general two-particle operator with power counting $\mathcal{O}(\LamQCD^3)$. Without loss of generality such operators can be written as
\begin{equation}
\mathcal{\hat{O}} = \frac{1}{\minus{v}}\sqrt{\frac{\minus{v}}{2}}\bar{h}_n \slashed{n}_+ f(D^\mu,v^\nu,\Gamma_\perp) \gamma^5 \xi_{sc} \sim \LamQCD^3 \times f(D^\mu,v^\nu,\Gamma_\perp)\,,
\end{equation}
where $f(D^\mu,v^\nu,\Gamma_\perp)$ is a 
dimensionless function of covariant-derivative components, 
$n_\pm v$ and a Dirac structure $\Gamma_\perp$. We used the fact that the only possible Dirac structure must contain one $\slashed{n}_+$ for the projection properties of the spinors, and further explicitly factored out a $\gamma^5$ to ensure non-zero overlap with pseudoscalar mesons. It follows that the remaining matrix $\Gamma_\perp$ cannot contain $\slashed{n}_+$ or $\slashed{n}_-$. With the chosen 
prefactor, the leading-power operators require that $f(D^\mu,v^\nu,\Gamma_\perp)
$ is type-III reparametrization-invariant and scales as $\mathcal{O}(1)$.

Since $f(D^\mu,v^\nu,\Gamma_\perp)$ must also be a Lorentz scalar, we are 
left with the two dimensionless and reparametrization-invariant building blocks
\begin{equation}
\label{eq:RPIIIIinvariants}
\minus{v}\frac{i \slashed{D}_\perp}{i \minus{D}} \sim 1\,, \qquad \frac{\minus{v}}{\plus{v}} \frac{i \plus{D}}{i \minus{D}} \sim 1\,,
\end{equation}
where we consider only derivatives on the light-quark field due to integration by parts, and neglect commutators of covariant derivatives since they generate three-particle operators. However, the second building block in~\eqref{eq:RPIIIIinvariants} is not independent, since it is related to two insertions of the first one by the equation of motion of the light-quark field,
\begin{equation}
\label{eq:lighteom}
    \slashed{n}_+ i\slashed{D}_\perp \frac{1}{i\minus{D}}i\slashed{D}_\perp \xi_{sc} = -\slashed{n}_+ i\plus{D} \xi_{sc}\,.
\end{equation}
Hence, the general $f(D^\mu,v^\nu,\Gamma_\perp)$ is an arbitrary power of the first structure in~\eqref{eq:RPIIIIinvariants}, resulting in the infinite tower of operators\footnote{Allowing for $v_\perp \neq 0$, type-I RPI transformations constrain the 
function $f$ to be $f(D^\mu,v^\nu,\Gamma_\perp) = (\slashed{v}_\perp - \minus{v} \frac{i \slashed{D}_\perp}{i \minus{D}})^k$.} 
\begin{equation}
\mathcal{\hat{O}}_k = \frac{1}{\minus{v}}\sqrt{\frac{\minus{v}}{2}}\bar{h}_n  \slashed{n}_+ \Bigl(\minus{v} \frac{i \slashed{D}_\perp}{i \minus{D}}\Bigr)^{\!k} \gamma^5 \xi_{sc} \sim \LamQCD^3\,.
\end{equation}
Now we analyze the operator $\mathcal{\hat{O}}_2$ and we apply the equation of motion \eqref{eq:lighteom} and the one for the heavy quark in the form
\begin{equation}
\minus{v}\;\bar{h}_n \slashed{n}_+ i\plus{\overleftarrow{D}} = -\frac{1}{\minus{v}}\;\bar{h}_n \slashed{n}_+ i\minus{\overleftarrow{D}}\,,
\end{equation}
so that, by using integration by parts, we get
\begin{align}
\mathcal{\hat{O}}_2 &= \sqrt{\frac{\minus{v}}{2}}\bar{h}_n\; \slashed{n}_+ \frac{\minus{v}}{i\minus{D}}i\slashed{D}_\perp \frac{1}{i\minus{D}}i\slashed{D}_\perp \gamma^5\; \xi_{sc}\nonumber\\
&= -\sqrt{\frac{\minus{v}}{2}}\bar{h}_n\; \slashed{n}_+  \frac{\minus{v}}{i\minus{D}}i\plus{D} \gamma^5\;\xi_{sc} = \mathcal{\hat{O}}_0\,.
\end{align}
Hence, the operator basis closes on $\hat{\mathcal{O}}_0, \hat{\mathcal{O}}_1$.
The second operator $\hat{\mathcal{O}}_1$ is related to the structure $\bar{h}_n \eta_{sc}$. Both are leading-power operators, which can be understood from the fact that in the rest frame they account for the two independent Dirac structures $\bar{h}_v \slashed{n}_+ q_s$ and $\bar{h}_v \slashed{n}_- q_s$.

\subsection{Matching Preliminaries}

This section serves as the starting point for the core of our computation: the matching of the LCDA, described in Sections~\ref{sec:Matching} and~\ref{sec:tail}.
Before starting we need to properly define our objects within the framework of SCET.

The two-particle operator that defines the QCD LCDA, is represented in 
massive SCET by the operator 
\begin{equation}
\label{eq:opSCETI}
\mathcal{O}_{C}(u) = \int \frac{dt}{2\pi} e^{-i u t \minus{p}} \bar{\chi}_C^{(Q)}(0)\,\slashed{n}_+ \gamma^5 \,\chi_C(tn_+) \,.
\end{equation}
We have taken the Fourier transform, and $p$, once a matrix element is taken, will be the heavy meson momentum $p_H$, or 
the sum of the parton momenta in partonic computations. We denoted the massive field with the $Q$ superscript. 
In SCET it is customary to work with collinear gauge-invariant building blocks for (hard-)collinear quark fields,
\begin{equation}\label{eq:chidef}
\chi_C(x) = W_C^\dagger(x) \xi_C(x)\,,
\end{equation}
where $W_C$ is a hard-collinear Wilson line (analogous definition for soft-collinear fields). The two Wilson lines in the fields in~\eqref{eq:opSCETI} combine to the standard finite-distance Wilson line $[0,tn_+]$ entering~\eqref{eq:QCDLCDA}.

The LCDA is defined such that when taking the matrix element between the heavy meson and vacuum states we get\footnote{We define the QCD LCDA such that the light anti-quark carries the light-cone momentum fraction $u$, while the usual definition assigns $u$ to the quark. We do this in order to establish a more intuitive relation with HQET where the light anti-quark carries momentum $\omega$.}
\begin{equation}
\label{eq:QCDLCDAdef}
\langle \B(\pB) | \mathcal{O}_C(u)|0\rangle = -i f_H \phi(u;\mu)\,.
\end{equation}
The shape of $\phi(u;\mu)$ depends on the renormalization scale $\mu$ at which the matrix element is evaluated. For very large scales $\mu \gg \mb$ the LCDA will tend to its asymptotic form, which is symmetric under the exchange of $u$ and $1-u$. On the other hand for scales $\mu \lesssim \mb$ it will develop an asymmetric peak due to the mass difference of its constituent quarks, and from now on we will focus on this situation, with the matching scale set at $\mu \sim \mb$, dropping it from the arguments of $\phi(u)$.
It is important to note that in the heavy meson the light anti-quark carries only a small fraction $\LamQCD/\mB$ of the total light-cone momentum at the matching scale.
This means that values of $u$ of $\mathcal{O}(1)$ must be suppressed in the LCDA. Therefore we expect the function to assume different scalings in the peak ($u \sim \g$) and in the tail ($u \sim 1$) regions,
implying an inhomogeneous function in the power counting parameter $\g$~\cite{Beneke:2000ry}
\begin{equation}
\label{eq:phiscalings}
\phi(u) \sim \begin{cases}
    \,\g^{-1} \,, &\text{ for $u\sim \g$ $\quad$(``peak")} \\[0.1cm]
    \,1 \,, &\text{ for $u\sim 1$ $\quad$(``tail")}
\end{cases}
\end{equation}
where the scalings are fixed by the normalization condition.
Hence, in order to perform a consistent calculation at leading power in $\g$, we match the two regions (peak and tail) of the function separately (Sections~\ref{sec:Matching} and~\ref{sec:tail}) and only at the end combine them in Section~\ref{sec:merging}.

It is instructive to look at the Feynman rules for the insertion of 
$\mathcal{O}_{C}(u)$, 
\begin{align}
\label{eq:feynruleV}
&\raisebox{-1.4cm}{\includegraphics[]{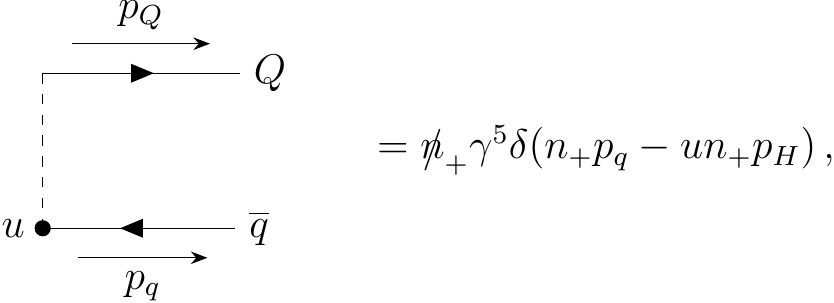}}\\
\label{eq:feynruleWQ}
&\raisebox{-1.4cm}{\includegraphics[]{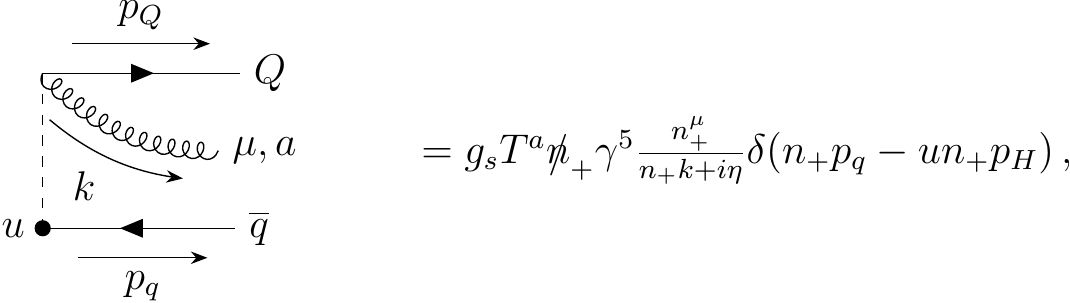}}\\
\label{eq:feynruleWq}
&\raisebox{-1.4cm}{\includegraphics[]{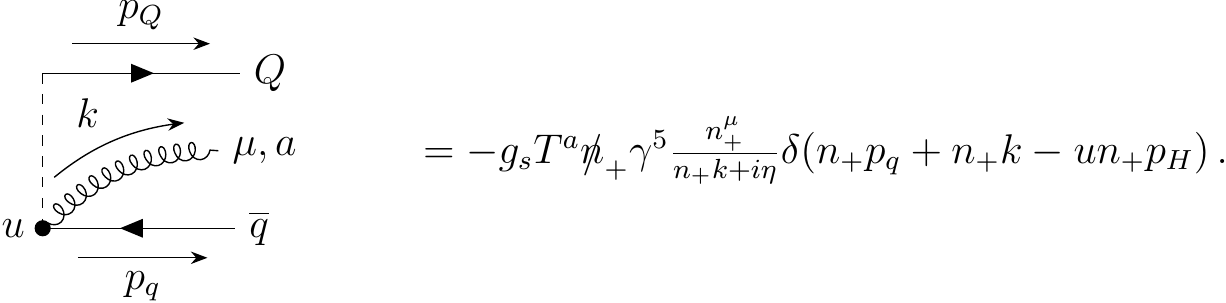}}
\end{align}
The gluon lines attached to the edges of the graph arise from the 
Wilson lines $W_C$ in \eqref{eq:chidef} associated with the heavy quark field $\bar{\chi}_C^{(Q)}(0)$ and the light anti-quark field $\chi_C(tn_+)$, respectively -- attachments of more than one gluon from the expansion of the Wilson line do not contribute 
to the one-loop matrix elements calculated below. The insertion of the 
non-local operator $\mathcal{O}_C(u)$ is represented with the dashed line, 
and the momentum fraction $u$ is injected at the vertex denoted by a dot. 
The delta functions enforce the  $n_-$ component of the sum of momenta 
flowing out from that vertex to take the value $u\minus{\pB}$. 


\section{Matching: Peak Region}
\label{sec:Matching}

The peak region is characterized by small momentum fractions of the light anti-quark in SCET, namely $u\sim \g$, and this scaling will be assumed for the rest of this section.
The matching equation will take the form of a convolution
\begin{equation}
\label{eq:matchingeq}
\mathcal{O}_C(u) = \int_0^\infty d\omega \,\mathcal{J}_p(u,\omega) \mathcal{O}_h(\omega)\,,
\end{equation}
with the bHQET non-local operator defined as 
\begin{equation}
\label{eq:bHQETbasis}
\mathcal{O}_{h}(\omega) \equiv \frac{1}{\mB}\int \frac{dt}{2\pi} e^{-i \omega t\minus{v}}\sqrt{\frac{\minus{v}}{2}}\bar{h}_n(0) W_{sc}(0) \,\slashed{n}_+ \gamma^5 \,\chi_{sc}(tn_+) \sim \frac{\LamQCD^2}{\mB}\,.
\end{equation}
Here $\omega \sim \LamQCD$ is the light-cone component of the spectator anti-quark momentum in the $\B$ rest frame. We note that this is simply the boosted version of the Fourier transform of the HQET operator in~\eqref{eq:phiplusdef}.
The Feynman rules for this operator are analogous to~\eqref{eq:feynruleV}-\eqref{eq:feynruleWq} with the replacement $u\minus{\pB}\to \omega \minus{v}$.
The matrix element of $\mathcal{O}_h(\omega)$ is related to the leading twist HQET LCDA $\varphi_+(\omega)$ at leading power\footnote{There are power corrections to this relation which are generated by our use of $h_n$ rather than $h_n^{\rm new}$ (see discussion at the end of Sec.~\ref{sec:ConstbHQET}), which therefore pose no problem.}
 in $\g$ due to the relation~\eqref{eq:hvhn}
\begin{equation}
\label{eq:HQETLCDAdef}
\langle \B(\pB)| \mathcal{O}_h(\omega)|0\rangle = -i \tilde{f}_H \varphi_+(\omega)\,,
\end{equation}
where $\tilde{f}_H$ denotes the scale-dependent bHQET decay constant given by the local limit of the Fourier transform of~\eqref{eq:bHQETbasis}, which will be studied in Section~\ref{sec:decayconst}. For external states defined in bHQET we use the standard QCD (as opposed to HQET) normalization convention.

It is important to mention that, in the peak region, there is only one bHQET operator contributing to the matching at leading power. Comparing with the bottom-up approach of Section~\ref{sec:systematicsbHQET} the operator $\mathcal{O}_h(\omega)$ corresponds to the non-local version of $\hat{\mathcal{O}}_0$, while the second operator $\hat{\mathcal{O}}_1$ would correspond to a non-local operator related to the subleading-twist HQET LCDA $\varphi_-(\omega)$ which is expected to contribute only through power corrections. This will be confirmed explicitly at the one-loop order through the following calculation.

Taking the hadronic matrix element of the matching equation~\eqref{eq:matchingeq} gives the QCD LCDA in the peak region $u\sim\g$, 
\begin{equation}
    \phi(u) = \frac{\tilde{f}_H}{f_H} \int_0^\infty d\omega\, \mathcal{J}_p(u,\omega) \varphi_+(\omega)\,.
\end{equation}
The long-distance physics described by the HQET LCDA is the same as in QCD, since the effective theory is built such that it reproduces the IR behaviour of the full theory. On the contrary, the UV physics is different, as one can easily notice from their different renormalization group equations (RGE). Both LCDAs are renormalization-scale dependent functions, however the QCD LCDA, which arises from a hard process, depends on a scale $\mu_h \gtrsim \mb$ (because the evolution is computed setting $m_Q=0$), 
while the HQET LCDA, describing the pure non-perturbative physics, should be evaluated at the scale $\mu_s > \mathcal{O}(\LamQCD)$ and $\mu_s \lesssim \mb$. 
We perform the matching at the hard-collinear scale $\mu \sim \mathcal{O}(\mb)$ such that no large logarithms $\ln \mu/\mb$ appear in the perturbative expansion of the hard-collinear matching function $\mathcal{J}_p$. The well-known evolution equations for the two LCDAs~\cite{Lepage:1979zb,Lange:2003ff} can then be used to relate the HQET LCDA at the 
low scale $\mu_s$ to the QCD LCDA at the scale $\mu_h$, thereby summing the large logarithms $\ln \mu_s/\mb$ and $\ln \mb/\mu_h$.

\begin{figure}
\centering
\includegraphics[width=0.75\textwidth]{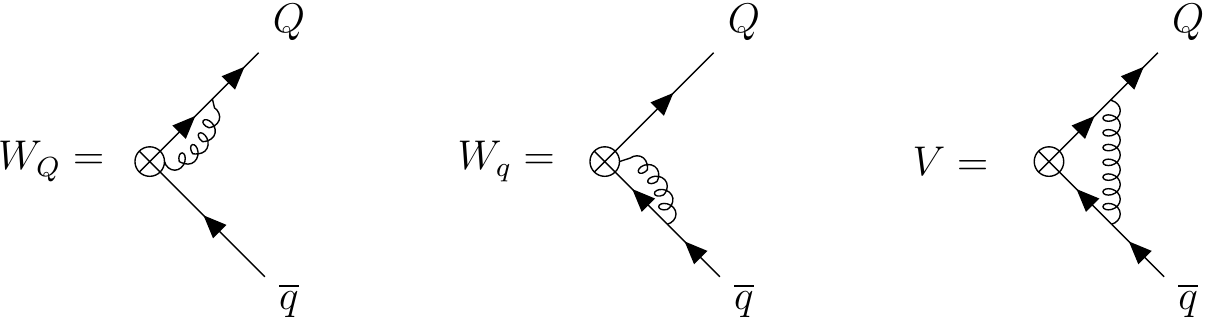}
\caption{\small Contributions to the matching: the operator insertion are represented with $\otimes$. $W_Q$ and $W_q$ stand for interactions between the Wilson lines and the heavy and light quark, respectively. The vertex diagram $V$ represents the gluon exchange between the two constituent quarks. The external field renormalization diagrams are not shown.}
\label{fig:names}
\end{figure}

We compute the perturbative jet function $\mathcal{J}_p(u,\omega)$ at one-loop order by taking the matrix element of the matching equation (\ref{eq:matchingeq}) between on-shell partonic states, namely
\begin{equation}
\label{eq:matchingLCDA}
\langle Q(p_Q) \bar{q}(p_q)| \mathcal{O}_{C}(u)|0\rangle_{\rm SCET} = \int_0^\infty d\omega \, \mathcal{J}_p(u,\omega) \langle Q(p_Q) \bar{q}(p_q)| \mathcal{O}_h(\omega)|0\rangle_{\rm bHQET}\,,
\end{equation}
and evaluating the matrix elements on both sides in the one-loop approximation. The three relevant diagrams are shown in Fig.~\ref{fig:names}, where $\otimes$ denotes the operator insertion, summing up the two Wilson lines contributions from~\eqref{eq:feynruleWQ} and~\eqref{eq:feynruleWq} in the first two diagrams. Defining the momentum fraction 
$s=n_+ p_q/n_+ (p_q+p_Q)\equiv n_+p_q/n_+ p_H$, 
the support of the SCET matrix element on 
the left-hand side implies that $0\leq s\leq 1$.

We begin with the SCET matrix element on the left. Tree-level
evidently implies $u=s$, hence we write to the one-loop order 
\begin{eqnarray}
\langle Q(p_Q) \bar{q}(p_q)| \mathcal{O}_{C}(u)|0\rangle_{\rm SCET} 
\nonumber \\
&& \hspace{-4cm} =\,\frac{1}{\minus{\pB}} \bar{u}(p_Q) \slashed{n}_+ \gamma^5 v(p_q)\,\biggl\{\delta(u-s) +\frac{\alpha_s C_F}{4\pi}M^{(1)}(u,s)\biggr\}\nonumber \\
&& \hspace{-3.6cm} + \,\frac{1}{\plus{\pB}} \bar{u}(p_Q) \slashed{n}_- \gamma^5 v(p_q)\;\;{\mbox{-- term}}\,,
\label{eq:SCET1}
\end{eqnarray}
where the superscript denotes the coefficient of $\as C_F/(4\pi)$ in the 
loop expansion.
In order to make the notation more compact, after performing the SCET computation we re-expressed the result in terms of full QCD spinors by using $\bar{\xi}_{n_-}(p_Q) = \bar{u}(p_Q)\frac{\slashed{n}_+\slashed{n}_-}{4}$ and $\xi_{n_-}(p_q) = \frac{\slashed{n}_-\slashed{n}_+}{4}v(p_q)$, 
in which case 
\begin{align}
    \frac{1}{\minus{\pB}}\bar{\xi}_{n_-}(p_Q) \slashed{n}_+\gamma^5 \xi_{n_-}(p_q) &= \frac{1}{\minus{\pB}}\bar{u}(p_Q) \slashed{n}_+ \gamma^5 v(p_q)\,,\nonumber\\
    \frac{1}{\mB}\bar{\xi}_{n_-}(p_Q) \slashed{n}_+ \frac{\slashed{p}_{q\perp}}{\minus{p_q}}\gamma^5 \xi_{n_-}(p_q) &= \frac{1}{\mB}\bar{u}(p_Q) \slashed{n}_+ \frac{\slashed{p}_{q\perp}}{\minus{p_q}}\gamma^5 v(p_q) = \frac{(-1)}{\plus{\pB}}\bar{u}(p_Q)\slashed{n}_-\gamma^5 v(p_q).
\end{align}
In the second line we also used the equations of motion of both quarks to simplify the Dirac structure (see further discussion in Appendix~\ref{sec:appSCETcomputation})
\begin{equation}
\label{eq:SCETeom}
   \bar{u}(p_Q)\slashed{n}_+ \frac{\slashed{p}_{q\perp}}{\minus{p_q}} \gamma^5 v(p_q) = -\frac{1}{\plus{v}}\bar{u}(p_Q)\slashed{n}_- \gamma^5 v(p_q)\,.
\end{equation}

We focus on the $\bar{u}(p_Q) \slashed{n}_+ \gamma^5 v(p_q)$ 
structure first. 
We regulate both ultraviolet (UV) and infrared (IR) divergences  
dimensionally in space-time dimension $d=4-2\epsilon$. In this case, 
the diagram $W_q$ is scaleless and vanishes. The renormalized one-loop 
on-shell matrix element is then the sum 
\begin{equation}
M^{(1)}(u,s) = V(u,s) + W_Q(u,s) +\frac{1}{2}Z^{OS(1)}_{\xi}\delta(u-s) + Z^{(1)}_{\mathcal{O}_C}(u,s)\,,
\end{equation}
of the remaining two diagrams, the heavy quark on-shell field-strength 
renormalization constant 
\begin{equation}
Z_{\xi}^{OS(1)} = -\frac{3}{\epsilon}-3 \ln \frac{\mu^2}{\mb^2}-4\,,
\end{equation}
and the contribution from the SCET operator renormalization kernel 
(in the $\overline{\rm MS}$ scheme), 
\begin{align}
\label{eq:ZOC}
    Z_{\mathcal{O}_C}^{(1)}(u,s) =& -\frac{2}{\epsilon}\biggl[\theta(s-u)\frac{u}{s}\biggl(1+\frac{1}{s-u}\biggr) + \theta(u-s)\frac{\bar{u}}{\bar{s}}\biggl(1+\frac{1}{u-s}\biggr)\biggr]_{s+}\nonumber \\
    &-\frac{1}{\epsilon}\delta(s-u)\Bigl(3+2\bar{s}\ln s +2s\ln \bar{s}\Bigr)\,,
\end{align}
where we use the standard notation $\bar{x} \equiv 1-x$ for momentum fraction variables. The plus distribution is defined as 
\begin{align}
\label{eq:uplus}
    &\int_0^1 du \,f(u) \Bigl[g(u,s) \Bigr]_{u+} = \int_0^1 du\, \bigl(f(u)-f(s)\bigr)g(u,s)\,,\nonumber\\
    &\int_0^1 ds \,f(s) \Bigl[g(u,s) \Bigr]_{s+} = \int_0^1 ds\, \bigl(f(s)-f(u)\bigr)g(u,s)\,,
\end{align}
where we indicate with a subscript the variable in which the integration is intended. The renormalization function $Z_{\mathcal{O}_C}(u,s)$ is the Efremov-Radyushkin-Brodsky-Lepage (ERBL) renormalization kernel~\cite{Efremov:1979qk,Lepage:1979zb,Lepage:1980fj} but written here as plus distributions with respect to the second argument $s$ instead of the more conventional first one,\footnote{The kernel in terms of a plus distribution in $u$ is provided in Appendix~\ref{sec:appSCETcomputation}.}  $u$. 

By computing the diagrams with a non-dimensional IR regulator, we checked 
that the above renormalization factors remove all UV divergences, as they 
should, such that $M^{(1)}(u,s)$ contains only IR divergences. We find
\begin{align}
M^{(1)}(u,s) = \phantom{+}&2\Biggl[\theta(s-u)\frac{u}{s}\Biggl(\biggl(\frac{1}{s-u}+1\biggr)\biggl(\ln \frac{\bar{s}\mu^2}{u(s-u)\mb^2}+i\pi\biggr) -1\Biggr) \Biggr]_{s+}\nonumber\\
+&2\Biggl[\theta(u-s)\frac{\bar{u}}{\bar{s}(u-s)}\biggl(2\ln \frac{\bar{s}\mu}{(u-s)\mb}+\frac{u}{s}\ln \frac{u \bar{s}}{u-s}\biggr)\Biggr]_{s+} \nonumber\\
+&\delta(u-s)\biggl\{-\frac{1}{\epsilon ^2}-\frac{1}{\epsilon}\biggl(2\ln \frac{\bar{s}\mu}{s\mb} +2 \pi i +\frac{5}{2}\biggr) \nonumber\\
-&2\ln^2 \frac{\mu}{s \mb}+2\ln s \ln \frac{\mu^2}{s \mb^2}+\ln \frac{\mu}{\mb}-4\pi i \ln \frac{\mu}{s \mb} +\frac{11}{12}\pi^2+2\nonumber\\
-&s \left(\ln^2 s + 2\Bigl(1-2\ln \frac{\mu}{s \mb} \Bigr)\ln \frac{\bar{s}}{s}+2\pi i \ln s+\frac{\pi^2}{3}\right)\nonumber\\
-&2 \ln \bar{s} \Bigl(2\ln \frac{\mu}{s \mb}-1\Bigr)+ 2(2-s) \text{Li}_2(s) \biggr\}\nonumber\\
+&2\Biggl[\theta(u-s)\Biggl(\frac{\bar{u}}{\bar{s}}\biggl(2\ln \frac{\bar{s}\mu}{(u-s)\mb}-1\biggr)+\frac{u}{\bar{s}}\ln \frac{u \bar{s}}{u-s}\Biggr) \Biggr]_{s+}\,.
\label{eq:fullSCETresult}
\end{align}
While the above result holds for $u$ and $s$ in the interval $[0,1]$, the 
matching equation \eqref{eq:matchingLCDA} in the peak region 
is valid when the external momentum $p_q$ is soft-collinear, which 
implies that $s\sim \lambda\ll 1$. Expanding \eqref{eq:fullSCETresult} 
for $u, s \sim \lambda$, $M^{(1)}(u,s)$ simplifies to
\begin{eqnarray}
M^{(1)}(u,s)\big|_{u,s\,\ll 1} &=&
2\Biggl[\frac{\theta(s-u)}{s-u}\frac{u}{s}\biggl(\ln \frac{\mu^2}{u(s-u)\mb^2}+i\pi\biggr)\Biggr]_{s+}\nonumber\\
&&\hspace*{-2cm} 
+\,2\Biggl[\frac{\theta(u-s)}{u-s}\biggl(2\ln \frac{\mu}{(u-s)\mb}+\frac{u}{s}\ln \frac{u}{u-s}\biggr)\Biggr]_{s+} \nonumber\\
&&\hspace*{-2cm} 
+\,\delta(u-s)\biggl\{-\frac{1}{\epsilon ^2}-\frac{1}{\epsilon}\biggl(2\ln \frac{\mu}{s\mb} +2 \pi i +\frac{5}{2}\biggr) \nonumber\\
&&\hspace*{-2cm} 
-\,2\ln^2 \frac{\mu}{s \mb}+2\ln s \ln \frac{\mu^2}{s \mb^2}+\ln \frac{\mu}{\mb}-4\pi i \ln \frac{\mu}{s \mb} +\frac{11}{12}\pi^2+2\biggr\}\,.
\label{eq:expandedSCETresult}
\end{eqnarray}
The expanded result could have been obtained directly by noting that when 
$u,s\ll 1$ are assumed from the start, the expansion of the loop integral 
in $\g$ is obtained from 
the sum of a hard-collinear and soft-collinear region in the sense 
of the region expansion \cite{Beneke:1997zp}. The region analysis is 
given in Appendix~\ref{sec:appcalc}, including the  
$\bar{u}(p_Q) \slashed{n}_- \gamma^5 v(p_q)$ 
structure in \eqref{eq:SCET1}, which turns out to be of higher order 
in $\lambda$. This substantiates our earlier statement that the 
non-local version of $\hat{\mathcal{O}}_1$ is suppressed in the peak 
region. 

Turning to the bHQET matrix element on the right-hand side of the 
matching equation in bHQET \eqref{eq:matchingLCDA}, we write it in the form 
\begin{align}\label{eq:bHQET1}
\langle Q(p_Q) \bar{q}(p_q)&| \mathcal{O}_h(\omega)|0\rangle_{\rm bHQET} \nonumber \\
& = \frac{1}{\minus{\pB}}\bar{u}(p_Q) \slashed{n}_+ \gamma^5 v(p_q)\biggl\{\delta\Bigl(\frac{\minus{p_q}}{\minus{v}} - \omega\Bigr) +\frac{\alpha_s C_F}{4\pi}N^{(1)}\Bigl(\omega,\frac{\minus{p_q}}{\minus{v}}\Bigr)\biggr\}\,.
\end{align}
The renormalized one-loop on-shell amplitude
\begin{equation}
N^{(1)}(\omega,\nu) = V_{\rm bHQET}(\omega,\nu) + {W_{Q}}_{\rm bHQET}(\omega,\nu) + Z_{\mathcal{O}_h}^{(1)}(\omega,\nu)\,,
\end{equation}
is again expressed in terms of the diagrams from Fig.~\ref{fig:names} and 
UV renormalization factors. The one-loop HQET on-shell field 
renormalization constant is scaleless, and the 
$\overline{\rm MS}$ operator renormalization kernel
\begin{equation}
Z_{\mathcal{O}_h}^{(1)}(\omega,\nu) = -\frac{2}{\epsilon}\biggl[\frac{\theta(\omega-\nu)}{\omega-\nu} + \frac{\omega}{\nu}\frac{\theta(\nu-\omega)}{\nu-\omega} \biggr]_{\nu+} + \delta(\omega-\nu)\biggl(\frac{1}{\epsilon^2}+\frac{2}{\epsilon}\ln \frac{\mu}{\nu}-\frac{5}{2\epsilon} \biggr)\,,
\end{equation}
coincides with the one for the HQET leading-twist LCDA $\varphi_+(\omega)$ 
\cite{Lange:2003ff}, as expected from boost invariance. We find
\begin{align}
N^{(1)}(\omega,\nu) = \phantom{+}&2\biggl[\frac{\theta(\nu-\omega)}{\nu-\omega}\frac{\omega}{\nu}\biggl(\ln \frac{\mu^2}{\omega(\nu-\omega)}+i\pi\biggr)\biggr]_{\nu+}\nonumber\\
+&2\biggl[\frac{\theta(\omega-\nu)}{\omega-\nu}\biggl(2\ln \frac{\mu}{\omega-\nu}+\frac{\omega}{\nu}\ln \frac{\omega}{\omega-\nu}\biggr)\biggr]_{\nu+}\nonumber\\
+&\delta(\omega-\nu)\biggl(-\frac{1}{\epsilon^2}-\frac{1}{\epsilon}\Bigl(2\ln \frac{\mu}{\nu} + 2\pi i +\frac{5}{2}\Bigr)-4\ln^2 \frac{\mu}{\nu}-4\pi i\ln \frac{\mu}{\nu}+\frac{5\pi^2}{6} \biggr)\,.
\label{eq:bHQET1loopresult}
\end{align}
Details on the individual diagrams with the full $\epsilon$ dependence are provided in Appendix~\ref{sec:appbHQETcomputation}, as well as the renormalization 
kernels with plus distributions with respect to $\omega$.

To extract the matching function $\mathcal{J}_p(u,\omega)$, we expand 
\eqref{eq:matchingLCDA} in $\alpha_s$. At tree level, \eqref{eq:matchingLCDA} 
reduces to
\begin{equation}
\delta\Bigl(\frac{\minus{p_q}}{\minus{\pB}} - u\Bigr) = \int_0^\infty d\omega \, \mathcal{J}_p^{(0)}(u,\omega) \delta\Bigl(\frac{\minus{p_q}}{\minus{v}}-\omega\Bigr)\,. 
\end{equation}
Identifying $p_H=p_Q+p_q$ with $m_H v$, which is legitimate up to power 
corrections, we obtain
\begin{equation}
\label{eq:Jtree}
\mathcal{J}^{(0)}_p(u,\omega) = \delta\Bigl(u-\frac{\omega}{\mB} \Bigr)\theta(\mB-\omega) \,,
\end{equation}
where the theta-function arises from the constraint $u\leq 1$ on the left-hand 
side. At the one-loop order, we then find
\begin{equation}
\label{eq:Joneloop_general}
\mathcal{J}^{(1)}_p(u,\omega) = \theta(\mB-\omega)\biggl\{M^{(1)}\Bigl(u,\frac{\omega}{\mB} \Bigr) - \mB N^{(1)}(u\mB,\omega) \biggr\}\,.
\end{equation}
Comparing \eqref{eq:bHQET1loopresult} and \eqref{eq:expandedSCETresult}, 
we note that their IR poles coincides, hence the matching coefficient 
is indeed short-distance. Moreover, since the simple replacement 
$[...]_{\omega/\mB+} \to [...]_{\omega+}$ holds, the plus-distribution 
terms also coincide and cancel in the difference \eqref{eq:Joneloop_general}, 
leaving 
\begin{equation}
M^{(1)}\Bigl(u,\frac{\omega}{\mB}\Bigr)-\mB N^{(1)}(u\mB,\omega) = \delta\Bigl(u-\frac{\omega}{\mB}\Bigr) \biggl(\frac{L^2}{2}+\frac{L}{2}+\frac{\pi^2}{12}+2\biggr)\,,
\label{eq:WWdiff}
\end{equation}
where we defined
\begin{equation}
L \equiv \ln \frac{\mu^2}{\mB^2}\,,
\end{equation}
identifying the quark mass $\mb$ with the meson mass $\mB$ since the difference is formally a power correction. The result for the one loop jet function in the peak region can be therefore written in the remarkably simple local (in momentum fraction) form
\begin{equation}
\label{eq:Jp}
    \mathcal{J}_p(u,\omega) = \theta(\mB-\omega) \delta\Bigl(u - \frac{\omega}{\mB}\Bigr) \biggl( 1 + \aCFopi \mathcal{J}^{(1)}_{\rm peak} + \mathcal{O}(\alpha_s^2)\biggr) \,,
\end{equation}
with
\begin{equation}
    \mathcal{J}^{(1)}_{\rm peak} \equiv \frac{L^2}{2}+\frac{L}{2}+\frac{\pi^2}{12}+2\,.
\end{equation}

The fact that the non-local terms drop out can be understood in terms of 
the region analysis. The matrix element $N^{(1)}$ accounts precisely for the soft-collinear region of $M^{(1)}$, such that only the hard-collinear region contributes to the matching function $\mathcal{J}_p^{(1)}(u,\omega)$.
Most of the terms will therefore cancel in~\eqref{eq:Joneloop_general}, since only logarithms of the hard-collinear scale $\mB^2$ can appear in the jet function. We find in fact that only one diagram, $W_Q$, contributes to the matching in the peak region, namely when the heavy quark couples to its own Wilson line, giving rise to a simple function proportional to the tree-level jet function~\eqref{eq:Jtree}.
The role of the delta functions in the Feynman rules \eqref{eq:feynruleV}-\eqref{eq:feynruleWq} and the fact that $u$ and $s$ are small in the peak region is crucial, because as soon as the loop momentum $k$ enters the delta function, it is forced to assume the same scaling as $u\minus{\pB}$ and $\minus{p_q}$, resulting in a soft-collinear contribution.
Therefore the only contributions to the jet function will come from diagrams where the hard-collinear loop momentum is not in the delta function argument, 
which happens only in the case of the diagram $W_Q$. 

\begin{figure}
    \centering
    \subfloat[Peak matching]{\includegraphics[width=0.23\textwidth]{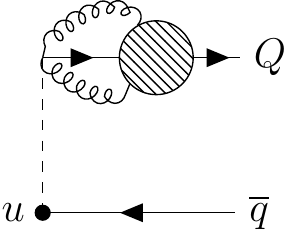}}$\qquad\qquad$
    \subfloat[Tail matching]{\includegraphics[width=0.23\textwidth]{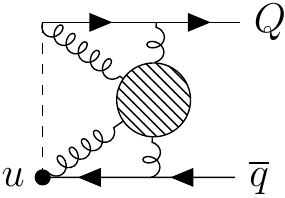}}
    \caption{\small Schematic diagrams for the all orders contributions to the jet function (only hard-collinear gluons), divided into the peak (left) and tail (right) part.}
    \label{fig:diagallord}
\end{figure}

We argue that the hard-collinear matching function is local in 
momentum fractions to all orders in the perturbative expansion, i.e. that 
it can be expressed as
\begin{equation}
\mathcal{J}_p(u,\omega) = \mathcal{J}_{\rm peak}\,
\delta\Bigl(u - \frac{\omega}{\mB}\Bigr) \,\theta(\mB-\omega)\,,
\end{equation}
where $\mathcal{J}_{\rm peak}$ is independent of $u$ and $\omega$. 
In the peak region $u \sim \g$, the only hard-collinear contributions that can arise are those related to diagrams where the spectator quark is not involved (so that $k$ is not in the argument of the delta functions), as shown in the left panel of Fig.~\ref{fig:diagallord}. These contributions are given by gluons emitted and absorbed by the heavy quark field and its local Wilson line, forcing $u$ to match the external momentum fraction $\omega/\mB$  through the delta function~\eqref{eq:feynruleV}. This makes 
it clear that $\mathcal{J}_{\rm peak}$ is in fact the quark-mass 
dependent short-distance coefficient in matching the massive SCET 
collinear quark field to the bHQET field by integrating out the 
hard-collinear modes:
\begin{equation}
\chi_C^{(Q)} = W_C^\dagger \xi_C^{(Q)} \to  \mathcal{J}_{\rm peak}(m_Q) 
\,\sqrt{\frac{\minus{v}}{2}}W_{sc}^\dagger h_n\,.
\end{equation}

This results in the following all-order form for the QCD LCDA in the peak region,
\begin{equation}
\label{eq:phipu}
    \phi_p(u) = \frac{\tilde{f}_H}{f_H} \mathcal{J}_{\rm peak} \mB \varphi_+(u\mB)\,,  \quad\quad {\rm for} \; u\sim \g\,,
\end{equation}
where the subscript $p$ means that this expression for $\phi(u)$ holds in the 
peak region only.
We emphasize that $\mathcal{J}_{\rm peak}$ is a momentum-fraction independent 
matching coefficient, i.e. a pure series in $\alpha_s$ free of large logarithms 
at the hard-collinear scale $\mu\sim \mb$, as is $\tilde{f}_H/f_H$.
Consistency requires that both sides of \eqref{eq:phipu} have the same scale dependence. We further discuss this in Section~\ref{sec:RGE}, where we also derive the RGE for $\mathcal{J}_{\rm peak}$.

\section{Matching: Tail Region}
\label{sec:tail}

We continue with the matching of the tail region of the LCDA, 
characterized by momentum fractions $u\sim 1$, which implies a different 
counting as compared to the peak region. As seen from~\eqref{eq:phiscalings} 
the tail is power suppressed, and of the same order as the LCDA normalization.
Therefore, the matrix element of $\mathcal{O}_C(u)$ will be of the same order 
of magnitude as the decay constant, which in QCD factorization of exclusive 
$B$ meson decays indeed contributes only through power corrections, the leading 
power involving the first inverse moment, $\lambda_B$. Momentum fractions $u\sim 1$ correspond to large values $\omega \gg \LamQCD$, in which case the tail of 
the HQET LCDA can be determined from an operator product expansion (OPE)~\cite{Lee:2005gza} and depends only on local non-perturbative inputs.
This amounts to an expansion in $\LamQCD/\omega$ with a tower of local operators in HQET. In terms of the matching of $\mathcal{O}_C(u)$, this translates into a fully perturbative determination of the $u$ dependence, with the decay constant $\tilde{f}_H$ as the only non-perturbative input at leading power. 
For this reason, it is instructive to first look at the simpler case of matching the local SCET operator instead of $\mathcal{O}_C(u)$, 
and to resume the non-local matching in Section~\ref{sec:nonlocaltail}.

\subsection{Local Matching: the Decay Constant}
\label{sec:decayconst}
As a starting point to understand some of the features present in the matching of the LCDA in the tail region, we proceed by matching the local SCET operator onto bHQET.
We first recall the main points of matching the QCD decay constant onto the HQET one in the heavy-meson rest frame.
The QCD and HQET decay constants are defined through the matrix elements
\begin{align}
\label{eq:decayQCDdef}
\langle \B(\pB)| \bar{Q}\gamma^\alpha \gamma^5  q|0\rangle &= -i f_H p_H^\alpha\,,\nonumber\\
\langle \B(\pB)| \bar{h}_v\gamma^\alpha \gamma^5 q_s|0\rangle &= -i f_H^{\rm HQET}(\mu) \mB v^\alpha\,,
\end{align}
where we are using $\pB = \mB v$ and the standard QCD normalization for the states, so that $F_{\rm stat}(\mu) = \sqrt{\mB} f_H^{\rm HQET}(\mu)$ for the mass-independent static decay constant.\footnote{Since fields are at 
$x=0$, we omit the position argument in this section.}
In HQET, the matching computation gives
\begin{equation}
\label{eq:fmatch}
\bar{Q}\gamma^\alpha \gamma^5 q = C_V(\mu)\, \bar{h}_v \gamma^\alpha  \gamma^5 q_s + C_S(\mu)\, v^\alpha \bar{h}_v \gamma^5 q_s\,,
\end{equation}
with
\begin{align}
\label{eq:CVCS}
C_V(\mu) &= 1- \aCFopi\biggl(\frac{3}{2}\ln \frac{\mu^2}{\mb^2} +4 \biggr) + \mathcal{O}(\as^2)\,,\nonumber\\
C_S(\mu) &= 2\aCFopi+ \mathcal{O}(\as^2)\,.
\end{align}
Multiplying~\eqref{eq:fmatch} by $v_\alpha$ and taking taking the matrix element of~\eqref{eq:fmatch} between vacuum and $\B$ meson states, we obtain the relation between the 
decay constants in QCD and HQET. Using 
\begin{equation}
\langle \B(\pB)| \bar{h}_v \gamma^5 q_s |0\rangle = \langle \B(\pB)| \bar{h}_v \slashed{v} \gamma^5 q_s |0\rangle = -if_H^{\rm HQET}(\mu) \mB \,,
\end{equation}
gives the well known result~\cite{Eichten:1989zv}
\begin{equation}
f_H =f_H^{\rm HQET}(\mu)\Bigl(C_V(\mu) + C_S(\mu) \Bigr) = f_H^{\rm HQET}(\mu)\biggl[1 - \aCFopi\biggl(\frac{3}{2}\ln \frac{\mu^2}{\mb^2}+2\biggr) + \mathcal{O}(\as^2) \biggr]\,.
\end{equation}

We next consider the boosted frame, with $\pB_\perp = 0$, and the matching relation 
for the bHQET decay constant, $\tilde{f}_H$. By contracting~\eqref{eq:decayQCDdef} with $n_{+\alpha}$ and $n_{-\alpha}$, respectively, the QCD decay constant can be expressed in two equivalent ways, as the matrix element of collinear currents in SCET~\cite{Hardmeier:2003ig}. This leads to the definition of the following two SCET operators, 
\begin{align}
    \mathcal{O}_+^C &= \frac{1}{\minus{\pB}}  \bar{\xi}^{(Q)}_C \slashed{n}_+ \gamma^5 \xi_C\,,\nonumber \\
    \mathcal{O}_-^C &= -\frac{1}{\plus{\pB}}  \bar{\xi}^{(Q)}_C \slashed{n}_+ \frac{\mb-i \overleftarrow{\slashed{D}}_\perp}{-i\minus{\overleftarrow{D}}}\frac{i\slashed{D}_\perp}{i \minus{D}} \gamma^5 \xi_C\,,
\end{align}
where we have used the expression for the suppressed $\eta_C$ spinor-field component given in Appendix~\ref{sec:appSCET}. By construction their hadronic matrix elements are  
\begin{equation}
\langle \B(\pB)|\mathcal{O}_\pm^C |0\rangle = -i f_H\,. 
\label{eq:bHQETfH}
\end{equation}
The SCET currents $\mathcal{O}_\pm^C$ are matched to the two dimensional bilinear-operator basis in bHQET, as derived in Section~\ref{sec:systematicsbHQET}, conveniently chosen as 
\begin{align}
\label{eq:localbHQETbasis}
    \mathcal{O}_+ &= \frac{1}{m_H \minus{v}}\sqrt{\frac{\minus{v}}{2}} \bar{h}_n \, \slashed{n}_+ \gamma^5 \xi_{sc}\,,\nonumber\\
    \mathcal{O}_- &= -\frac{1}{m_H}\sqrt{\frac{\minus{v}}{2}} \,\bar{h}_n \, \slashed{n}_+ \frac{i \slashed{D}_\perp}{i\minus{D}}\gamma^5 \xi_{sc}\,,
\end{align}
such that the tree level matching is diagonal
\begin{equation}
\begin{pmatrix}
\mathcal{O}_+^C \\
\mathcal{O}_-^C
\end{pmatrix}
 = \biggl[1 + \aCFopi \begin{pmatrix}
 C_{++}^{(1)} &  C_{+-}^{(1)} \\
 C_{-+}^{(1)} & C_{--}^{(1)}
 \end{pmatrix}+\mathcal{O}(\alpha_s^2)\biggr]
 \begin{pmatrix}
 \mathcal{O}_+\\
 \mathcal{O}_-
 \end{pmatrix}.
\end{equation}
In analogy with HQET, we choose to define the bHQET decay constant $\tilde{f}_H(\mu)$ as the matrix element of 
\begin{equation}
\label{eq:OpmMatEl}
    \langle \B(\pB)| \mathcal{O}_+ |0\rangle = -i \tilde{f}_H(\mu)\,.
\end{equation}
At tree level\footnote{In the following we only need the tree-level matrix element, however, the relation should hold to all orders due to RPI,
as well as the link with the corresponding operator in the rest frame.} the matrix element $\langle \B(\pB)|\mathcal{O}_-|0\rangle$ also equals $-i\tilde{f}_H$ since
\begin{align}
\label{eq:OminMatel}
    &\langle \B(\pB) |\mathcal{O}_-|0\rangle = (1+ \mathcal{O}(\alpha_s))\langle \B(\pB)| \mathcal{O}_-^C |0\rangle = (1+ \mathcal{O}(\alpha_s))\langle \B(\pB)| \mathcal{O}_+^C |0\rangle \nonumber\\
    &= (1+ \mathcal{O}(\alpha_s))\langle \B(\pB)| \mathcal{O}_+ |0\rangle = -i\tilde{f}_H(\mu)(1+ \mathcal{O}(\alpha_s))\,.
\end{align}
To determine the relation between $f_H$ and $\tilde{f}_H(\mu)$ at the one-loop order, it is sufficient to match only one of the 
SCET operators, and we choose the simpler case $\mathcal{O}_+^C$. 
We obtain
\begin{equation}
\label{eq:boostdecaymatch}
 \mathcal{O}_+^C = \biggl(1-\aCFopi\biggl(\frac{3}{2}\ln \frac{\mu^2}{\mb^2}+3\biggr)\biggr)\mathcal{O}_+ + \aCFopi\mathcal{O}_- \,.
\end{equation}
The one-loop coefficients are different from the ones in the rest frame~\eqref{eq:CVCS} because we choose a different, more convenient basis for the boosted frame.
Taking the hadronic matrix element of~\eqref{eq:boostdecaymatch} leads to the relation between the decay constants
\begin{equation}
\label{eq:reldec}
f_H = \tilde{f}_H(\mu)\biggl[1 - \aCFopi\biggl(\frac{3}{2}\ln \frac{\mu^2}{\mb^2}+2\biggr) +\mathcal{O}(\as^2)\biggr]\,,
\end{equation}
which implies $\tilde{f}_H(\mu) = f^{\rm HQET}_H(\mu)$, as was to be expected as a natural consequence of boost invariance. For later use, we define the ratio between the HQET and QCD decay constants as
\begin{equation}
d(\mu) \equiv \frac{\tilde{f}_H(\mu)}{f_H}= 1 + \aCFopi \biggl(\frac{3}{2}L+2 \biggr) + \mathcal{O}(\alpha_s^2) \,,
\label{eq:fbmatch}
\end{equation}
again identifying $\mB$ and $\mb$.

\subsection{Non-Local Matching: the LCDA Tail}
\label{sec:nonlocaltail}

In matching the non-local operator $\mathcal{O}_C(u)$ we will use the same operator basis~\eqref{eq:localbHQETbasis} that we used for the decay constant matching, yielding to the matching equation 
\begin{equation}
\label{eq:matchingtail}
    \mathcal{O}_C(u) = \mathcal{J}_+(u)\mathcal{O}_+ + \mathcal{J}_-(u)\mathcal{O}_-\,.
\end{equation}
The analysis of Section~\ref{sec:systematicsbHQET} ensures that there cannot be other two-particle operators at this order in $\lam$. 
Eq.~\eqref{eq:matchingtail} holds to all orders in $\as$, since the hard-collinear contributions in the tail region are independent on the light anti-quark soft-collinear momentum $p_q$, which would contribute only through power corrections. The right panel in Fig.~\ref{fig:diagallord} shows this diagrammatically. When a hard-collinear gluon is emitted from the spectator quark field $\chi_C(tn_+)$, the momentum fraction $u$ is forced by~\eqref{eq:feynruleV} or~\eqref{eq:feynruleWq} to scale as $u \sim 1$, and the internal lines of the diagram would become insensitive to the external soft-collinear momentum.

At tree level the partonic matrix elements of the local operators read 
\begin{equation}
    \langle Q(p_Q)\bar{q}(p_q)|\mathcal{O}_\pm|0\rangle_{\rm bHQET}=\frac{1}{n_\pm p_H}\bar{u}(p_Q)\slashed{n}_\pm \gamma^5 v(p_q)\,.
\end{equation}
The SCET renormalized matrix element in the tail region does not contain any soft-collinear physics (explicitly checked in Appendix~\ref{sec:apptail}) and starts at the one-loop level
\begin{equation}
    \langle Q(p_Q)\bar{q}(p_q)|\mathcal{O}_C(u) |0\rangle_{\rm SCET} = \aCFopi \sum_\pm M^{(1)}_\pm(u)\langle Q(p_Q)\bar{q}(p_q)|\mathcal{O}_\pm|0\rangle\,.
\end{equation}
In the tail region $M^{(1)}_\pm(u)$ is obtained by expanding the 
left-hand side for small $s$ but not $u$. In this case, we obtain from 
\eqref{eq:fullSCETresult} the $s$-independent expression
\begin{equation}
M_+^{(1)}(u\sim 1,s\sim \g) = \frac{2 \bar{u}}{u}\biggl(2(1+u)\ln \frac{\mu}{u\mb} -2u + 1\biggr)\,,
\label{eq:tailfromfull}
\end{equation}
where we used that the delta function $\delta(u-s)$ never contributes 
when $s\ll u$ and the theta functions become trivial. 

Alternatively, the renormalized one-loop amplitude, when $u \sim 1$ and $p_q \ll p_Q$ (that is, $s\sim \lambda$), is given by the hard-collinear ($hc$) region of the full result. The leading-power term is obtained by setting $p_q$ to zero. We find 
\begin{align}
    M_+^{(1)}(u) &= V_+(u)\Bigl|_{hc} + W_Q(u)\Bigl|_{hc} + Z^{(1)}_{\mathcal{O}_C}(u)\Bigl|_{hc}\,, \nonumber\\
    M_-^{(1)}(u) &= V_-(u)\Bigl|_{hc}\,,
\end{align}
where the contribution proportional to the Dirac structure~\eqref{eq:SCETeom} comes only from the $V$ diagram in Fig.~\eqref{fig:names}. It is UV and IR finite, hence there are also no counterterms. Results for individual diagrams are provided in Appendix~\ref{sec:apptail}.

Taking the matrix element between on-shell partonic states of the matching equation~\eqref{eq:matchingtail} and including now $\mathcal{O}_-$ leads to
\begin{equation}
    \mathcal{J}_\pm(u) = \aCFopi M^{(1)}_\pm(u) + \mathcal{O}(\alpha_s^2)\,,
\end{equation}
with 
\begin{align}
\mathcal{J}^{(1)}_+(u) &= \frac{2\bar{u}}{u}\biggl((1+u)\left[L-2\ln u\right] -2u +1 \biggr)\,,\nonumber\\
\mathcal{J}^{(1)}_-(u) &= 2\bar{u}\,.
\end{align}
The expression for $\mathcal{J}^{(1)}_+(u)$ agrees with \eqref{eq:tailfromfull} as it should.

Because of~\eqref{eq:OminMatel}, when taking the matrix element of~\eqref{eq:matchingtail} between the heavy meson and vacuum states, $\mathcal{J}^{(1)}_+$ and $\mathcal{J}^{(1)}_-$ will be summed, we can hence write
\begin{equation}
   \phi_t(u)= \frac{\tilde{f}_H}{f_H} \mathcal{J}_{\rm tail}(u)\,,\quad\quad\quad {\rm for} \;u\sim 1\,,
\end{equation}
where 
\begin{equation}
    \mathcal{J}_{\rm tail}(u) \equiv \mathcal{J}_+(u) + \mathcal{J}_-(u)=\aCFopi \frac{2\bar{u}}{u}\biggl((1+u)[L-2\ln u] -u +1 \biggr) + \mathcal{O}(\as^2)\,.
\label{eq:Jtailresult}
\end{equation}
The subscript $t$ is a reminder that this expression for $\phi(u)$ holds in the 
tail region only. We emphasize that $\phi(u)$ in the tail region needs no 
information from the HQET LCDA $\varphi_+(\omega)$, since it is related to 
the leading local term in the OPE, which involves only the decay constant, 
which is factored out in the definition of $\phi(u)$.

At this point it is worth remarking that the LCDAs are boost invariant quantities. Therefore the matching can be carried out in the heavy meson rest frame, from QCD to standard HQET.
However, in a physical process involving the production of an energetic heavy meson, the hard matching is naturally performed in the boosted frame, yielding a $\Q$-dependent hard function and a SCET operator. Therefore we performed the matching in the boosted frame using bHQET. As a consistency check, we also performed the matching calculation in the rest frame finding complete agreement with the one in the boosted frame.


\section{Merging of  the Peak and Tail}
\label{sec:merging}

From the peak and the tail matching we obtained 
\begin{equation}
\label{eq:resultofmatch}
\phi(u) =  
\begin{cases}
\phi_p(u) = 
\displaystyle \frac{\tilde{f}_H}{f_H}\mathcal{J}_{\rm peak} \mB \varphi_+(u \mB) \,, \quad\quad\quad {\rm for} \;u\sim\g\,,\\[0.4cm]
\phi_t(u) = 
\displaystyle \frac{\tilde{f}_H}{f_H} \mathcal{J}_{\rm tail}(u)\,, \qquad\qquad\qquad\qquad\hspace*{0.0cm} {\rm for} \;u\sim1\,.
\end{cases}
\end{equation}
Since $\phi(u)$ should be a continuous function on $[0,1]$, the two approximations must 
be equal in the overlap region $\lambda\ll u\ll 1$, which requires
\begin{equation}
\label{eq:overlapregion}
\mathcal{J}_{\rm peak} \mB \varphi_+^{\rm asy}(u \mB) \stackrel{!}{=} 
\mathcal{J}_{\rm tail}(u)\Bigl|_{u \ll 1}\,.
\end{equation}
Here $\varphi_+^{\rm asy}(\omega)$ is the asymptotic form of 
$\varphi_+(\omega)$ for $\omega\gg \LamQCD$, which is independent of 
hadronic parameters at leading power in $\LamQCD/\omega$ (which is implied here 
by the term ``asymptotic form''), and can be computed 
perturbatively~\cite{Lee:2005gza}. This is a crucial property without which 
\eqref{eq:overlapregion} could never hold.

Let us check that \eqref{eq:overlapregion} is satisfied at order $\as$. 
At this order~\cite{Lee:2005gza},
\begin{equation}
\label{eq:phiasy}
\varphi_+^{\rm asy}(\omega) \equiv \frac{\alpha_s C_F}{2\pi \omega}\biggl( \ln \frac{\mu^2}{\omega^2}+1\biggr)\,,
\end{equation}
hence we can use $\mathcal{J}_{\rm peak}=1+\mathcal{O}(\alpha_s)$ in 
\eqref{eq:overlapregion} to obtain for the left-hand side 
\begin{equation}
\mB \varphi_+^{\rm asy}(u\mB) = \frac{\alpha_s C_F}{2\pi u}(L-2\ln u +1)\,,
\end{equation}
which indeed coincides with $\mathcal{J}_{\rm tail}(u)\Bigl|_{u \ll 1}$ from 
\eqref{eq:Jtailresult}.

This shows that $\phi(u)$ is parametrically continuous in the overlap region. It is  
therefore consistent to make it technically continuous by applying a 
``merging function'' $\vartheta(u; \delta,\sigma)$ satisfying 
\begin{equation}
    \vartheta(u;\delta,\sigma)\Bigl|_{u\ll \delta} = 1\,, \qquad \vartheta(u;\delta,\sigma)\Bigl|_{u\gg \delta} = 0\,,
\end{equation}
and to add both expressions as
\begin{equation}
\label{eq:phiu}
\phi(u) = \vartheta(u; \delta,\sigma) \phi_p(u) + \bigl(1-\vartheta(u;\delta,\sigma)\bigr)\phi_t(u)\,,
\end{equation}
where the parameter $\delta$ defines the centre of the overlap region ($\g \ll \delta \ll 1$), and $\sigma$ its width. 
We adopt the form 
\begin{equation}
\vartheta(u;\delta,\sigma) = \frac{1}{1+e^{\frac{u-\delta}{\sigma}}}\,,
\end{equation}
allowing us to choose the limiting case
\begin{equation}
    \vartheta(u;\delta,0) = \theta(\delta-u)\,,
\label{eq:sharpmerging}
\end{equation}
in analytic computations. We set $\sigma \sim \mathcal{O}(10^{-2})$ in the numerical analysis of 
Section~\ref{sec:numerics}.


\section{QCD LCDA and its Properties}
\label{sec:lcdaprop}

We now check that the peak-tail merged expression~\eqref{eq:phiu} for the QCD LCDA is consistent with 
the endpoint behaviour, normalization and evolution equation of the QCD LCDA.

\subsection{Endpoint Behaviour}

The QCD LCDA is expected to vanish linearly at the endpoints $u = 0, 1$. 
The limit $u\to 0$ is evident from~\eqref{eq:phiu} as $\mathcal{J}_{\rm peak}$ 
is independent of $u$, while $\varphi_+(\omega)$ is proportional to $\omega$ 
for small $\omega$~\cite{Grozin:1996pq}. For $u\to 1$, 
we find from \eqref{eq:Jtailresult}
\begin{equation}
\phi(u) \underset{u \to 1}{\longrightarrow} \frac{\alpha_s C_F}{\pi}\bar{u} L \to 0\,,
\end{equation}
proving (at $\mathcal{O}(\alpha_s)$)
that the LCDA vanishes linearly at $u=1$ as it should.

\subsection{Normalization}
\label{sec:norm}

We next verify that the integral of~\eqref{eq:phiu} on $[0,1]$ is correctly 
normalized to 1. For this purpose, we use the merging function 
\eqref{eq:sharpmerging}.\footnote{In this case, $\phi(u)$ is not technically 
continuous, but the discontinuity is power-suppressed in the expansion 
parameters, and hence parametrically absent.}

To calculate the normalization integral, we first recall the expression for the zeroth cut-off moment of the 
HQET LCDA~\cite{Lee:2005gza},
\begin{equation}
\label{eq:M0}
M_0(\Lambda_{\rm UV}) = \int_0^{\Lambda_{\rm UV}} d\omega\, \varphi_+(\omega) = 1 - \aCFopi \biggl(2\ln^2 \frac{\mu}{\Lambda_{\rm UV}} + 2\ln \frac{\mu}{\Lambda_{\rm UV}} + \frac{\pi^2}{12} \biggr) + \mathcal{O}\Bigl(\alpha_s \frac{\LamQCD}{\Lambda_{\rm UV}}, \as^2\Bigr)\,,
\end{equation}
valid for $\Lambda_{\rm UV}\gg \LamQCD$. Since $m_H\delta\gg \LamQCD$, we 
can write  
\begin{eqnarray}
\int_0^1 du\, \phi(u) &=& \int_0^\delta du\, \phi_p(u) + \int_\delta^1 du\, \phi_t(u)
\nonumber \\
&=& \frac{\tilde{f}_H}{f_H} \mathcal{J}_{\rm peak} M_0(m_H\delta)
+\frac{\tilde{f}_H}{f_H} \int_{\delta}^1 du \, \mathcal{J}_{\rm tail}(u)
\nonumber \\
&=&
1+\aCFopi \biggl[M_0^{(1)}(\mB\delta ) + \mathcal{J}^{(1)}_{\rm peak} + d^{(1)} + \int_{\delta}^1 du \, \mathcal{J}^{(1)}_{\rm tail}(u) \biggr] + \mathcal{O}(\alpha_s^2) 
\nonumber \\
&=& 1 + \mathcal{O}\Bigl(\delta,\frac{\g}{\delta}\Bigr) +\mathcal{O}(\alpha_s^2)\,,
\label{eq:norm}
\end{eqnarray}
where $d^{(1)}$ is the coefficient of $\alpha_s C_F/(4\pi)$ in \eqref{eq:fbmatch}. This
shows that the LCDA is indeed properly normalized at leading power in $\delta$ and $\g/\delta$, which constitutes a non-trivial check of our matching procedure.

\subsection{Evolution Equation}
\label{sec:RGE}

We check that $\phi(u)$ from~\eqref{eq:phiu} satisfies the correct ERBL evolution 
equation \cite{Lepage:1979zb,Lepage:1980fj,Efremov:1979qk} of the full QCD LCDA at $\mathcal{O}(\alpha_s)$, 
that is 
\begin{equation}
\mu \frac{d\phi(u)}{d\mu} = -\frac{\alpha_s C_F}{\pi}\int_0^1 dv\, V_{\rm ERBL}(u,v) \phi(v)\equiv \frac{\alpha_s C_F}{\pi} R(u)\,,
\label{eq:ERBLcheck}
\end{equation}
with the kernel easily derived from~\eqref{eq:ZOC} as
\begin{equation}
    V_{\rm ERBL}(u,v) = \frac{\epsilon}{2}Z_{\mathcal{O}_C}^{(1)}(u,v)\,.
\end{equation}
Due to the overall coupling constant, the integral $R(u)$ on the right-hand side 
is needed only at $\mathcal{O}(\alpha_s^0)$, giving
\begin{equation}
R(u) = - \int_0^\delta dv\, \mB V_{\rm ERBL}(u,v) \varphi_+(\mB v)\,,
\end{equation}
with no contribution from the tail at this order. Therefore $v \sim \g$, but $u$ 
can still be small or of order 1. We therefore multiply the kernel by 
$1=\theta(\delta-u) + \theta(u-\delta)$ and approximate it accordingly 
in the two regions:
\begin{eqnarray}
V_{\rm ERBL}(u,v) &\approx&  \theta(\delta-u)\biggl\{-u\biggl[\frac{\theta(v-u)}{v(v-u)} + \frac{\theta(u-v)}{u(u-v)} \biggr]_{v+}- \delta(u-v)\Bigl(\ln u + \frac{3}{2} \Bigr)\biggr\}\nonumber \\
&&-\,\theta(u-\delta) \biggl[\bar{u}\Bigl(1+\frac{1}{u} \Bigr) \biggr]_{v+}\,.
\end{eqnarray}
The convolution of the $u>\delta$ term is simple and results in 
\begin{eqnarray}
&& \theta(u-\delta)\bar{u}\frac{1+u}{u} \int_0^\delta dv \mB\, (\varphi_+(v \mB)-\varphi_+(u\mB)) 
\nonumber\\ 
&&\approx \theta(u-\delta)\bar{u}\frac{1+u}{u} \,M_0(\mB\delta)
\approx \theta(u-\delta)\bar{u}\frac{1+u}{u}\,,
\label{eq:erlblexpanded}
\end{eqnarray}
where we used the fact that $\g \ll \delta \ll 1$, allowing us to use the tree-level expression of $M_0(\mB \delta)$ (which equals 1), as well as neglecting $\varphi_+(u\mB)\ll\varphi_+(v \mB)$ (since $u \gg v$).

To compute the $u<\delta$ contribution to $R(u)$, we recall the evolution equation for the HQET LCDA~\cite{Lange:2003ff}
\begin{equation}
\label{eq:phiplusRGE}
\mu \frac{d\varphi_+(\omega)}{d\mu} = -\frac{\alpha_s C_F}{\pi} \int_0^\infty d\nu\, \Gamma_{\rm LN}(\omega,\nu)\varphi_+(\nu)\,,
\end{equation}
with
\begin{equation}
\Gamma_{\rm LN}(\omega,\nu) = -\omega\biggl[\frac{\theta(\nu-\omega)}{\nu(\nu-\omega)}+\frac{\theta(\omega-\nu)}{\omega(\omega-\nu)}\biggr]_{\nu +} + \delta(\omega-\nu) \Bigl(\ln \frac{\mu}{\omega}-\frac{1}{2} \Bigr)\,.
\end{equation}
Comparing to \eqref{eq:erlblexpanded}, we see that the low-$u$ part of the ERBL kernel can be expressed as
\begin{equation}
\label{eq:VtoGammaLimit}
\theta(\delta-u) V_{\rm ERBL}(u,s) = \theta(\delta-u)\biggl[\mB \Gamma_{\rm LN}(u\mB,v\mB) - \delta(u-v)\Bigl(\ln \frac{\mu}{\mB}+1 \Bigr)\biggr]\,.
\end{equation}
Putting both regions together, we find
\begin{eqnarray}
R(u) &=& \theta(\delta-u)\mB\biggl[-\int_0^{\delta \mB}d\nu \,\Gamma_{\rm LN}(u\mB,\nu)\varphi_+(\nu) + \varphi_+(u\mB)\Bigl(\ln\frac{\mu}{\mB} + 1\Bigr)\biggr]\nonumber\\ 
&&+ \,\theta(u-\delta)\bar{u}\frac{1+u}{u}\,.
\label{eq:Rrhs}
\end{eqnarray}
Now we study the left-hand side of \eqref{eq:ERBLcheck}. Keeping in mind that the derivative of the coupling constant with respect to $\mu$ counts as $\mathcal{O}(\alpha_s^2)$ and therefore can be neglected, we obtain 
\begin{align}
\frac{\pi}{\as C_F} \frac{d\phi(u)}{d\ln \mu} =& \;\theta(\delta-u)\mB \biggl[-\int_0^\infty d\nu\, \Gamma_{\rm LN}(u\mB,\nu)\varphi_+(\nu) \nonumber\\
&+ \frac{\varphi_+(u\mB)}{4}\biggl(\frac{d\mathcal{J}_{\rm peak}^{(1)}}{d\ln \mu} + \frac{d}{d\ln\mu}d^{(1)}\biggr) \biggr]+ \frac{1}{4}\theta(u-\delta) \frac{d\mathcal{J}_{\rm tail}^{(1)}(u)}{d\ln\mu}\nonumber\\
=&\;\theta(\delta-u)\mB \biggl[-\int_0^\infty d\nu\, \Gamma_{\rm LN}(u\mB,\nu)\varphi_+(\nu) + \varphi_+(u\mB)\Bigl(\ln\frac{\mu}{\mB} + 1 \Bigr) \biggr]\nonumber\\
&+ \theta(u-\delta) \,\bar{u}\frac{1+u}{u}\,.
\end{align}
which agrees with the right-hand side of \eqref{eq:ERBLcheck}, 
provided the upper limit of the $\nu$-integral in \eqref{eq:Rrhs} can be 
extended up to infinity.  This is indeed allowed, because 
$\mB \delta \gg \LamQCD$ and including the tail adds only a 
suppressed contribution.

The relation~\eqref{eq:VtoGammaLimit} shows that in the peak region the evolution of the QCD LCDA is the same as the evolution equation in HQET, except by a local term, as was already noted in~\cite{Beneke:2021pkl}.
This turns out to be useful in deriving the RGE for $\mathcal{J}_{\rm peak}$ from~\eqref{eq:phipu}
\begin{equation}
    \frac{d}{d\ln \mu}\mathcal{J}_{\rm peak} = \biggl(\frac{1}{\phi_p(u)} \frac{d\phi_p(u)}{d\ln \mu} - \frac{1}{\varphi_+(u\mB)}\frac{d\varphi_+(u\mB)}{d\ln\mu} -\frac{1}{d(\mu)}\frac{d}{d\ln\mu}d(\mu)\biggr)\mathcal{J}_{\rm peak}\,,
\end{equation}
since the $u$-dependence cancels  between the first two terms. The RGE 
for the matching function in the peak region is indeed local, 
\begin{equation}
    \frac{d}{d\ln \mu}\mathcal{J}_{\rm peak} = \aCFopi \biggl[4\ln \frac{\mu}{m_H}+1\biggr]\mathcal{J}_{\rm peak}  + \mathcal{O}(\as^2) \,,
\end{equation}
as required by consistency. 

\subsection{Comparison with Previous Work}
\label{sec:comparison}

At this point we compare our result for the QCD LCDA 
with the previous work~\cite{Ishaq:2019dst}, which also relates it 
to the universal HQET LCDA, but obtained a substantially different and more 
complicated expression for the analogue of the matching function, 
called $\mathcal{J}_{\rm peak}$ and $\mathcal{J}_{\rm tail}$ here. 

A major difference of the set-up in~\cite{Ishaq:2019dst} is that 
the matching is not split into a peak and tail region, where the 
LCDA at the matching scale has different power counting. Instead, 
the intervals $u \in [0,1] \to \omega \in [0,\infty]$ are smoothly 
mapped by assigning the tree-level matching function
\begin{equation}
\label{eq:DYtreelevel}
\mathcal{J}^{(0)}(u,\omega) = \delta\Bigl(u - \frac{\omega}{\omega + \mb}\Bigr)\,.
\end{equation}
The difficulty with this expression is that the HQET variable 
$\omega$ does not know about $m_Q$, hence the combination 
$\omega+m_Q$ has inevitably inhomogeneous power counting, making 
a systematic scale separation difficult. The inhomogeneity is 
inherited by the one-loop matching function, $Z^{(1)}$ in 
Eq.~(15) of~\cite{Ishaq:2019dst}. The benefits of strict scale 
separations become evident when comparing \eqref{eq:Jp} and 
\eqref{eq:Jtailresult} to  $Z^{(1)}$. 

The interpretation of  $Z^{(1)}$ in terms of the results of 
the present work is made more difficult, since $Z^{(1)}$ appears to include 
the evolution of the HQET LCDA from the soft scale $\mu_H$. 
Furthermore, unlike the present approach, which allows us to 
construct the QCD LCDA itself, as shown in the subsequent section, 
the distributions in  $Z^{(1)}$ require integration in 
both variables  $u$ and $\omega$, and hence allow only the 
computation of moments rather than $\phi(u)$ itself.

\section{QCD LCDA of the $\bar{B}$ and $D$ Meson}
\label{sec:numerics}

In this section, we determine the QCD LCDA of $\bar{B}$ and $D$ mesons from 
\eqref{eq:phiu} in terms of the universal HQET LCDA at the soft scale, 
and evolve it to scales $\mu\gg \mB$. This involves three steps: 1) specifying 
the HQET LCDA input at the soft scale $\mu_s = 1~\si{GeV}$, 
and evolving it to the matching scale of order $m_H$. 2) Matching the QCD 
LCDA to the HQET one at the scale $m_H$, which is the main result of 
this work. c) Evolving the QCD LCDA from $m_H$ to the scale 
$Q$ of the hard scattering process with the standard evolution 
equation.

\subsection{Evolution of $\varphi_+$ from $\mu_s$ to $\mb$}
\label{sec:HQETevol}

For the non-perturbative function $\varphi_+(\omega)$, at the soft scale $\mu_s = 1~\si{GeV}$, we employ different models which need to satisfy the known properties of the HQET LCDA.
Of particular importance for our analysis are the cut-off zeroth moment~\eqref{eq:M0}, the asymptotic behaviour~\eqref{eq:phiasy} and the inverse moment $\lambda_B(\mu_s)$:
\begin{align}
    \int_0^{\Lambda_{\rm UV}} d\omega \, \varphi_+(\omega;\mu_s) &= M_0(\Lambda_{\rm UV})\,,\nonumber\\
    \varphi_+(\omega;\mu_s) &\underset{\omega \gg \LamQCD}{\longrightarrow} \varphi_+^{\rm asy}(\omega;\mu_s)\,,\nonumber\\
    \int_0^\infty \frac{d\omega}{\omega}\varphi_+(\omega;\mu_s) &= \lambda_B^{-1}(\mu_s)\,.
\end{align}
These properties are satisfied by gluing continuously a radiative tail $\varphi_+^{\rm asy}$~\cite{Lee:2005gza} onto normalized models $\varphi_+^{\rm mod}$ as
\begin{equation}
\label{eq:phimus}
    \varphi_+(\omega;\mu_s) = \biggl(1+\frac{\as(\mu_s)C_F}{4\pi} \biggl[\frac{1}{2}-\frac{\pi^2}{12}\biggr] \biggr)\varphi_+^{\rm mod}(\omega;\mu_s) + \theta(\omega-\sqrt{e}\mu_s)\varphi_+^{\rm asy}(\omega;\mu_s) \,,
\end{equation}
with the following properties
\begin{align}
    \int_0^{\infty} d\omega \, \varphi_+^{\rm mod}(\omega;\mu_s) &= 1\,,\nonumber\\
    \omega \varphi_+^{\rm mod}(\omega;\mu_s) &\underset{\omega \to \infty}{\longrightarrow} 0\,,\nonumber\\
    \int_0^\infty \frac{d\omega}{\omega}\varphi_+^{\rm mod}(\omega;\mu_s) &= \frac{1}{\omega_0}\,,
\end{align}
where the relation between $\lambda_B(\mu_s)$ and the parameter $\omega_0$ can be derived from~\eqref{eq:phimus}
\begin{equation}
    \omega_0 = \lambda_B\biggl(1 + \frac{\as(\mu_s)C_F}{4\pi}\biggl( \frac{1}{2} - \frac{\pi^2}{12} -\frac{4}{\sqrt{e}}\frac{\lambda_B}{\mu_s}\biggr) \biggr)\,.
\label{eq:om0}
\end{equation}
We adopt by default the exponential model~\cite{Grozin:1996pq} for $\varphi_+^{\rm mod}$, 
\begin{equation} 
\label{eq:phiexp}
    \varphi_+^{\rm exp}(\omega,\omega_0;\mu_s)=\frac{\omega}{\omega_0^2}e^{-\omega/\omega_0}\,.
\end{equation}
We emphasize that adding the tail to the models at the low scale $\mu_s$ is 
required, because the asymptotic form~\eqref{eq:phiasy} must hold at any scale. 
The exponential model~\eqref{eq:phiexp} without the 
radiative tail is unphysical and not suitable for our purpose.

In order to estimate the systematic uncertainties due to the model dependence in Sec.~\ref{sec:WtoBgamma} below, we employ the three two-parameter models \cite{Beneke:2018wjp}\footnote{The definition \eqref{eq:om0} of $\omega_0$ 
here differs from the one in \cite{Beneke:2018wjp}. In the absence of the 
$\alpha_s$ correction in \eqref{eq:om0}, the present $\omega_0$ coincides 
with $\lambda_B$.}
\begin{align}
\varphi_+^{\rm (I)}(\omega;\mu_s) &= \biggl[1-\beta+ \frac{\beta}{2-\beta} \frac{\omega}{\omega_0} \biggr]\varphi_+^{\rm exp}\bigl(\omega,(1-\beta/2)\omega_0;\mu_s\bigr)\,,\qquad \text{for } \phantom{-}0\leq\beta\leq1\,,\nonumber\\
\varphi_+^{\rm (II)}(\omega;\mu_s) &= \frac{(1+\beta)^\beta}{\Gamma(2+\beta)}\biggl(\frac{\omega}{\omega_0}\biggr)^{\beta}\varphi_+^{\rm exp}\Bigl(\omega,\frac{\omega_0}{1+\beta};\mu_s\Bigr)\,,\hspace*{2.36cm}\text{for }-\frac{1}{2}<\beta<1\,,\nonumber\\
\varphi_+^{\rm (III)}(\omega;\mu_s) &= \frac{\sqrt{\pi}}{2\Gamma(3/2+\beta)}U\Bigl(-\beta,\frac{3}{2}-\beta,(1+2\beta)\frac{\omega}{\omega_0}\Bigr)\nonumber\\
&\phantom{-}\times\varphi_+^{\rm exp}\Bigl(\omega,\frac{\omega_0}{1+2\beta};\mu_s\Bigr)\,,\hspace*{4.87cm}\text{for } \phantom{-}0\leq\beta<\frac{1}{2}\,,
\label{eq:phimodels}
\end{align}
where $U(a,b,z)$ is the confluent hypergeometric function of the second kind.
The three models are generalizations of the exponential model and reduce to it 
for $\beta=0$.

The exponential model is then evolved with leading-logarithmic (LL) evolution\footnote{Strictly speaking, one-loop matching should be combined with NLL evolution. The implementation of NLL evolution is beyond the scope of this work, since the largest uncertainty is anyway from initial condition for the HQET LCDA.}
to the matching scale $\mu$, with analytic solution~\cite{Beneke:2018wjp,Beneke:2022msp}
\begin{equation}
\label{eq:phiexpevol}
\varphi_+^{\rm exp-LL}(\omega;\mu) = e^{V+2a\gamma_E} \Gamma(2+a)\biggl( \frac{\mu_s}{\omega_0}\biggr)^a \frac{\omega}{\omega_0^2}\phantom{,}_1F_1\Bigl(2+a,2,-\frac{\omega}{\omega_0}\Bigr)\,,
\end{equation}
with~\cite{Lee:2005gza}
\begin{align}
a \equiv a(\mu,\mu_s) &= -\int_{\as(\mu_s)}^{\as(\mu)} \frac{d\alpha}{\beta(\alpha)}\Gamma_{\rm cusp}(\alpha) = \frac{\Gamma_0}{2\beta_0}\ln r + \mathcal{O}(\as)  \,,\nonumber\\
V \equiv V(\mu,\mu_s) &= -\int_{\as(\mu_s)}^{\as(\mu)} \frac{d\alpha}{\beta(\alpha)}\biggl[\Gamma_{\rm cusp}(\alpha)\int_{\as(\mu_s)}^{\alpha}\frac{d\alpha'}{\beta(\alpha')}+\gamma_+(\alpha) \biggr]\nonumber\\
&=\frac{\Gamma_0}{4\beta_0^2}\biggl[\frac{4\pi}{\as(\mu_s)}\biggl(-\ln r+1-\frac{1}{r}\biggr)+ \frac{\beta_1}{2\beta_0}\ln^2 r +\frac{2\gamma_0}{\Gamma_0}\beta_0 \ln r \nonumber\\
&\quad+\biggl(\frac{\Gamma_1}{\Gamma_0}-\frac{\beta_1}{\beta_0} \biggr)(\ln r -r+1)\biggr] + \mathcal{O}(\as)\,,
\label{eq:aVdef}
\end{align}
where $r=\as(\mu)/\as(\mu_s)$ and $\gamma_+(\as) =\gamma_0 \as C_F/(4\pi)+\mathcal{O}(\as^2)$ with $\gamma_0=-2C_F$, while the QCD beta function and the cusp anomalous dimensions are defined as
\begin{equation}
\label{eq:betacuspdef}
    \beta(\as) = \mu\frac{d\as}{d\mu} = -2\as \sum_{n=0}^{\infty} \beta_n \biggl(\frac{\as}{4\pi}\biggr)^{n+1}\,, \qquad \Gamma_{\rm cusp}(\as) = \sum_{n=0}^\infty \Gamma_n \biggl(\frac{\as}{4\pi}\biggr)^{n+1}\,,
\end{equation}
with $\beta_0 = 11-\frac{2}{3}n_f$, $\beta_1 = 102-\frac{38}{3}n_f$ and $\Gamma_0=4C_F$, $\Gamma_1 = 4C_F(\frac{67}{3}-\pi^2-\frac{10}{9}n_f)$.

\begin{figure}
    \centering
    \subfloat{\includegraphics[width=0.33\textwidth]{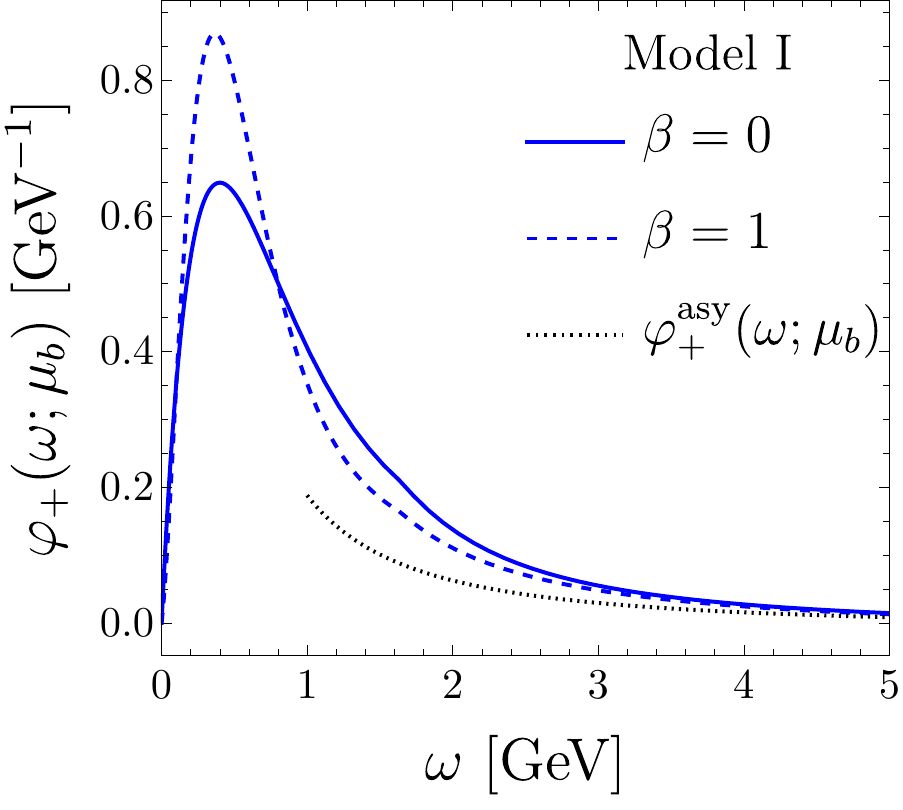}}
    \subfloat{\includegraphics[width=0.33\textwidth]{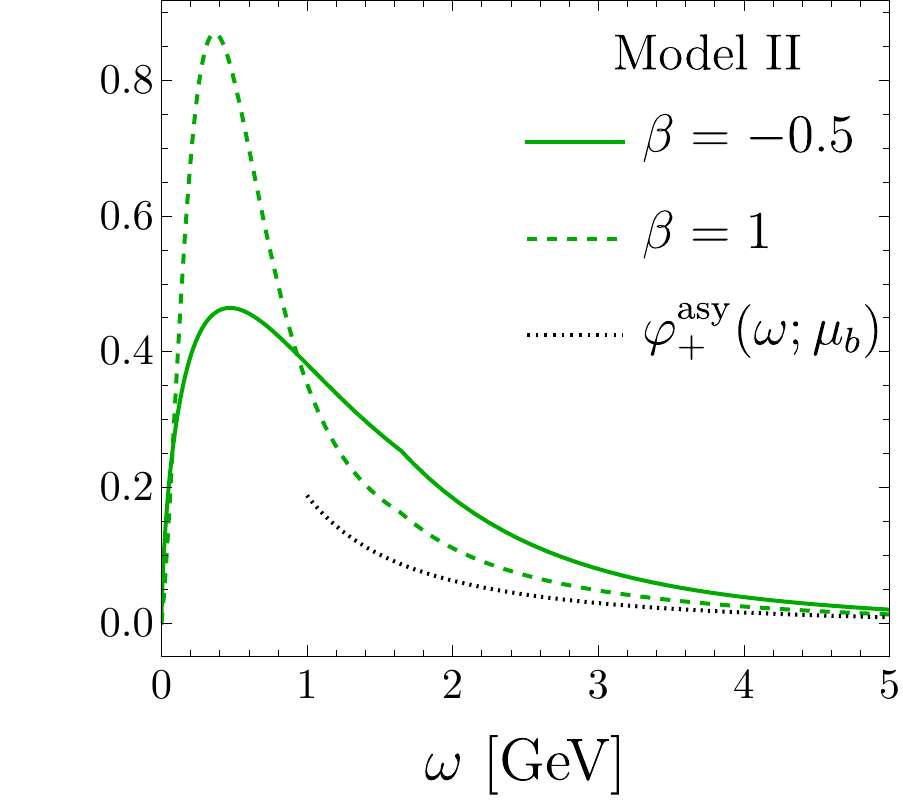}}
    \subfloat{\includegraphics[width=0.33\textwidth]{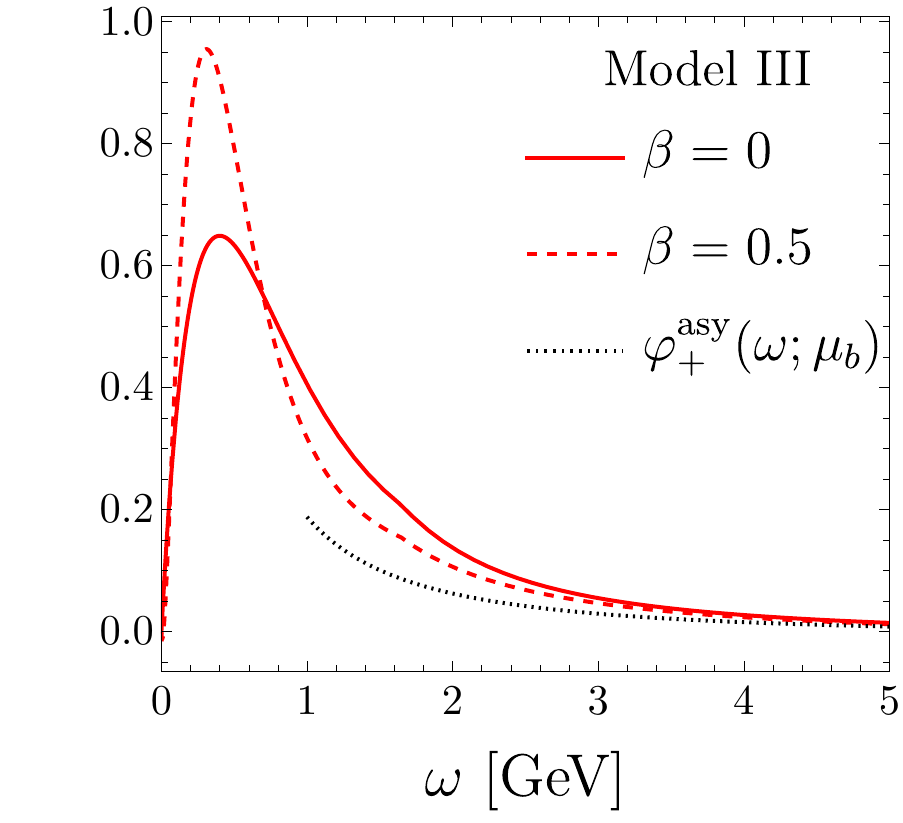}}
    \caption{\small The solid (dashed) curves show three models of $\varphi_+(\omega;\mu_b)$~\eqref{eq:phimodels} at the lower (upper) $\beta$ value. The dotted curve is the asymptotic form of the HQET LCDA~\eqref{eq:phiasy}.}
    \label{fig:HQETmodels}
\end{figure}

Analogous analytical solutions can be found for the three models 
in~\eqref{eq:phimodels}~\cite{Beneke:2018wjp,Beneke:2022msp}.
We do not evolve the radiative tail $\varphi_+^{\rm asy}(\omega;\mu_s)$ (as well as the $\mathcal{O}(\as)$ term) in~\eqref{eq:phimus}, since this would be formally an NLL effect.\footnote{Evolving the tail would result in an asymptotic behaviour for $\varphi_+(\omega;\mu)$ that does not match $\varphi_+^{\rm asy}(\omega;\mu)$.}
In Figure~\ref{fig:HQETmodels} we show for the limiting values of the parameter $\beta$ that the resulting functions $\varphi_+(\omega;\mu_b)$ at the matching scale $\mu_b=4.8~\si{GeV}$ have the correct asymptotic behaviour~\eqref{eq:phiasy}. 
Here and below the inverse moment $\lambda_B(\mu_s)$, which sets the scale-dependent parameter $\omega_0$ of the HQET LCDA, is taken to 
be in the conservative range 
$(350\pm150)~\si{MeV}$~\cite{Beneke:2011nf}. Eq.~\eqref{eq:om0} 
then yields $\omega_0 = 329.5\pm 134.8~\si{MeV}$.

\subsection{Initial Condition of the QCD LCDA at $\mu=\mb$}

The main result of this work consists of constructing the 
initial condition of the QCD LCDA at the matching scale of order of the 
heavy meson mass from \eqref{eq:resultofmatch} and \eqref{eq:phiu} 
in terms of the HQET model~\eqref{eq:phimus} evolved to this scale as 
discussed in the previous subsection.
The numerical illustrations in this and the following subsection 
refer to the simple exponential model, $\beta=0$.

For our numerical studies, we use the meson mass values $m_B = 5.279~\si{GeV}$ and $m_D = 1.870~\si{GeV}$~\cite{Workman:2022ynf}, and fix the matching scales $\mu$ to $\mu_b = 4.8~\si{GeV}$ and $\mu_c = 1.6~\si{GeV}$ for the $\bar{B}$ and the $D$ meson, respectively. The required values for the strong coupling constant are $\alpha_s^{(n_f=5)}(\mu_b)=0.215$ and $\alpha_s^{(n_f=4)}(\mu_c)=0.334$, where three-loop running (from \texttt{RunDec}~\cite{Herren:2017osy}) is used for the evolution from $\as(m_Z) = 0.1179$. We use the meson masses in the matching functions instead of the heavy quark masses for simplicity, since the difference is a power 
correction beyond the leading-power accuracy of the treatment.  
In QCD, we decouple the charm and bottom quarks in the running of $\as$ at $\mu_{\rm{dec},c}=1.279~\si{GeV}$ and $\mu_{\rm{dec},b}=4.163~\si{GeV}$, respectively. In HQET, always $n_f=3$ and $n_f=4$ is employed for the $D$ and $\bar{B}$ meson, respectively. 
We consider the cases of a $\bar{B}$ meson, a $D$ meson and for illustration 
a hypothetical meson $M_{15}$ with mass of 15 GeV, for which we use the matching scale $\mu_{15}=15$ GeV. 
The value of $\delta$ 
entering the merging function~\eqref{eq:phiu} must be chosen such that it 
satisfies $\LamQCD/\mb \ll \delta \ll 1$,
\begin{equation}
    \delta(\bar{B}) = 0.45\,, \qquad \delta(D) = 0.65\,, \qquad \delta(M_{15}) = 0.30\,.
\end{equation}
In the following we will vary $\delta$ by $\pm 15\%$ to obtain a systematic uncertainty on the shape of $\phi(u)$.
The smoothing parameter $\sigma \sim \mathcal{O}(10^{-2})$ is chosen such that it gives a reasonable shape of the function $\vartheta(u;\delta,\sigma)$. Our default values are
\begin{equation}
\label{eq:sigmavals}
     \sigma(\bar{B}) = 0.05\,, \qquad \sigma(D) = 0.05\,, \qquad \sigma(M_{15}) = 0.02\,.
\end{equation}

\begin{figure}[t]
\centering
\subfloat{\includegraphics[width=0.45\textwidth]{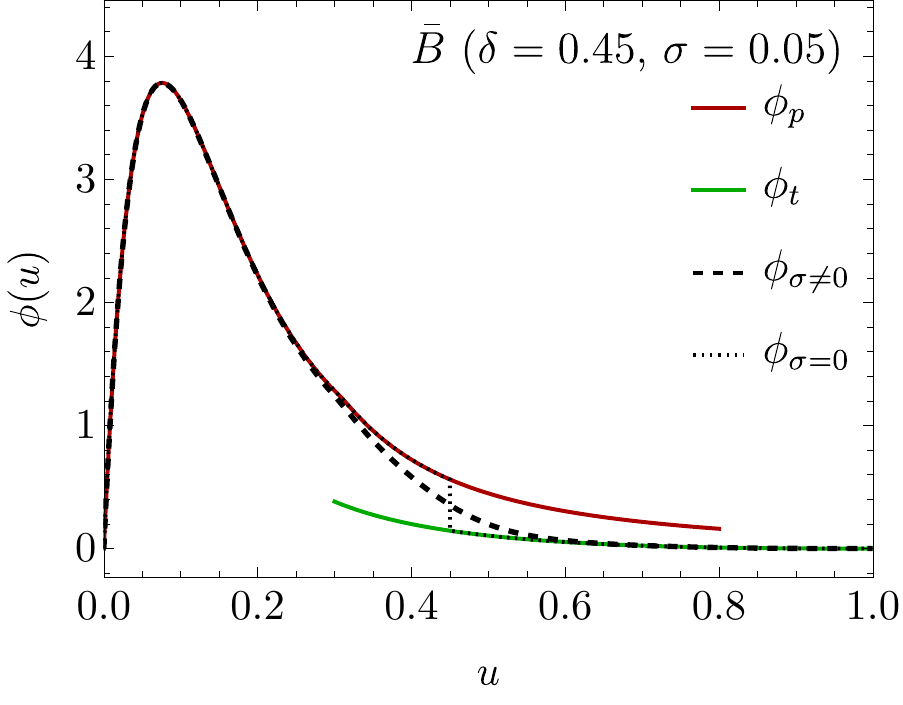}}
\subfloat{\includegraphics[width=0.47\textwidth]{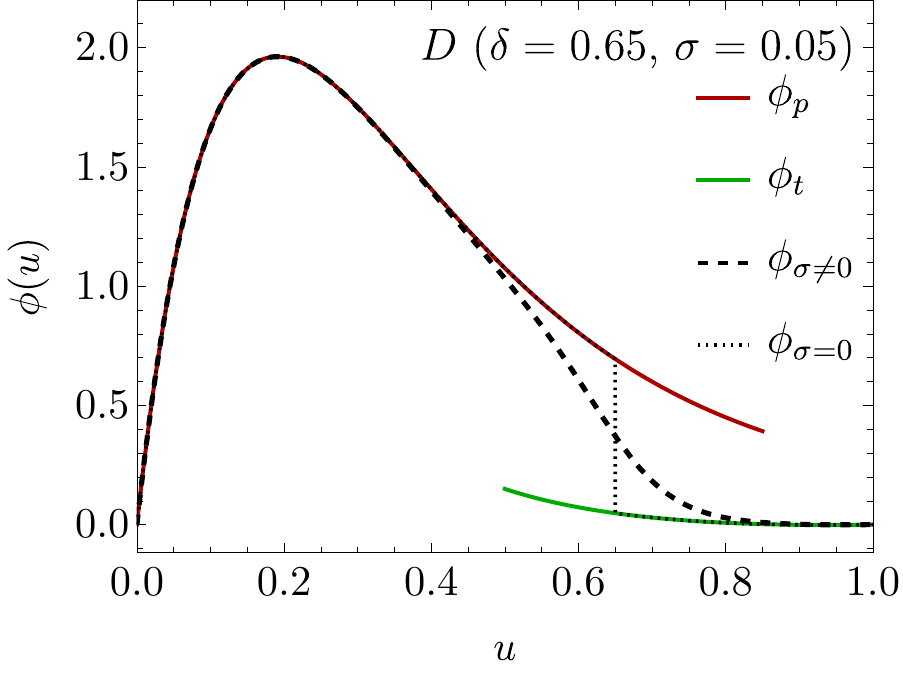}}
\vspace*{0.3cm}
\subfloat{\includegraphics[width=0.46\textwidth]{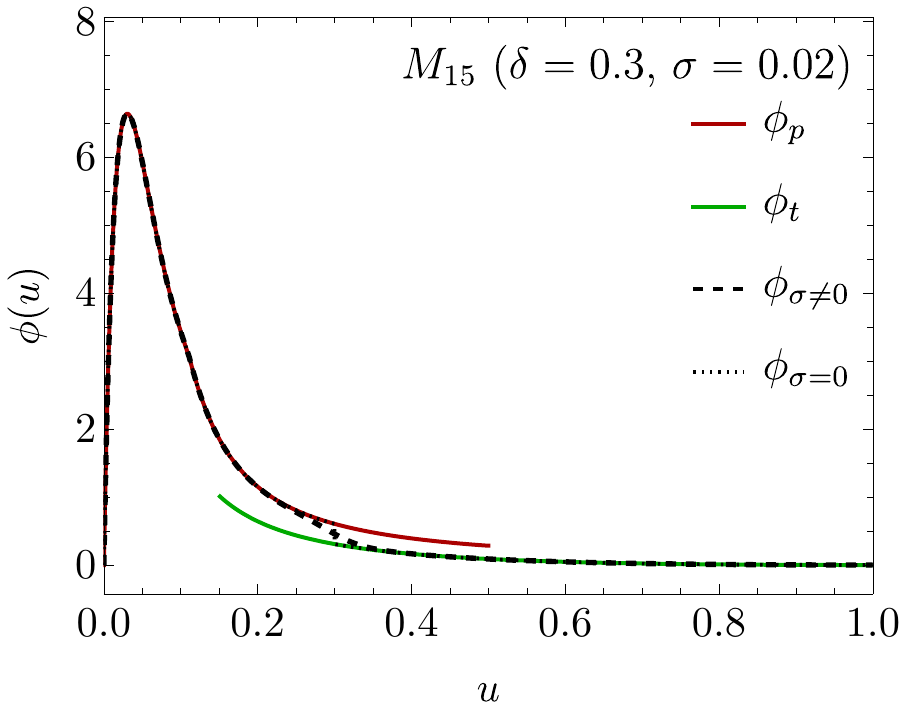}}
\caption{\small The QCD LCDA $\phi(u)$ obtained from~\eqref{eq:phiu} with $\sigma \neq 0$  (dashed) and $\sigma=0$ (dotted) for the cases of the 
$\bar{B}$, $D$ and 15~GeV meson. The peak (red) and tail (green) functions are shown as a reference.}
\label{fig:smoothing}
\end{figure}

The quality of merging the peak and tail for the QCD LCDA at 
the matching scale is shown in Fig.~\ref{fig:smoothing}.
We notice that for the $\bar{B}$ meson, the two peak and tail functions 
$\phi_p(u)$ and $\phi_t(u)$, respectively, do not overlap perfectly, which 
we attribute to the effect of power corrections and higher order terms in the perturbative expansion.\footnote{Although the discontinuity is strongly model dependent, as shown in Figure~\ref{fig:HQETmodels}.}
This interpretation is corroborated by the lower panel, which 
displays the LCDA of the hypothetical 15~GeV mass meson with a smaller gap between the two functions around $u \sim 0.3$  due to smaller power corrections. However the overlap is not perfect due to unknown higher-order corrections in the $\as$ expansion, as expected. 
The merging does not work well for the $D$ meson, which is not surprising, 
since the heavy-quark expansion is known to receive large corrections 
for charmed mesons. In the figure, we also display the LCDAs obtained with 
$\sigma=0$ to better visualize the jump between the two curves at the point 
$\delta$.

We recall that the constructed QCD LCDA is normalized to unity parametrically. 
However, due to power corrections, the actual normalization of the function obtained from \eqref{eq:phiu} falls 
short of 1 by about $10-15\%$.
We therefore rescale $\phi(u)$ such that its normalization is exactly 1, and will 
always use this rescaled version of the LCDA in the following.
While this procedure is adequate for the cases of the $\bar{B}$ and 
$D$ mesons, there is a subtle complication in the formal 
large meson-mass limit. In order to take this limit, one needs to control a series of corrections to $M_0$ in powers of $\mathcal{O}(\as \ln \frac{\mu_s}{\LamUV})$ arising from the evolution of the HQET LCDA from the low scale $\mu_s$, which are not summed by the standard RGE (which deals with logarithms of $\mu/\mu_s$ only). This issue has been already studied in~\cite{Feldmann:2014ika} in dual space. In Appendix~\ref{sec:appM0}, we show that when evolving the HQET LCDA with the procedure described in~\cite{Feldmann:2014ika} (i.e.~resumming both logarithms $\ln \frac{\mu}{\mu_s}$ and $\ln \frac{\mu_s}{\LamUV}$), we find numerical agreement with the fixed-order result~\eqref{eq:M0} for $M_0(\LamUV)$ within a few percent, and the QCD LCDA normalization approaches 1 in the heavy-quark limit as required. For the $\bar{B}$ and $D$ mesons the improved resummation gives essentially the same result as the one used in this section, since the power corrections are larger than 
the unresummed $\ln \frac{\mu_s}{\LamUV}$ terms (see Appendix). 

A useful and common way of parametrizing the QCD LCDA expresses $\phi(u)$ as a series in Gegenbauer polynomials,
\begin{equation}
\label{eq:gegephi}
\phi(u) = 6u(1-u)\biggl[1+\sum_{n=1}^{\infty}a_n(\mu)C^{(3/2)}_n(2u-1) \biggr]\,,
\end{equation}
which automatically ensures that the LCDA is normalized to 1.
The Gegenbauer moments $a_n(\mu)$ are defined through
\begin{equation}
\label{eq:gegemom}
a_n(\mu) = \frac{2(2n+3)}{3(n+1)(n+2)}\int_0^1 du\, C^{(3/2)}_n(2u-1)\phi(u)\,,
\end{equation}
and they are expected to decrease for increasing $n$ such that the series can be truncated. 

\begin{figure}[t]
\centering
\includegraphics[width=0.65\textwidth]{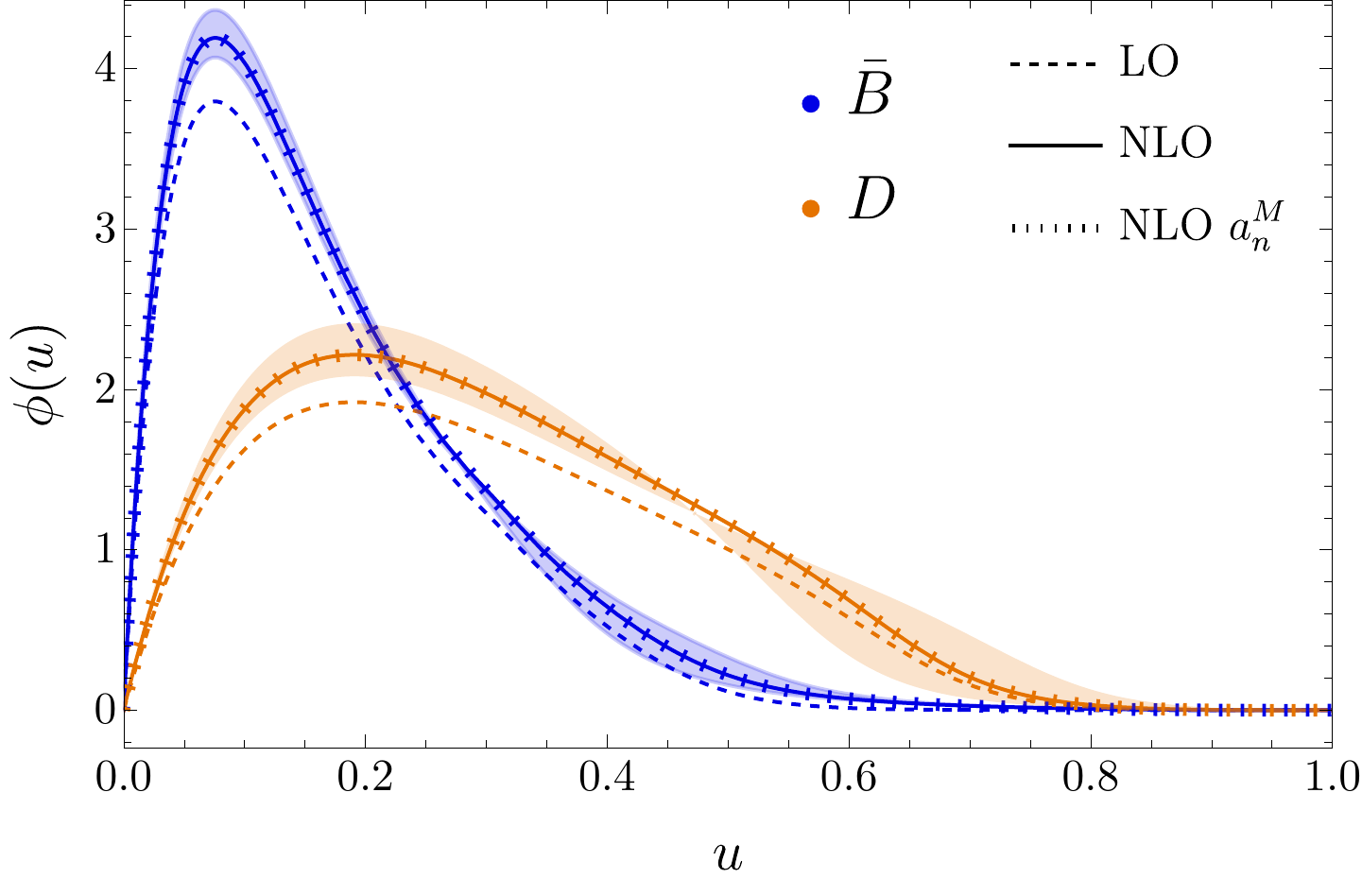}
\caption{\small QCD LCDA $\phi(u)$ at the heavy quark scale for a $\bar{B}$ (blue) and a $D$ (orange) meson at tree level (dashed) and including NLO corrections (solid). 
The ``transverse-dashed'' curves are the results for the Gegenbauer expansion and the shaded areas is the uncertainty obtained by varying $\delta$ by $\pm 15\%$.}
\label{fig:phiu}
\end{figure}

We use the merging function with values~\eqref{eq:sigmavals} and keep the first 20 Gegenbauer moments for both $\bar{B}$ and $D$ meson (and all HQET models).\footnote{We checked that in evaluating numerically the integrals in~\eqref{eq:gegemom} the effects of varying $\sigma \in [0,0.05]$ are negligible with respect to the variation of $\delta$.
However for $\sigma \simeq 0$ the expansion needs to be truncated before it starts to resolve the discontinuity of the LCDA for $\sigma=0$, which happens at different orders for different models and mesons. For this reason, we use the values~\eqref{eq:sigmavals}. } 
For the $\bar{B}$ meson, we find for $n=1,\ldots$ 
\begin{equation}
\begin{split}
a_n^{\bar{B}}(\mu_b) = \{-1.082, 0.826, -0.513, 0.288, -0.157, 0.078, -0.030, 0.008, \dots\}\,,
\end{split}
\end{equation}
which exhibits a good convergence. The dots stand for the higher moments, 
which have modulus smaller than $0.005$. In the case of the $D$ meson, 
\begin{equation}
a_n^D(\mu_c) = \{-0.659, 0.206, -0.057, 0.036, -0.004, -0.007, \dots\}\,.
\end{equation}
The Gegenbauer series for the $D$ meson converges faster than the one for the $\bar{B}$, as it should, because $\phi_{D}(u)$ is ``closer" to the asymptotic form $6u\bar{u}$ than  $\phi_{\bar{B}}(u)$. The final results for the QCD LCDA for the $\bar{B}$ and $D$ mesons at the heavy quark scale are shown in Fig.~\ref{fig:phiu}.\footnote{To show the difference between NLO and LO, we normalize to 1 the NLO LCDA and use the same normalization factor for the LO approximation.}
The shaded band is the systematic uncertainty obtained by varying $\delta$ by $\pm 15\%$. 
The figure shows that both 
methods of expressing $\phi(u)$---with the merging function $\vartheta(u;\delta,\sigma)$ or by expanding it in Gegenbauer polynomials---yield indistinguishable results, justifying the truncation of the Gegenbauer expansion. The figure also 
demonstrates that the NLO matching correction is important in the peak 
region.\footnote{We recall that in the tail region NLO is technically the 
leading term, to which the correction is not available.} 

\subsection{Evolution to the Hard Scale}
\label{sec:evolHard}

Lastly, in order to evolve the QCD LCDA to higher scales, it is convenient to express it in terms of Gegenbauer polynomials as in~\eqref{eq:gegephi}. 
The one-loop kernel relevant to LL evolution is diagonal for the Gegenbauer 
moments $a_n$ with anomalous dimension $\gamma_n$, hence the evolution from the matching scale $\mu$ to the high scale $\mu_h$ takes the simple form
\begin{equation}
\label{eq:gegerun}
\frac{a_n(\mu_h)}{a_n(\mu)} = \biggl(\frac{\alpha_s(\mu_h)}{\alpha_s(\mu)}\biggr)^{\!\frac{\gamma_n}{2\beta_0}} \,,
\end{equation}
where the anomalous dimension is
\begin{equation}
\gamma_n = 2C_F\biggl(4\sum_{k=1}^{n+1}\frac{1}{k} - \frac{2}{(n+1)(n+2)}-3\biggr)\,.
\end{equation}
We evolve the LCDA up to the hard scale $\mu_h = m_W = 80.377 \pm 0.012~\si{GeV}$, finding
\begin{align}
\label{eq:gege80GeV}
  a_n^{\bar{B}}(m_W) &= \{-0.826, 0.542, -0.302, 0.156, -0.079, 0.037, -0.014, \dots \}\,,\nonumber\\
  a_n^{D}(m_W) &=\{-0.416, 0.100, -0.023, 0.013, \dots\}\,.
\end{align}

\begin{figure}[t]
\centering
\subfloat{\includegraphics[width=0.495\textwidth]{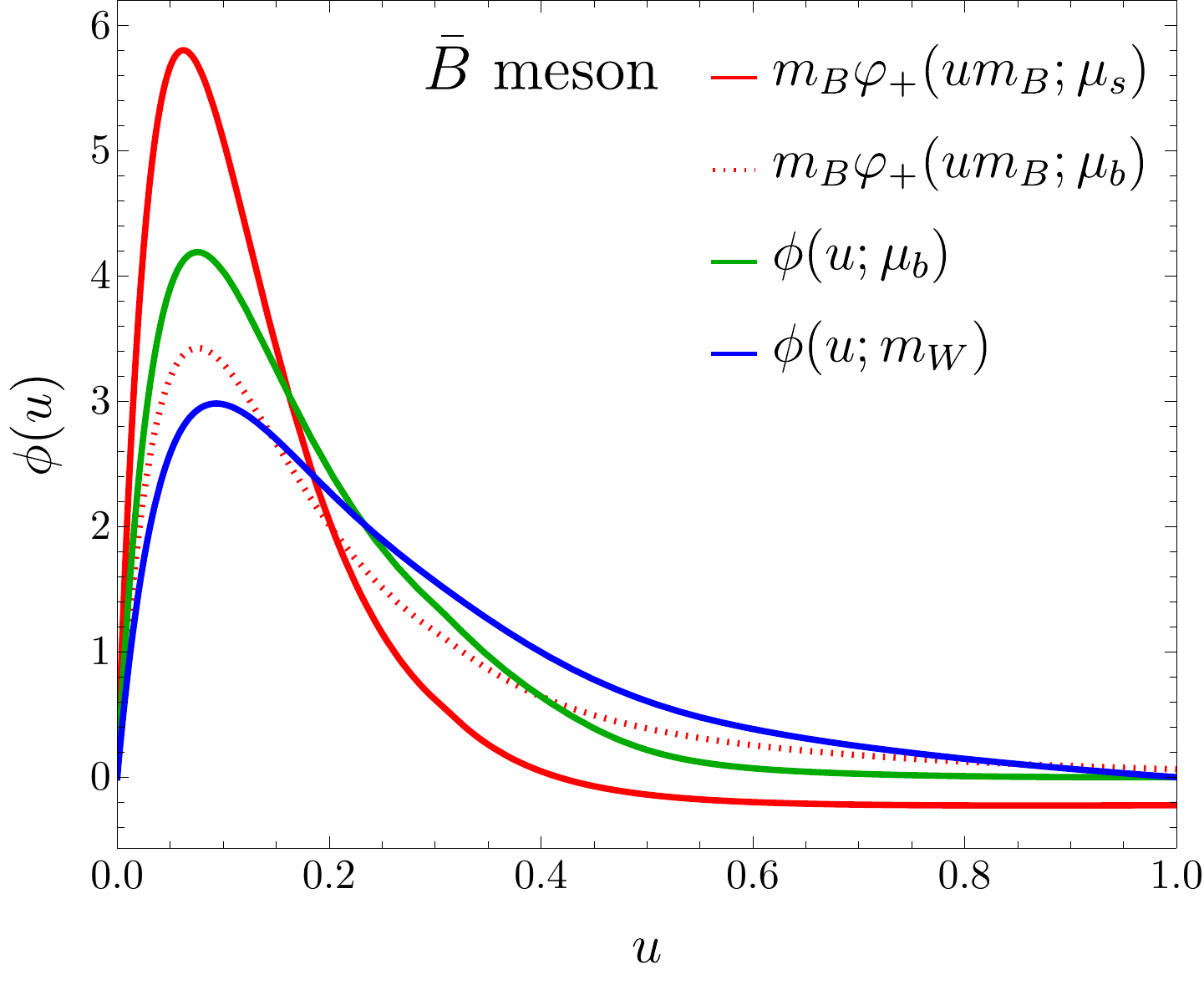}}
\subfloat{\includegraphics[width=0.505\textwidth]{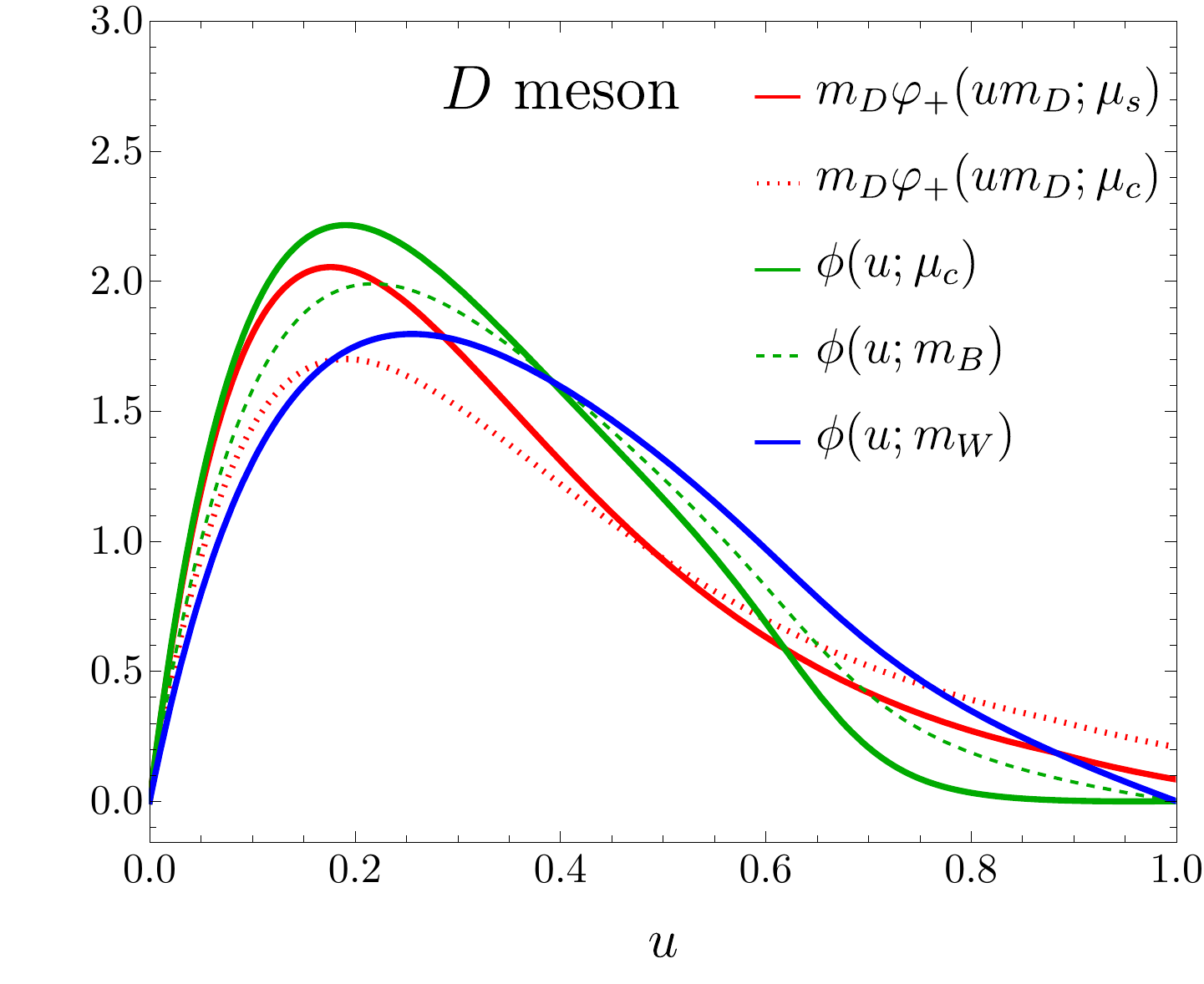}}
\caption{\small Evolution of the LCDA from the initial HQET condition (red) to the hard scale $m_W$ (blue) and the LCDA at the matching scale for the 
$\bar B$ (left) and $D$ meson (right).}
\label{fig:finalplot}
\end{figure}

In Fig.~\ref{fig:finalplot} we show all functions appearing in the three steps discussed in this 
section, starting with the HQET LCDA at the soft scale $\mu_s=1$ GeV (red, solid), evolved in HQET to the 
matching scale $\mu$ (red, dotted). This is matched to $\phi(u)$ 
(green, solid) and then evolved in QCD to the hard scale $m_W$ (blue, solid). 
For the $D$ meson, we also show the LCDA at the 
$\bar{B}$ mass scale (green, dashed). 

\section{Branching Fraction of the $W^- \to B^- \gamma$ Decay}
\label{sec:WtoBgamma}

The QCD LCDA of a heavy meson is the relevant hadronic matrix element for the 
decay $W^- \to B^- \gamma$, in which a highly boosted $\bar{B}$ meson is 
produced with typical energy of order $\Q = m_W \gg m_b$. In computing the 
decay amplitude in QCD factorization the LCDA should be evaluated at this 
scale. The results of this work allow us to derive a factorization formula that 
reduces the hadronic input to the universal leading-twist HQET LCDA 
and resums large logarithms between $\LamQCD$ and $m_W$, by matching the HQET 
LCDA to the QCD one at the intermediate scale $\mu_b \sim \mathcal{O}(m_b)$.

A previous study of this process within QCD factorization was undertaken 
in~\cite{Grossman:2015cak}. The authors used for the $\bar{B}$ meson QCD LCDA at the low scale $\mu_s=1~$GeV a model inspired by the HQET exponential model, and evolved it 
up to the $W$ mass employing the light meson LCDA evolution equation. 
This approach correctly resums the large logarithms between the scales $m_b$ and $m_W$, but misses the $b$-quark mass effects in the evolution between the hadronic scale $\LamQCD$ and $m_b$. Our strategy is to follow the work of Ref.~\cite{Grossman:2015cak} in performing the hard matching at the scale $m_W$ at leading power\footnote{In practice this corresponds to considering a massless meson.} in $m_b/m_W$, but then we shall employ the QCD LCDA at the matching scale $\mu_b$ of order 
$m_b$ derived earlier in the present work, which sums correctly the logarithms 
from the evolution from 1 GeV to $\mu_b$.

\begin{figure}
    \centering
    \subfloat[$1/x$ contribution\label{fig:Wa}]{\includegraphics[width=0.25\textwidth]{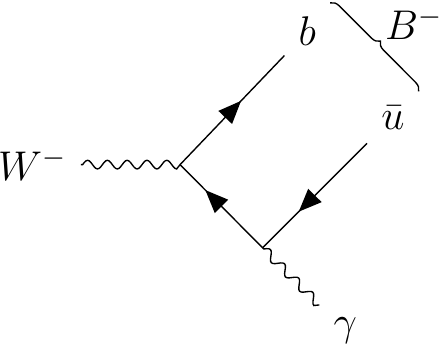}}$\qquad\quad$
    \subfloat[$1/\bar{x}$ contribution\label{fig:Wb}]{\includegraphics[width=0.25\textwidth]{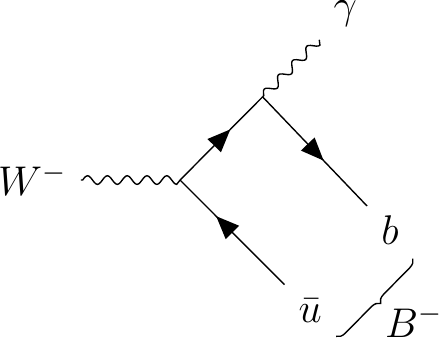}}$\qquad\quad$
    \subfloat[Local contribution\label{fig:Wc}]{\includegraphics[width=0.25\textwidth]{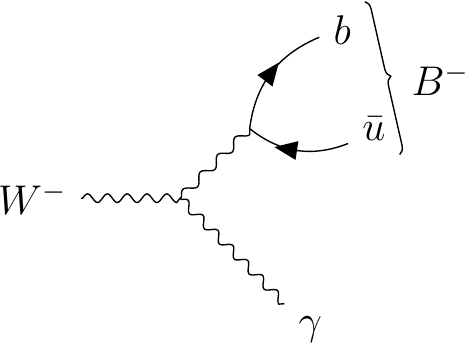}}
    \caption{\small Tree-level diagrams contributing to the process $W^- \to \bar{B}^- \gamma$.}
    \label{fig:Wdiagrams}
\end{figure}

The tree-level contributions to this decay originate from the three diagrams in Figure~\ref{fig:Wdiagrams}, where $x$ denotes the momentum fraction of the $\bar{u}$ anti-quark and $\bar{x}=1-x$ the one of the $b$ quark. The first two diagrams are symmetric under $x \leftrightarrow \bar{x}$ except for the electric charges of the quarks.  Including the hard-scale one-loop QCD corrections, they lead to the following convolutions with the $\bar{B}$ meson LCDA, 
\begin{equation}
\label{eq:hardconv}
I_\pm^B =\int_0^1 dx \, H_\pm(x,\mu_h) \phi_B(x;\mu_h)\,, \qquad \bar{I}_\pm^B = \int_0^1 dx \, H_\pm(\bar{x},\mu_h) \phi_B(x;\mu_h)\,,
\end{equation}
where $H_\pm$ are two hard-scattering kernels computed in Ref.~\cite{Grossman:2015cak} and listed for completeness in Appendix~\ref{sec:appcrosscheck}. 
At tree level, $H_\pm^{(0)}(x) = 1/x$. In Fig.~\ref{fig:Wc} the meson is produced from a local interaction, which adds a constant term to the decay amplitude. 
Following~\cite{Grossman:2015cak}, the amplitude is parametrized in terms of two form factors 
as  
\begin{equation}
\label{eq:Wamp}
    i \mathcal{A}(W^- \to \bar{B}^- \gamma) = \frac{e g_2 f_B}{4\sqrt{2}}V_{ub}^*\biggl(i \epsilon_{\mu\nu\alpha\beta}\frac{p_B^\mu q^\nu \varepsilon_W^\alpha \varepsilon_\gamma^{*\beta}}{p_B \cdot q}F_1^B- \varepsilon_W^\perp \cdot \varepsilon_\gamma^{\perp *} F_2^B \biggr)\,,
\end{equation}
where $e$ is the positron charge, $g_2$ the SU(2) coupling constant, $\varepsilon_W$ ($\varepsilon_\gamma$) the polarization vector of the $W$ boson (photon), and $p_B$ ($q$) the momentum of the $\bar{B}$ meson (photon) in the final state.
We use the short hand notation
\begin{equation}
    \varepsilon_W^\perp \cdot \varepsilon_\gamma^{\perp *} = \varepsilon_W \cdot \varepsilon_\gamma^{*} - \frac{q\cdot \varepsilon_W \;p_B \cdot \varepsilon_\gamma^*}{p_B \cdot q}\,.
\end{equation}
The form factors $F_{1,2}^B$ in~\eqref{eq:Wamp} are expressed in terms of the convolutions integrals in~\eqref{eq:hardconv} as
\begin{align}
    F_1^B &= Q_u I_+^B + Q_d \bar{I}_+^B\,,\nonumber\\
    F_2^B &= 2(Q_u-Q_d) - Q_u I_-^B + Q_d \bar{I}_-^B\,,
\label{eq:ffs}
\end{align}
and the constant term $2(Q_u-Q_d)=2$ in $F_2^B$ comes from the local contribution of Fig.~\ref{fig:Wc}. $Q_u=2/3$ and $Q_d=-1/3$ are the quark electric charges in units of the positron charge. The form factors are thus perturbative series in $\as$, and contain the Gegenbauer moments of the $\bar{B}$ meson LCDA. 

Squaring the amplitude~\eqref{eq:Wamp} and dividing by the total $W$-boson decay width $\Gamma_W$ gives the branching ratio 
\begin{equation}
    \text{Br}(W \to B\gamma) = \frac{\Gamma(W \to B \gamma)}{\Gamma_W} = \frac{\alpha_{\rm em} m_W f_B^2}{48 v^2 \Gamma_W}|V_{ub}|^2 \Bigl(|F_1^B|^2 + |F_2^B|^2\Bigr)\,,
\end{equation}
which also holds for the CP-conjugated $W^+ \to B^+ \gamma$ decay.
As numerical inputs we use $v = 246.22~\si{GeV}$~\cite{Workman:2022ynf} for the Higgs vacuum expectation value, $f_B = 190.0\pm 1.3~\si{MeV}$~\cite{FlavourLatticeAveragingGroup:2019iem} for the $B$ meson decay constant, $\Gamma_W=2.085 \pm 0.042~\si{GeV}$~\cite{Workman:2022ynf} for the $W$ boson total width and $\alpha_{\rm em} = 1/137.036$ for the fine-structure constant, evaluated at $q^2=0$. We use the exclusive determination of $|V_{ub}|$ from~\cite{Leljak:2021vte}, $|V_{ub}|=(3.77 \pm 0.15)\cdot 10^{-3}$.
To evaluate the form factors $F_{1,2}^B$, we employ the QCD LCDA from the matching with HQET evolved to the hard scale $\mu_h=m_W$, as derived in the previous section. The convolutions~\eqref{eq:hardconv} can be found in closed form as functions of the Gegenbauer moments $a_n^{\bar{B}}(\mu_h)$~\cite{Grossman:2015cak}. Using the first 20 Gegenbauer moments at $\mu_h$ from~\eqref{eq:gege80GeV} our result for the branching ratio reads 
\begin{equation}
\label{eq:resumresult}
\text{Br}(W\to B \gamma) = (2.58 \pm 0.21_{\rm in}\,^{+0.05}_{-0.08}\,_{\mu_h}\,^{+0.05}_{-0.08}\,_{\mu_b}\,^{+0.18}_{-0.13}\,_{\delta}\,^{+0.61}_{-0.34}\,_{\beta}\,^{+2.95}_{-0.98}\,_{\lambda_B}) \cdot 10^{-12} \,.
\end{equation}
Since the factorization holds at the level of the amplitude~\eqref{eq:Wamp}, we kept the $\mathcal{O}(\as^2)$ terms from the square of the form factors to obtain the above numbers. If we instead truncate the expansion at first order the central value would be $\text{Br}^{\rm trunc}=2.54\cdot 10^{-12}$ with similar uncertainties. 
We divided the uncertainty budget into the different sources: 
1) input uncertainties ($f_B$, $\Gamma_W$, $m_W$, mainly $|V_{ub}|$), 
2) hard-scale variation in the range $\mu_h \in [m_W/2, 2m_W]$, 
3) matching scale variation in the range $\mu \in[\mu_b/2,2\mu_b]$,
4) variation of the peak-tail merging point $\delta$ by $\pm 15\%$,
5) HQET LCDA model-shape dependence by varying $\beta$ for the three 
models within its respective domain,
6) varying $\lambda_B=(350\pm 150)$~MeV within its uncertainty.

\begin{figure}
    \centering
    \includegraphics[width=0.42\textwidth]{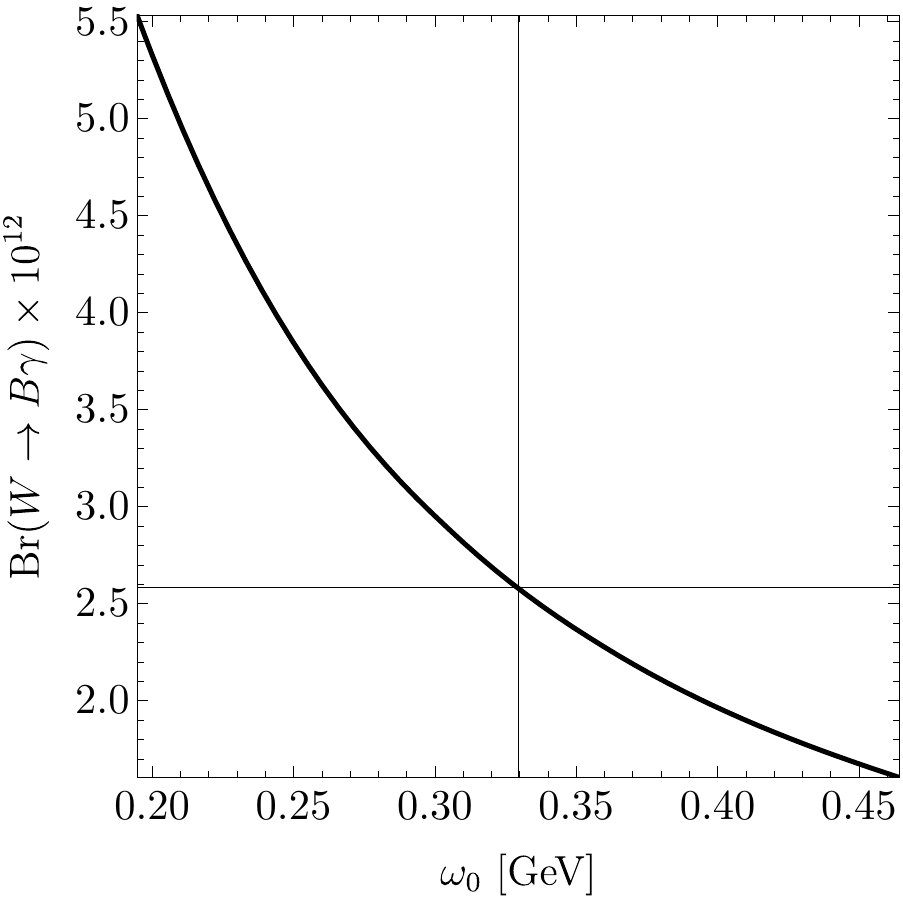}
    \includegraphics[width=0.43\textwidth]{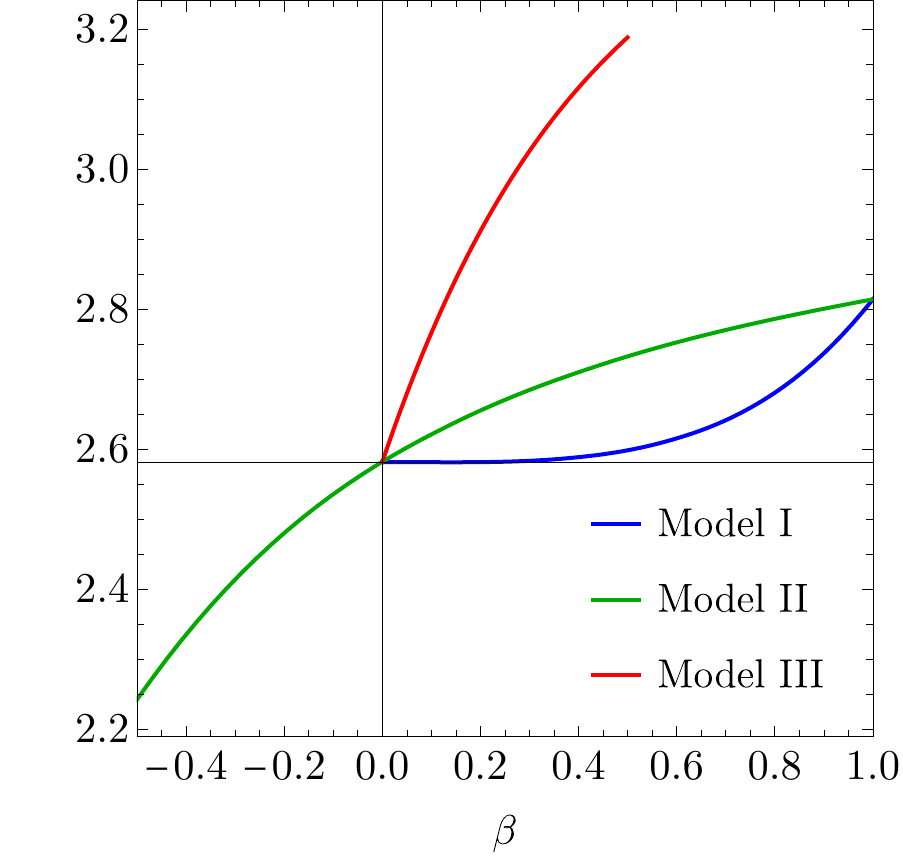}
    \caption{\small Branching ratio as a function of the parameters in the HQET LCDA models~\eqref{eq:phimodels}.}
    \label{fig:BrDep}
\end{figure}

The numerical result highlights the present large uncertainty arising 
from the HQET LCDA at the low scale. Therefore, we show the dependence of the branching ratio on the parameters $\omega_0$ and $\beta$ of~\eqref{eq:phimodels} in Fig.~\ref{fig:BrDep}. 
The branching ratio is particularly sensitive to the lower bound on $\lambda_B$, as can be seen in the left panel, since smaller values of $\lambda_B$ shift the peak of the LCDA to smaller values of $x$, where the tree-level hard-scattering kernel is enhanced by the $1/x$-behaviour.

We can now compare with the approach of ~\cite{Grossman:2015cak}. Updating their result for our numerical inputs gives
\begin{equation}
\label{eq:gros}
\text{Br}(W\to B \gamma)\Bigl|_{\rm exp.~model} = (1.99\pm 0.17_{\rm in}\,^{+0.03}_{-0.06}\,_{\mu_h}\,^{+2.48}_{-0.80}\,_{\lambda_B}) \cdot 10^{-12} \,,
\end{equation}
which again is close to its truncated expansion $\text{Br}^{\rm trunc}|_{\rm exp.~model} = 1.96 \cdot 10^{-12}$.

Comparing with our new result~\eqref{eq:resumresult}, we notice that correctly resumming the logarithms $\ln m_b/\LamQCD$ within HQET instead of using the ERBL evolution between $\mu_s$ and $m_b$ as was done to obtain \eqref{eq:gros}, enhances the branching ratio by almost 30\%, while the relative uncertainties are roughly the same.

Another approach to the $W\to B\gamma$ decay rate has been adopted 
in~\cite{Ishaq:2019zki}, where the hard process was computed at fixed 
order and evaluated at $\mu_b \sim \mathcal{O}(m_b)$ in terms of a 
convolution with the HQET LCDA at this scale. This approach consistently takes into account the heavy meson LCDA evolution from $\LamQCD$ to $\mu_b$, but treats 
$m_W \sim \mathcal{O}(m_b)$, and therefore does not resum the large collinear logarithms $\ln m_b/m_W$. In the framework of HQET factorization at leading power in $\LamQCD/m_b$ only the first diagram in Fig.~\ref{fig:Wdiagrams} contributes, as indeed the convolutions $I_+^B \simeq I_-^B \sim m_b/\LamQCD$ are the dominant ones. Neglecting the $\bar{I}^B_\pm$ and local 
terms in \eqref{eq:ffs}, the form factors $|F_{1,2}^B|$ are equal, and given by  
\begin{equation}
\label{eq:F12HQET}
    F_{1,2}^B\Bigl|_{\rm HQET} = \pm Q_u \frac{\tilde{f}_B(\mu_b)}{f_B}\int_0^\infty d\omega\, T(\omega,m_b,m_W,\mu_b) \varphi_+(\omega;\mu_b)\,,
\end{equation}
where $T(\omega,m_b,m_W,\mu_b)$ is the HQET hard-scattering kernel, and the upper (lower) sign refers to $F_1^B$ ($F_2^B$).
We show in Appendix~\ref{sec:appcrosscheck} that expanding to fixed-order one-loop the convolutions~\eqref{eq:hardconv} of the QCD hard scattering kernels with the matched and evolved LCDA  
reproduces the HQET factorization result~\cite{Ishaq:2019zki} expanded at leading power in $m_b/m_W$.
Using the matching functions $\mathcal{J}_{\rm peak}$ and  $\mathcal{J}_{\rm tail}$
and the evolution kernel $f_{\rm ERBL}$ introduced in Appendix~\ref{sec:appcrosscheck}, this can be schematically written as
\begin{equation}
    T(\omega,m_b,m_W,\mu_b)\Bigl|_{m_b \ll m_W} = H(x,m_W,\mu_h) \otimes_x f_{\rm ERBL}(x,u,\mu_h,\mu_b) \otimes_u \mathcal{J}_p(u,\omega, m_b,\mu_b)\,,
\end{equation}
which is a strong check of our factorization theorem for the QCD LCDA.

Equation~\eqref{eq:F12HQET} allows us to compute the branching fraction with the fixed-order HQET result and to estimate of the effect of resumming the logarithms $\ln m_b/m_W$. We find  
\begin{equation}
   \text{Br}(W\to B \gamma)\Bigl|_{\rm HQET} = (2.61\pm 0.22_{\rm in}\,^{+0.19}_{-0.71}\,_{\mu_b}\,^{+0.50}_{-0.42}\,_{\beta}\,^{+3.09}_{-1.03}\,_{\lambda_B}) \cdot 10^{-12} \,,
\end{equation}
where the second uncertainty was obtained by varying $\mu_b/2 <\mu<2\mu_b$.
The difference in the central value with respect to the resummed result~\eqref{eq:resumresult} is not sizeable, but the poor convergence of the fixed-order result is reflected into the large uncertainty from the scale variation. Indeed, truncating the expansion of the form factors squared after the $\mathcal{O}(\alpha_s)$ term, one would get $\text{Br}^{\rm trunc}|_{\rm HQET} = 2.11\cdot 10^{-12}$ with a negative uncertainty deriving from the scale variation of more than 100\%. This is due to a negative one-loop correction of about 50\% with respect to the tree-level result, making manifest the importance of resumming the logarithms of $m_b/m_W$.


\section{Conclusion}
\label{sec:conclusion}

HQET is a well-known tool to factorize the heavy-quark 
mass scale from the strong interaction scale and exploit 
the flavour and spin-symmetry of heavy-quark interactions 
in the infrared. The vacuum-to-heavy-meson matrix elements 
of local heavy-light quark current, defining the heavy-meson 
decay constant, have been among the first applications 
of HQET. Surprisingly, the matrix elements of the 
corresponding non-local operators at light-like 
separation, the so-called light-cone distribution amplitudes 
have never been studied, with the exception 
of~\cite{Ishaq:2019dst}. The heavy-meson LCDA in 
QCD is the relevant hadronic quantity whenever a heavy 
meson is produced ultra-relativistically in a hard 
process. 

In this work, we established a relation between 
the leading-twist QCD LCDA of a heavy meson and the 
leading-twist, heavy-quark mass independent, universal 
HQET LCDA in the form of a convolution of the latter 
with a perturbative, quark-mass dependent matching 
function, which takes a rather simple form at the 
one-loop order considered here. The expression allows 
one to express the LCDA for heavy mesons of different 
masses in terms of the universal HQET function 
$\varphi_+(\omega)$. Although not discussed here, the 
LCDA of vector mesons $H^*$ can also be matched to 
the same $\varphi_+(\omega)$ due to the spin-symmetry. 

We constructed explicitly the QCD LCDAs for the 
$\bar{B}$- and $D$-meson. For the latter case, power 
corrections of $\LamQCD/m_c$ 
are sizeable. In the case of the $\bar{B}$-meson, the 
present poor knowledge of the HQET LCDA are inherited 
by the QCD LCDA. Nevertheless, the improved evaluation 
of the $W^-\to B^-\gamma$ decay rate shows an increase 
of about 50\% relative to previous calculations, 
related to the consistent summation of logarithms 
of the various scales involved. 

While the rates for the exclusive production 
of $B$ mesons at high energies are very small, although 
not immeasurable, another potential application 
of the framework presented here constitutes factorization 
theorems for charmed $D$ mesons, when the charm 
mass is considered as an intermediate scale. 
We leave this investigation to future work.

\subsubsection*{Acknowledgements}
We thank Matthias K\"onig, Philipp B\"oer and Yao Ji for useful discussions. MB thanks the particle theory group at UC Berkeley and LBL, and GF the Department of Physics and Astronomy of the University of South Carolina for hospitality while this work was completed. This research was supported in part by the Deutsche
Forschungsgemeinschaft (DFG, German Research Foundation) through
the Sino-German Collaborative Research Center TRR110 “Symmetries
and the Emergence of Structure in QCD” (DFG Project-ID 196253076, NSFC Grant No. 12070131001, - TRR 110).

\appendix


\section{bHQET from SCET}
\label{sec:appSCET}

In the main text we derived the bHQET Lagrangian starting from HQET, see~\eqref{eq:bHQETLfromHQET}. In this appendix, we derive the same Lagrangian, but starting from the massive SCET Lagrangian. The full theory collinear field $Q(x)$ 
\begin{equation}
Q(x) = \xi(x) + \eta(x)\,,
\end{equation}
is written in terms of its large and small components 
\begin{equation}
\xi(x) = \frac{\slashed{n}_-\slashed{n}_+}{4}Q(x)\,, \qquad \eta(x) = \frac{\slashed{n}_+\slashed{n}_-}{4}Q(x)\,,
\end{equation}
respectively. The small spinor $\eta$ is integrated out, giving the relation
\begin{equation}\label{eq:etaeom}
\eta(x) = \frac{i \slashed{D}_\perp + \mb}{i \minus{D}}\frac{\slashed{n}_+}{2}\,\xi(x) \sim \mathcal{O}(\lam)\xi(x)\,.
\end{equation}
Using the definition of the bHQET field,
\begin{equation}
\label{eq:hnxi}
\xi(x) = \sqrt{\frac{\minus{v}}{2}} e^{-i \mb v\cdot x}h_n(x)\,,
\end{equation}
we can find the inverse relation of~\eqref{eq:hndefinition}
\begin{equation}
\label{eq:psihn}
Q(x) = \sqrt{\frac{\minus{v}}{2}} e^{-i\mb v \cdot x} \biggl(1 + \frac{\mb +\mb \slashed{v}_\perp + i \slashed{D}_\perp}{\mb \minus{v} + i \minus{D}}\frac{\slashed{n}_+}{2} \biggr)h_n(x)\,.
\end{equation}
which is used in order to obtain~\eqref{eq:hvhnExact} in the main text.\footnote{If instead the alternative definition \eqref{eq:newhn} of the 
boosted HQET field was used, the relation would read
\[
    Q(x) = e^{-i\mb v\cdot x}\biggl[1+\frac{i\slashed{D}-\slashed{v}i v\cdot D}{2\mb+iv \cdot D} \biggr]h_v(x) = \sqrt{\frac{\minus{v}}{2}}e^{-i\mb v\cdot x}\biggl[1+\frac{i\slashed{D}-\slashed{v}i v\cdot D}{2\mb+iv \cdot D} \biggr]\biggl[1+\frac{1+\slashed{v}_\perp}{\minus{v}}\frac{\slashed{n}_+}{2} \biggr]h_n^{\rm new}(x)\,.
\]
}
Inserting the relation~\eqref{eq:hnxi} into 
the collinear part of the massive SCET Lagrangian,
\begin{equation}
\mathcal{L}_{\rm SCET} = \bar{\xi}(x) \biggl[i\plus{D} + (i\slashed{D}_\perp -\mb)\frac{1}{i\minus{D}}(i\slashed{D}_\perp + \mb) \biggr]\frac{\slashed{n}_+}{2}\xi(x)\,,
\end{equation}
and expanding in powers of $\g$, we find
\begin{align}
\mathcal{L}_{\rm SCET} =&\, \frac{\minus{v}}{2}\bar{h}_n(x) \biggl[ \mb \plus{v} + i\plus{D}\nonumber \\
&+(\mb(\slashed{v}_\perp-1))+i\slashed{D}_\perp)\frac{1}{\mb \minus{v} + i\minus{D}}(\mb(\slashed{v}_\perp +1)+i\slashed{D}_\perp)\biggr]\frac{\slashed{n}_+}{2}h_n(x)\nonumber\\
=& \bar{h}_n(x)\biggl[\mb \frac{\minus{v}\plus{v}}{2}+i\plus{D}\frac{\minus{v}}{2}\nonumber\\
&-\frac{\mb \minus{v}\plus{v}}{2}\biggl(1-\frac{i\minus{D}}{\mb \minus{v}}\biggr)+i v_\perp \cdot D_\perp +\mathcal{O}(\lam\g^2 \Q)\biggr]\frac{\slashed{n}_+}{2}h_n(x)\nonumber\\
=& \,\bar{h}_n(x) iv \cdot D \frac{\slashed{n}_+}{2}h_n(x)\Bigl(1+\mathcal{O}(\g) \Bigr)\,,
\end{align}
leading to the same Lagrangian as in~\eqref{eq:bHQETLfromHQET}.


\section{Details of the Calculation}
\label{sec:appcalc}

\subsection{SCET Matrix Element Computation}
\label{sec:appSCETcomputation}

We provide more details on the SCET computation of $\langle Q(p_Q) \bar{q}(p_q)|\mathcal{O}_{C}(u)|0\rangle$.  The external state momenta are
\begin{equation}
p_Q^\mu =\mb v^\mu = \bar{s} \minus{\pB} \frac{n_-^\mu}{2} + \frac{\mb^2}{\bar{s}\minus{\pB}}\frac{n_+^\mu}{2}\,, \qquad p_q^\mu = \minus{p_q}\frac{n_-^\mu}{2}=s \,\minus{\pB} \frac{n_-^\mu}{2} \,,
\end{equation}
where $\minus{\pB} = \mb \minus{v} + \minus{p_q}$ and $\bar{s}\equiv 1-s$. We do not (yet) assign any scaling to $\minus{p_q}$ (hence $s$) nor $u$. For this reason we refer to the results as ``full" in the sense that they contain simultaneously the peak and the tail regions. In this section we identify $\mB$ with $\mb$ which can be done at leading power.

There is a technical subtlety in the computation of the diagram $V$ of Fig.~\ref{fig:names} in SCET: there are two Dirac structures contributing, but the second one does not vanish only if we keep $p_{q\perp} \neq 0$ due to the equation of motion of the suppressed spinor $\eta$ in SCET (see \eqref{eq:etaeom}). We deal with this by keeping $p_{q\perp} \neq 0$ at an early stage of the computation, then substitute the $p_{q\perp}$ dependent Dirac structure with~\eqref{eq:SCETeom} and afterwards remove all the left-over power-suppressed terms (such as $p_{q\perp}/{n_+ p_Q}$) 
by sending $p_{q\perp} \to 0$.
In this way we can write for diagram $V$
\begin{equation}
    \langle Q(p_Q) \bar{q}(p_q)| \mathcal{O}_C(u)|0\rangle\Bigl|_{V} = \aCFopi \sum_\pm V_\pm(u,s)\frac{1}{n_\pm \pB}\bar{u}(p_Q)\slashed{n}_\pm \gamma^5 v(p_q) \,.
\end{equation}

The results of the three diagrams of Fig.~\ref{fig:names} computed in dimensional regularization ($d=4-2\epsilon$) are 
\begin{align}
\label{eq:W1SCET}
W_Q(u,s) &= \biggl(\frac{\mu^2}{\mb^2}\biggr)^\epsilon 2e^{\epsilon \gamma_E}\Gamma(\epsilon)\,\biggl\{\frac{\delta(u-s)}{2\epsilon (1-2\epsilon)}+\frac{\bar{u}}{\bar{s}^{1-2\epsilon}}\frac{\theta(u-s)}{(u-s)^{1+2\epsilon}} \biggr\}\,,\\
W_q(u,s) &= 0\,,\\
\label{eq:VSCET}
V_+(u,s) &= \biggl(\frac{\mu^2}{\mb^2} \biggr)^\epsilon 2e^{\epsilon \gamma_E}\Gamma(\epsilon)\,\biggl\{\theta(s-u) \,e^{i \pi \epsilon}\frac{u}{s}\biggl(\frac{\bar{s}}{u(s-u)}\biggr)^\epsilon\biggl(1-\epsilon +\frac{1}{s-u} \biggr)\nonumber\\
&\hspace*{-0.5cm} + \theta(u-s)\biggl(1-\frac{\bar{u}}{\bar{s}}\biggr)^{-2\epsilon}\,\biggl[(1-\epsilon)\frac{\bar{u}}{\bar{s}} -\frac{u}{s} \biggl(1-\epsilon +\frac{1}{s-u}\biggr)\biggl(1-\Bigl(\frac{u-s}{u \bar{s}}\Bigr)^\epsilon\biggr) \biggr]\biggr\}\,,\\
\label{eq:Vmin}
V_-(u,s) &=-\biggl(\frac{\mu^2}{\mb^2}\biggr)^\epsilon 2e^{\epsilon \gamma_E}\Gamma(\epsilon)(1-\epsilon)\frac{u^{1-\epsilon}}{s \bar{s}^{1-\epsilon}}\,\biggl\{e^{i\pi \epsilon}\frac{\theta(s-u)}{(s-u)^\epsilon} + \frac{\theta(u-s)}{(u-s)^{\epsilon}}\biggl(1-\Bigl(\frac{\bar{s}u}{u-s}\Bigr)^\epsilon\biggr)\biggr\}\,.
\end{align}
The $W$-diagrams contribute only to the $+$ term.

In order to extract the UV poles we need to perform the expansion in $\epsilon$, which requires the introduction of the plus distribution defined as in~\eqref{eq:uplus}. By keeping a small off-shellness to regulate the IR divergences we find
\begin{align}
W_Q^{\rm UV}(u,s) &=\frac{2}{\epsilon}\biggl[\frac{\bar{u}}{\bar{s}}\frac{\theta(u-s)}{u-s}\biggr]_{u+} \,,\\
W_q^{\rm UV}(u,s) &=\frac{2}{\epsilon}\biggl[\frac{u}{s}\frac{\theta(s-u)}{s-u} \biggr]_{u+} \,,\\
V^{\rm UV}_+(u,s) &= \frac{2}{\epsilon}\biggl(\frac{u}{s}\,\theta(s-u) + \frac{\bar{u}}{\bar{s}}\,\theta(u-s) \biggr)\,,\\
V_-^{\rm UV}(u,s) &= 0\,.
\end{align}
Adding the $\overline{\rm MS}$ field strength renormalization, we get
\begin{equation}
Z_{\mathcal{O}_C}^{(1)}(u,s) = -\frac{2}{\epsilon}\biggl[\theta(s-u)\frac{u}{s}\biggl(1+\frac{1}{s-u}\biggr) + \theta(u-s)\frac{\bar{u}}{\bar{s}}\biggl(1+\frac{1}{u-s}\biggr)\biggr]_{u+}\,,
\end{equation}
which agrees with the standard ERBL evolution kernel~\cite{Efremov:1979qk, Lepage:1979zb, Lepage:1980fj}, as it should be. 

\subsection{Peak Region}
\label{sec:apppeak}

In this section, we provide the peak region ($u\sim \g$ and $s \sim \g$) expressions for the individual diagrams and then compare them to the full results presented in Appendix~\ref{sec:appSCETcomputation}.
In the peak region the leading-power contribution scales as $1/\g$, hence $V_-$ 
from~\eqref{eq:Vmin} is a power correction, and the operator basis reduces to a single operator~\eqref{eq:bHQETbasis}. 
Diagram $V$ contains only the soft-collinear region $k \sim \g(1,\lam,\lam^2)\Q$ 
and reduces to 
\begin{equation}
V_+(u,s)\Bigl|_{sc} = \biggl(\frac{\mu^2}{\mb^2}\biggr)^\epsilon 2e^{\epsilon \gamma_E} \Gamma(\epsilon) \frac{u}{s}\biggl\{e^{i\pi \epsilon} u^{-\epsilon}\frac{\theta(s-u)}{(s-u)^{1+\epsilon}} + \frac{\theta(u-s)}{(u-s)^{1+\epsilon}} \Bigl((u-s)^{-\epsilon} - u^{-\epsilon} \Bigr) \biggr\}\,,
\end{equation}
which coincides with the bHQET result $\mb V_{\rm bHQET}(\omega,\nu)$ in~\eqref{eq:VbHQET} of Appendix~\ref{sec:appbHQETcomputation} with $\omega = u \mb$ and $\nu = s \mb$.
In diagram $W_Q$ both the hard- and soft-collinear regions contribute, giving, respectively
\begin{align}
\label{eq:W1hcsmallu}
W_Q(u,s)\Bigl|_{hc} &= \biggl(\frac{\mu^2}{\mb^2}\biggr)^\epsilon 2e^{\epsilon \gamma_E}\Gamma(\epsilon)\frac{\delta(u-s)}{2\epsilon (1-2\epsilon)}\,,\\
W_Q(u,s)\Bigl|_{sc} &= \biggl(\frac{\mu^2}{\mb^2}\biggr)^\epsilon 2e^{\epsilon \gamma_E}\Gamma(\epsilon) \frac{\theta(u-s)}{(u-s)^{1+2\epsilon}}\,.
\end{align}
The soft-collinear region coincides with the bHQET result $\mb {W_Q}_{\rm bHQET}(\omega,\nu)$ in~\eqref{eq:W1bHQET} of Appendix~\ref{sec:appbHQETcomputation} with $\omega = u \mb$ and $\nu = s \mb$.

These are the only two regions that contribute since expanding to leading power the full expressions for $V_+(u,s)$ and $W_Q(u,s)$ given in~\eqref{eq:VSCET} and~\eqref{eq:W1SCET} of Appendix~\ref{sec:appSCETcomputation} we find
\begin{align}
V_+(u,s) &\underset{u \sim \g}{\longrightarrow} V_+(u,s)\Bigl|_{sc} \,,\nonumber\\
W_Q(u,s) &\underset{u\sim \g}{\longrightarrow} W_Q(u,s)\Bigl|_{sc} + W_Q(u,s)\Bigl|_{hc} \,.
\end{align}
This shows that in the peak region the one-loop SCET amplitude is not entirely dominated by soft-collinear modes, but still contains perturbative information from the hard-collinear scale. These hard-collinear contributions will be part of the matching function $\mathcal{J}_p(u,\omega)$ defined in~\eqref{eq:matchingeq}.

\subsection{Tail Region}
\label{sec:apptail}

In the tail region  ($u\sim1$ and $s \sim \g$) the leading-power contribution scales as 1, thus suppressed by one order of $\g$ with respect to the peak region~\ref{sec:apppeak}, as expected from~\eqref{eq:phiscalings}.
The results for the diagram $V$ are
\begin{align}
V_+(u)\Bigl|_{hc} &= \biggl(\frac{\mu^2}{u^2 \mb^2}\biggr)^\epsilon 2e^{\epsilon \gamma_E}\Gamma(1+\epsilon) \bar{u}\biggl(\frac{(1-\epsilon)^2}{\epsilon} +\frac{1}{u} \biggr)\,,\nonumber\\
V_-(u)\Bigl|_{hc} &= \biggl(\frac{\mu^2}{u^2\mb^2}\biggr)^{\epsilon}2e^{\epsilon \gamma_E} \Gamma(1+\epsilon)\bar{u}(1-\epsilon)\,,
\end{align}
with UV poles
\begin{align}
V_+^{\rm UV}(u)\Bigl|_{hc} &= \frac{2\bar{u}}{\epsilon}\,,\nonumber\\
V_-^{\rm UV}(u)\Bigl|_{hc} &= 0\,.
\end{align}
The corresponding results for the diagram $W_Q$ are
\begin{align}
W_{Q}(u)\Bigl|_{hc} &= \biggl(\frac{\mu^2}{\mb^2}\biggr)^\epsilon 2e^{\epsilon \gamma_E}\Gamma(\epsilon)\frac{\bar{u}}{u^{1+2\epsilon}}\,,\\
W_Q^{\rm UV}(u)\Bigl|_{hc} &= \frac{2}{\epsilon}\frac{\bar{u}}{u}\,.
\end{align}
The renormalization kernel can then be inferred from the UV poles to be
\begin{equation}
    Z^{(1)}_{\mathcal{O}_C}(u)\Bigl|_{hc} = -\frac{2}{\epsilon}\bar{u}\Bigl(1+\frac{1}{u}\Bigr)\,.
\end{equation}
Again, expanding the full result for $V$ and $W_Q$ in~\eqref{eq:VSCET},~\eqref{eq:Vmin} and~\eqref{eq:W1SCET} to leading power we find the above computed regions
\begin{align}
V_+(u,s) &\underset{u\sim 1}{\longrightarrow} V_+(u)\Bigl|_{hc} \,,\nonumber\\
V_-(u,s) &\underset{u\sim 1}{\longrightarrow} V_-(u)\Bigl|_{hc} \,,\nonumber\\
W_Q(u,s) &\underset{u\sim 1}{\longrightarrow} W_Q(u)\Bigl|_{hc} \,,\nonumber\\
Z^{(1)}_{\mathcal{O}_C}(u,s) &\underset{u\sim 1}{\longrightarrow} Z^{(1)}_{\mathcal{O}_C}(u)\Bigl|_{hc}\,,
\end{align}
showing that in the tail region only the hard-collinear modes contribute, and 
verifying explicitly that the result is independent on the power-suppressed external momentum fraction $s\ll u$.

\subsection{bHQET Matrix Element Computation}
\label{sec:appbHQETcomputation}

The bHQET matrix element $\langle Q(p_Q)\bar{q}(p_q)|\mathcal{O}_h(\omega)|0\rangle$ is computed with external momenta parameterized as
\begin{equation}
p_Q^\mu = \mb v^\mu\,, \qquad v^\mu = \minus{v} \frac{n_-^\mu}{2} + \frac{1}{\minus{v}}\frac{n_+^\mu}{2}\,, \qquad p_q^\mu = \nu \minus{v} \frac{n_-^\mu}{2}\,,
\end{equation}
where we have rescaled $\nu$ and $\omega$ by $\minus{v}$ such that they correspond to the variables in the $\B$ rest frame.
The results for the bare, dimensionally regulated bHQET diagrams are given by
\begin{align}
\label{eq:W1bHQET}
W_{Q,\text{bHQET}}(\omega,\nu) &= \mu^{2\epsilon}\, 2e^{\epsilon \gamma_E} \Gamma(\epsilon) \frac{\theta(\omega -\nu)}{(\omega -\nu)^{1+2\epsilon}}\,,\\ 
W_{q,\text{bHQET}}(\omega,\nu) &= 0\,,\\
\label{eq:VbHQET}
V_{\text{bHQET}}(\omega,\nu) &= \mu^{2\epsilon} 2e^{\epsilon \gamma_E}\Gamma(\epsilon) \frac{\omega}{\nu}\,\biggl\{e^{i\pi\epsilon}\omega^{-\epsilon}\frac{\theta(\nu-\omega)}{(\nu-\omega)^{1+\epsilon}} \nonumber\\
&\quad+ \frac{\theta(\omega-\nu)}{(\omega-\nu)^{1+\epsilon}}\bigl((\omega-\nu)^{-\epsilon}-\omega^{-\epsilon}\bigr) \biggr\}\,.
\end{align}
Their UV poles are obtained by regulating the IR divergences by an off-shellness $p_q^2\neq 0$ and found to be
\begin{align}
W_{Q,\text{bHQET}}^{\rm UV}(\omega,\nu) &=\frac{2 \omega}{\epsilon}\biggl[\frac{\theta(\omega-\nu)}{\omega(\omega-\nu)} \biggr]_{\omega+} - \delta(\omega-\nu)\biggl(\frac{1}{\epsilon^2}+\frac{2}{\epsilon}\ln \frac{\mu}{\nu} \biggr)\,,\\
W_{q,\text{bHQET}}^{\rm UV}(\omega,\nu) &= \frac{2\omega}{\epsilon}\biggl[\frac{\theta(\nu-\omega)}{\nu(\nu-\omega)}\biggr]_{\omega+} +\frac{2}{\epsilon}\delta(\omega-\nu)\,,\\
V_{\rm bHQET}^{\rm UV}(\omega,\nu) &= 0\,.
\end{align}
The plus distribution is defined in analogy to~\eqref{eq:uplus} as
\begin{align}
    &\int_0^\infty d\omega\, f(\omega) \Bigl[g(\omega,\nu) \Bigr]_{\omega+} = \int_0^\infty d\omega\,\bigl(f(\omega)-f(\nu)\bigr)g(\omega,\nu)\,,\nonumber\\
    &\int_0^\infty d\nu\, f(\nu) \Bigl[g(\omega,\nu) \Bigr]_{\nu+} = \int_0^\infty d\nu\,\bigl(f(\nu)-f(\omega)\bigr)g(\omega,\nu)\,.
\end{align}

The renormalization kernel of the bHQET matrix element when expressed as distribution in the first argument 
$\omega$ is then obtained as
\begin{equation}
    Z_{\mathcal{O}_h}^{(1)}(\omega,\nu) = -W_{Q,\text{bHQET}}^{\rm UV}(\omega,\nu)-W_{q,\text{bHQET}}^{\rm UV}(\omega,\nu)-\frac{\delta(\omega-\nu)}{2}\bigl(Z^{(1)}_\xi+Z^{(1)}_{h_n}\bigr)\,,
\end{equation}
which leads to~\cite{Lange:2003ff}
\begin{equation}
Z_{\mathcal{O}_h}^{(1)}(\omega,\nu) = -\frac{2\omega}{\epsilon}\biggl[\frac{\theta(\omega-\nu)}{\omega(\omega-\nu)} + \frac{\theta(\nu-\omega)}{\nu(\nu-\omega)} \biggr]_{\omega+} +\delta(\omega-\nu)\biggl(\frac{1}{\epsilon^2}+\frac{1}{\epsilon} \left[2\ln \frac{\mu}{\nu}-\frac{5}{2}\right] \biggr)\,,
\end{equation}
where we added the $\overline{\text{MS}}$ renormalization constants of the fields $Z_{h_n}^{(1)} = 2/\epsilon$ and $Z_{\xi}^{(1)}=-1/\epsilon$.


\section{Cross-check with HQET Factorization}
\label{sec:appcrosscheck}

The running of $\phi(u)$ can be also expressed in the form of a convolution of the LCDA at a lower scale with an evolution kernel 
\begin{equation}
\label{eq:phievolv}
\phi(x;\mu_h) = \int_0^1 du\, f_{\rm ERBL}(x,u,\mu_h,\mu) \phi(u;\mu)\,,
\end{equation}
which turns out useful in order to compute analytically the fixed order large collinear logs $\ln \Q/\mb$ from our factorization of the LCDA at the scale $\mu$ 
as done for the $W\to B \gamma$ decay in Section~\ref{sec:WtoBgamma}.
Using~\eqref{eq:gegephi},~\eqref{eq:gegemom} and~\eqref{eq:gegerun}, we obtain the evolution function $f_{\rm ERBL}(x,u,\mu_h,\mu)$ as an expansion in Gegenbauer polynomials
\begin{equation}
\label{eq:frunning}
f_{\rm ERBL}(x,u,\mu_h,\mu) = 6 x\bar{x} \sum_{n=0}^\infty N_n \biggl(\frac{\alpha_s(\mu_h)}{\alpha_s(\mu)}\biggr)^{\frac{\gamma_n}{2\beta_0}} C_n^{(3/2)}(2x-1)C_n^{(3/2)}(2u-1) \,,
\end{equation}
where 
\begin{equation}
    N_n \equiv \frac{2(2n+3)}{3(n+1)(n+2)}\,.
\end{equation}
This evolution function has the following properties
\begin{equation}
\label{eq:flim}
f_{\rm ERBL}(x,u,\mu_h,\mu) \underset{\mu_h\to \infty}{\longrightarrow} 6x \bar{x}\,, \qquad f_{\rm ERBL}(x,u,\mu,\mu) = \delta(x-u)\,.
\end{equation}
Re-expanding in $\alpha_s$, we find 
\begin{equation}
\label{eq:exprunning}
\biggl(\frac{\alpha_s(\mu_h)}{\alpha_s(\mu)}\biggr)^{\frac{\gamma_n}{2\beta_0}} = 1 + \gamma_n \frac{\alpha_s(\mu)}{8\pi} \ln \frac{\mu^2}{\mu_h^2} + \mathcal{O}\Bigl(\alpha_s(\mu)^2\Bigr)\,,
\end{equation}
and by inserting~\eqref{eq:exprunning} into~\eqref{eq:frunning} we find the logarithmic term  in the evolution function at $\mathcal{O}(\alpha_s)$, 
\begin{equation}
f_{\rm ERBL}(x,u,\mu_h,\mu) = \delta(x-u) +\frac{\alpha_s(\mu)}{8\pi}\ln \frac{\mu^2}{\mu_h^2}6x\bar{x}\sum_{n=0}^\infty \gamma_n N_n C_n^{(3/2)}(2x-1)C_n^{(3/2)}(2u-1)\,.
\end{equation}
Finally, this results in the evolved LCDA in terms of the LCDA at the scale $\mu$ given by
\begin{equation}
\label{eq:phiatmuQ}
\phi(x;\mu_h) = \phi(x;\mu) + \frac{\alpha_s(\mu)}{8\pi}\ln \frac{\mu^2}{\mu_h^2}6x\bar{x}\sum_{n=0}^\infty \gamma_n N_n C_n^{(3/2)}(2x-1) \int_0^1 du\, C_n^{(3/2)}(2u-1)\phi(u;\mu)\,.
\end{equation}

We can expand at fixed-order in $\alpha_s$ the convolutions~\eqref{eq:hardconv} to recover the ``HQET factorization"~\cite{Ishaq:2019zki} result.
This leads to the LCDA~\eqref{eq:phiatmuQ} with $\mu=\mu_b$, which has to be inserted into the convolutions~\eqref{eq:hardconv}.
The hard-scattering kernels take the form
\begin{equation}
\label{eq:hardW}
H_\pm(x,\mu_h) = \frac{1}{x}\Bigl(1+\frac{\as(\mu_h)C_F}{4\pi} h_\pm(x,\Q,\mu_h)+\mathcal{O}(\alpha_s^2)\Bigr)\,,
\end{equation}
with the perturbative functions~\cite{Grossman:2015cak}
\begin{equation}
h_\pm (x, \Q,\mu_h) = -(2\ln x+3)\Bigl(\ln \frac{\mu_h^2}{\Q^2}+i\pi \Bigr)+\ln^2 x -9 +(\pm 1-2)\frac{x\ln x}{1-x}\,.
\end{equation}
The convolution of $H_\pm(\bar{x},\mu_h)$ with the LCDA of a heavy meson gives a subleading contribution, as do $\mathcal{O}(x)$ terms in $h_\pm$, since the LCDA is 
concentrated at small $x$. Neglecting these suppressed terms, the two hard scattering kernels are the same, $H_+(x,\mu_h) = H_-(x,\mu_h) \equiv H(x,\mu_h)$ (analogous definition for $h(x,Q,\mu_h)$).

We will extract the HQET hard-scattering kernel $T(\omega,\mu_b)$ by requiring that the QCD and HQET factorization give the same amplitude, namely
\begin{equation}
f^{\rm HQET}_B(\mu_b)\int_0^\infty d\omega\, T(\omega,\mu_b) \varphi_+(\omega;\mu_b) = f_B \int_0^1 dx\, H(x,\mu_h)\phi_B(x;\mu_h)\,,
\end{equation}
when expanded to fixed order.
We then take the convolution of the hard-scattering kernel~\eqref{eq:hardW} at the scale $\mu_h$ with the LCDA~\eqref{eq:phiatmuQ}, and apply~\eqref{eq:phiu} to $\phi_B(u;\mu_b)$ with $\sigma=0$:
\begin{eqnarray}
\label{eq:hardexp}
&&\int_0^\infty d\omega \, T(\omega,\mu_b) \varphi_+(\omega;\mu_b) = \frac{f_B}{f^{\rm HQET}_B(\mu_b)}\int_0^1 dx\, H(x,\mu_h) \phi_B(x;\mu_h)
\nonumber\\
&&= \int_0^{\delta m_b} d\omega\, \frac{m_b}{\omega} \varphi_+(\omega;\mu_b)\biggl[1 + \aCFopi\biggl( h\Bigl(\frac{\omega}{m_b},\Q,\mu_h\Bigr) + \mathcal{J}_{\rm peak}^{(1)}(m_b,\mu_b)\biggr) \biggr] \\
&&\hspace*{0.5cm}+ \frac{\as}{8\pi} \ln \frac{\mu_b^2}{\mu_h^2}\int_0^{\delta m_b} d\omega\, \varphi_+(\omega;\mu_b)\sum_{n=0}^\infty \Bigl[\gamma_n N_n \int_0^1 dx\, 6 \bar{x}\,C_n^{(3/2)}(2x-1)\Bigr] C_n^{(3/2)}\Bigl(2\frac{\omega}{m_b}-1\Bigr)\,,
\nonumber
\end{eqnarray}
where we neglected the subleading contribution from the tail
\begin{equation}
\aCFopi \int_\delta^1 dx\, \frac{1}{x}\mathcal{J}^{(1)}_{\rm tail}(x) \propto \frac{1}{\delta} \ll \frac{m_b}{\LamQCD}\,.
\end{equation}
The last line of~\eqref{eq:hardexp} can be simplified with the help of
\begin{equation}
\int_0^1 dx \, 6 \bar{x} \,C_n^{(3/2)}(2x-1) = 3(-1)^n\,,
\end{equation}
which follows from the relation between Gegenbauer polynomials and Legendre polynomials $P_n(x)$
\begin{equation}
    C^{(3/2)}_n(z) = \frac{d}{dz}P_{n+1}(z)\,.
\end{equation}
Then the Gegenbauer series can be summed to
\begin{equation}
\sum_{n=0}^\infty (-1)^n 3 \frac{\gamma_n}{2C_F} N_n C_n^{(3/2)}\Bigl(2\frac{\omega}{m_b}-1\Bigr) =  -\frac{m_b}{\omega}\Bigl(2\ln \frac{\omega}{m_b}+3 \Bigr)\,,
\end{equation}
which follows from expanding the right-hand side in Gegenbauer moments and 
employing the generating function of the Gegenbauer polynomials
\begin{equation}
    \frac{1}{(1-2 z t +t^2)^{\frac{3}{2}}}= \sum_{n=0}^{\infty}C^{(3/2)}_n(z) t^n\,.
\end{equation}
Putting these results into~\eqref{eq:hardexp}, we can write
\begin{align}
\label{eq:hardexp2}
&\int_0^\infty d\omega \, T(\omega,\mu_b) \varphi_+(\omega;\mu_b) = \int_0^{\delta m_b} d\omega\, \frac{m_b}{\omega} \varphi_+(\omega;\mu_b)\nonumber\\
&\times\biggl[1 + \aCFopi\biggl( h\Bigl(\frac{\omega}{m_b},\Q,\mu_h\Bigr) + \mathcal{J}_{\rm peak}^{(1)}(m_b,\mu_b) -\ln \frac{\mu_b^2}{\mu_h^2}\Bigl(2\ln \frac{\omega}{m_b}+3\Bigr) \biggr) \biggr]\,.
\end{align}
Thus, we identify the HQET hard-scattering kernel with  
\begin{align}
\label{eq:Tresult}
T^{(0)}(\omega) &= \frac{m_b}{\omega}\,,\nonumber\\
T^{(1)}(\omega,m_b,\mu_b) &= \biggl(h\Bigl(\frac{\omega}{m_b},\Q,\mu_h\Bigr) + \mathcal{J}_{\rm peak}^{(1)}(m_b,\mu_b) -\ln \frac{\mu_b^2}{\mu_h^2}\Bigl(2\ln \frac{\omega}{m_b}+3\Bigr) \biggr)\,T^{(0)}(\omega)\,,
\end{align}
The $\mu_h$ dependence of $h$ cancels with the last term from the LCDA 
evolution, as required, and the final result is
\begin{align}
T^{(1)}(\omega,m_b,\mu_b) =& \biggl[\frac{1}{2}\ln^2 \frac{\mu_b^2}{m_b^2}-2 \ln \frac{\mu_b^2}{m_b^2} \ln \frac{\omega}{m_b}+\ln ^2\frac{\omega}{m_b}-\frac{5}{2}\ln \frac{\mu_b^2}{m_b^2}\nonumber\\
&-\left(\ln \frac{m_b^2}{Q^2}+i\pi\right) \left(2 \ln\frac{\omega}{m_b}+3\right)+\frac{\pi ^2}{12} -7\biggr]T^{(0)}(\omega)\,,
\end{align}
in agreement with Eq.~(39) of~\cite{Ishaq:2019zki}.
\section{On the Normalization of the QCD LCDA}
\label{sec:appM0}

In Section~\ref{sec:norm} of the main text we proved that 
the QCD LCDA resulting from the matching to HQET is properly normalized to 1 at the one-loop order up to power 
corrections in the parameters $\delta$ and 
$\lambda/\delta$. For the $D$ and $\bar{B}$ mesons, we 
observed a negative deviation of $\sim 10-15\%$ from unity, 
which can naturally be attributed to power corrections. 
However, increasing the 
meson mass has the unexpected effect of enlarging the 
normalization deficit, implying that the procedure 
of Section~\ref{sec:numerics} does not yield a well-defined 
heavy-quark limit. In this appendix, we identify the 
origin of this effect and explain the solution. While 
of conceptual importance for the construction of the QCD 
LCDA, we also find that the modified procedure discussed 
in this appendix is not required for the relevant cases of the $D$ and $\bar{B}$ mesons.

\subsection{Log Analysis of the Cut-off Moment}

Inspecting the fixed-order result~\eqref{eq:norm} for the QCD LCDA normalization,
we notice that the only quantity which is computed differently between the analytical evaluation~\eqref{eq:norm} and the numerical analysis in Section~\ref{sec:numerics} is the HQET LCDA cut-off moment $M_0(\LamUV,\mu)$.
Indeed, using the OPE prediction 
\begin{equation}
\label{eq:M0OPE}
    M_0^{\rm OPE}(\LamUV,\mu) = 1 -\frac{\as(\mu)C_F}{4\pi}\biggl(2\ln^2 \frac{\mu}{\LamUV} + 2\ln \frac{\mu}{\LamUV}+ \frac{\pi^2}{12}\biggr)\,,
\end{equation}
for $M_0$ in~\eqref{eq:norm} implies that the QCD LCDA normalization is affected only by power corrections in $\delta = \Lambda_{\rm UV}/m_H$ and $\lambda/\delta$, which decrease with increasing meson mass.
Therefore it is clear that the problem with the heavy-quark limit of the numerical evaluation after evolving the HQET LCDA from the initial scale $\mu_s$ to the matching scale $\mu=\mathcal{O}(m_Q)$ must arise from
\begin{equation}
\label{eq:intneqOPE}
    \int_0^{\LamUV} d\omega \, \varphi_+(\omega;\mu) \neq M_0^{\rm OPE}(\LamUV,\mu)\,.
\end{equation}

For the exponential model examined in the main text, the LL-evolved HQET LCDA can be written in closed form as
\begin{equation}
\label{eq:phimu}
\varphi_+(\omega;\mu) = \varphi_+^{\rm exp-LL}(\omega;\mu)+\frac{\as(\mu_s)C_F}{4\pi} \biggl[\frac{1}{2}-\frac{\pi^2}{12}\biggr]\varphi_+^{\rm exp}(\omega;\mu_s) + \theta(\omega-\sqrt{e}\mu_s)\varphi_+^{\rm asy}(\omega;\mu_s) \,,
\end{equation}
with $\varphi_+^{\rm exp-LL}(\omega;\mu)$ given in \eqref{eq:phiexpevol}. The left-hand side of~\eqref{eq:intneqOPE} can therefore be computed analytically, resulting in  
\begin{align}
\label{eq:phinorm}
N_{\int}(\LamUV,\mu) &\equiv \int_0^\LamUV d\omega\, \varphi_+(\omega;\mu)= N_{\int}^{\rm{exp-LL}}(\LamUV,\omega_0,\mu,\mu_s)\nonumber \\
&-\frac{\as(\mu_s)C_F}{4\pi}\biggl(2\ln^2 \frac{\mu_s}{\LamUV}+ 2\ln\frac{\mu_s}{\LamUV} + \frac{\pi^2}{12} + \mathcal{O}(e^{-\frac{\LamUV}{\omega_0}}) \biggr)\,.
\end{align}
The exponentially small corrections in the $\mathcal{O}(\as)$ term can be safely neglected. $N_{\int}^{\rm{exp-LL}}$ denotes the analytical integral of the evolved exponential model, which therefore depends on the cut-off $\LamUV$, $\mu$ and $\mu_s$, as well as the hadronic LCDA parameter $\omega_0$. The explicit expression reads
\begin{eqnarray}
\label{eq:Nint}
N_{\int}^{\rm{exp-LL}}(\LamUV,\omega_0,\mu,\mu_s) &=& e^{V + 2 \gamma_E a} \Gamma (a+2) \left(\frac{\mu_s}{\LamUV}\right)^a \frac{1}{2x^{2+a}}\phantom{,}_1F_1\Bigr(2+a,3,-\frac{1}{x}\Bigl)\nonumber\\
&=& e^{V + 2 \gamma_E a} \frac{\Gamma (a+2)}{\Gamma(1-a)} \left(\frac{\mu_s}{\LamUV}\right)^a + \mathcal{O}(x)\,,
\end{eqnarray}
where $x\equiv \omega_0/\LamUV \sim \lambda/\delta$ and $a$ is defined in \eqref{eq:aVdef}.
The second line of \eqref{eq:phinorm} originates from the 
second and third terms in \eqref{eq:phimu}.

The leading power in the expansion of~\eqref{eq:phinorm} in $x$ is model-independent, i.e. independent of the initial condition for the LCDA, and free of power corrections, and can be compared to $M_0^{\rm OPE}$:\footnote{The superscript LL$-\mu_s$ on the left-hand side indicates that the cut-off moment has been computed from the HQET LCDA evolved from the low scale $\mu_s$.}
\begin{align}
\label{eq:M0evolLLmu0}
M_0^{\rm{LL}-\mu_s}(\LamUV,\mu) &= \frac{e^{V + 2 \gamma_E a} \Gamma (a+2)}{\Gamma (1-a)} \left(\frac{\mu_s}{\LamUV}\right)^a -\frac{\as(\mu_s)C_F}{4\pi} \biggl[2 \ln ^2\frac{\mu_s}{\LamUV}+2\ln \frac{\mu_s}{\LamUV}+\frac{\pi ^2}{12}\biggr]\nonumber\\
&=M_0^{\rm OPE}(\LamUV,\mu) + \mathcal{O}\Bigl(\as^2 \ln^2 \frac{\mu_s}{\LamUV} \ln^2 \frac{\mu_s}{\mu}\Bigr) \,.
\end{align}
We note that re-expanding this result in $\alpha_s$, the one-loop order reproduces the fixed-order result~\eqref{eq:M0OPE} and the $\mu_s$-dependence drops out.
However $M_0^{\rm{LL}-\mu_s}$ suffers from higher-order corrections with large logarithms $\ln \mu_s/\LamUV$ which are not resummed by evolving $\varphi_+$ from $\mu_s$ to $\mu$.

\begin{figure}
    \centering
    \includegraphics[width=\textwidth]{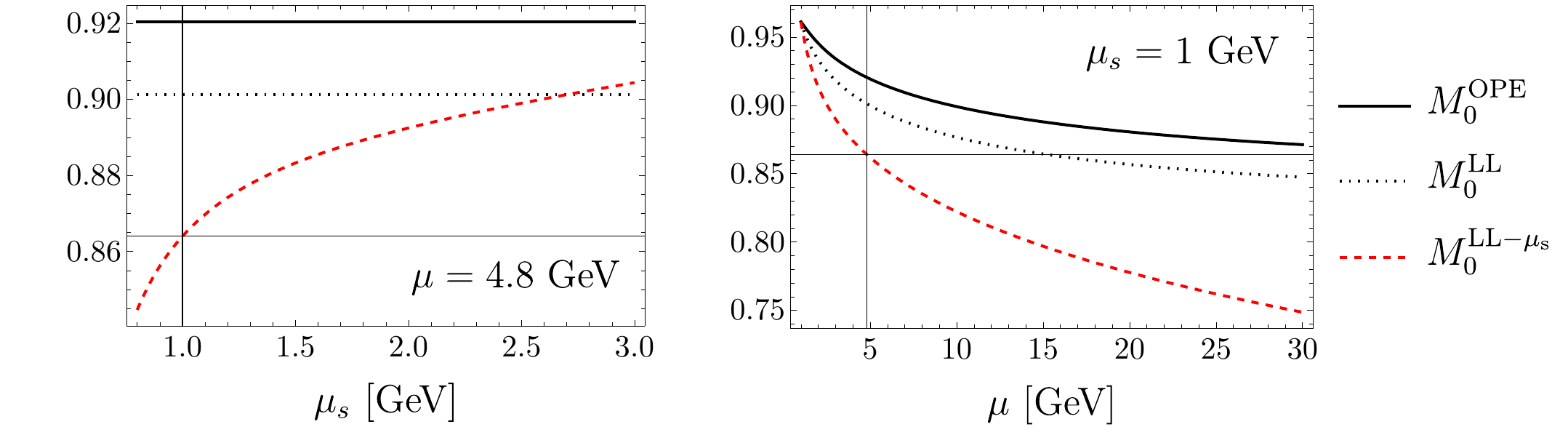}
    \caption{\small Numerical comparison of $M_0^{\rm OPE}$, $M_0^{\rm{LL}-\mu_s}$ and $M_0^{\rm LL}$, as functions of the low scale $\mu_s$ (left) and the matching scale $\mu$ (right). The cut-off $\LamUV$ is set to $\Lambda_\mu$. The solid lines correspond to the default values for the $\bar{B}$ meson.}
    \label{fig:M0intro}
\end{figure}

We show the numerical difference between $M_0^{\rm OPE}$ and $M_0^{\rm{LL}-\mu_s}$ from these higher-order corrections in 
Figure~\ref{fig:M0intro}. 
When taking the heavy-quark limit we require the hierarchy $\mu_s\sim \LamQCD \ll \LamUV \ll \mu\sim m_Q$, and increase $\LamUV$ with $\mu$ as much as possible in order to decrease the power corrections. 
We adopt
\begin{equation}
\LamUV = \sqrt{0.01 \mu^2 + \mu\cdot 1\,\mbox{GeV}} \equiv 
\Lambda_{\mu}\,.    
\end{equation}
With this choice\footnote{In this way the QCD LCDA normalization~\eqref{eq:norm} would still suffer in the heavy-quark limit from corrections $\mathcal{O}(\as \delta/(4\pi)) \lesssim 1\%$ that survive in this limit. The natural way of eliminating them would be to chose $\LamUV = \sqrt{\mu \cdot 1\,\rm{GeV}}$. This, however, generates power corrections to $M_0$ decreasing with the inverse square root of the heavy meson mass (once setting $\mu =m_H)$.
Since the focus of this section is on $M_0$ (while $\delta$ is an artificial parameter) we defined $\Lambda_\mu$ such that the power corrections in HQET are still linear.} $\LamUV$ approaches $\mu/10$ for large values of $\mu$.
The left panel shows, for fixed $\mu=\mu_b=4.8\,$GeV, the 
residual $\mu_s$-dependence of $M_0^{\rm{LL}-\mu_s}$ (dashed red), which indicates the effects of the higher-order corrections, against $M_0^{\rm OPE}$ (solid black), which is $\mu_s$-independent. 
The right panel shows, for the default value $\mu_s=1~\rm{GeV}$, the $\mu$-dependence of the different evaluations of $M_0$. Since $\mu\sim m_Q$, large values of $\mu$ correspond to the heavy-meson limit. It is apparent that the evaluation of the zeroth moment $M_0^{\rm{LL}-\mu_s}$ from the evolved HQET LCDA deviates significantly from the OPE prediction, has 
a sizeable residual $\mu_s$ dependence, and, importantly, 
the wrong heavy-quark limit, as the difference to  $M_0^{\rm OPE}$ increases with $\mu$.

To exclude the possibility that the model-independent OPE evaluation \eqref{eq:M0OPE} is inaccurate due to large logarithms of $\mu/\LamUV$, we consider a third determination of $M_0$, by resumming the logarithms $\ln \mu/\LamUV$ in~\eqref{eq:M0OPE} to LL accuracy. The RGE of $M_0$ follows from 
\begin{align}
\label{eq:dM0dlnmu}
\frac{d M_0}{d\ln \mu} =& \int_0^\LamUV d\omega\, \frac{d \varphi_+(\omega;\mu)}{d\ln \mu} = -\biggl[\Gamma_{\rm cusp}\ln \frac{\mu}{\LamUV} + \frac{\as}{4\pi}(\Gamma_0 + \gamma_0) + \mathcal{O}(\as^2) \biggr]M_0(\LamUV,\mu)\,,
\end{align}
where we employed the RGE for $\varphi_+$~\eqref{eq:phiplusRGE} and the knowledge of its asymptotic form.
The solution of \eqref{eq:dM0dlnmu} reads
\begin{equation}
    M_0^{\rm LL}(\LamUV,\mu)= e^{V_M(\mu,\LamUV)}M_0^{\rm OPE}(\LamUV,\LamUV)\,,
\end{equation}
with
\begin{equation}
    V_M(\mu,\LamUV) = -\int_{\as(\LamUV)}^{\as(\mu)} \frac{d\alpha}{\beta(\alpha)} \biggl[\Gamma_{\rm cusp}(\alpha) \int_{\as(\LamUV)}^\alpha \frac{d\alpha'}{\beta(\alpha')} + \frac{\alpha}{4\pi}(\gamma_0 + \Gamma_0) + \mathcal{O}(\alpha^2) \biggr]\,.
\end{equation}
$M_0^{\rm LL}$ is shown in black-dotted in 
Figure~\ref{fig:M0intro}, which demonstrates that the 
higher-order logarithms are numerically not important. 
This corroborates that the problem with 
$M_0^{\rm{LL}-\mu_s}$ arises from the uncancelled 
logarithms of $\mu_s/\LamUV$ in higher-orders, as 
suggested by the $\mu_s$ dependence in the left panel of Figure~\ref{fig:M0intro}.

\subsection{Improved Evolution}

This points to a problem with the standard evolution of 
the HQET LCDA for the cut-off moments. Its origin is that 
the LCDA is secretly a two-scale object, with logarithms of the form
\begin{equation}
\ln \frac{\mu}{\LamQCD} \,, \qquad \ln \frac{\mu}{\omega}\,,
\end{equation} 
where $\LamQCD$ is represented by the low scale $\mu_s$ at 
which the initial condition is set, while $\omega$ can be 
much larger than $\LamQCD$ in the asymptotic region. 
When computing the convergent inverse moment, which does not 
need to be cut off at $\LamUV$, the integration in 
$\omega$ covers the whole domain from 0 to $\infty$, 
making $\ln \mu /\omega$ effectively scaleless.
On the other hand, when computing moments with a 
$\omega$-cut-off $\LamUV \gg \LamQCD$, a UV scale is introduced, and new logarithms from $\ln \mu/\omega \to \ln \mu/\LamUV$ develop, which are not summed by the standard RGE that 
deals with the collinear logarithms $\ln \mu/\mu_s$.

The authors (FLW) of \cite{Feldmann:2014ika} proposed an ``improved evolution'', the idea of which is to modify the natural initial scale of the LCDA evolution in order to capture 
the different logarithms in different regions of $\omega$. 
These are better distinguished in ``dual space''~\cite{Bell:2013tfa}, where the evolution is diagonal in the dual space variable. We briefly review the essential definitions and relations in dual space that will be needed in the following to implement the improved evolution.

The dual function $\rho_+(\eta;\mu)$ is defined as the 
Bessel-function transformation
\begin{equation}
\label{eq:gotodual}
    \rho_+(\eta;\mu) = \int_0^\infty \frac{d\omega}{\omega} \sqrt{\frac{\omega}{\eta}} J_1\!\left(2\sqrt{\frac{\omega}{\eta}}\right) \varphi_+(\omega;\mu)
\end{equation}
of $\varphi_+$, 
which diagonalizes the one-loop evolution kernel such that
\begin{equation}
\label{eq:rhoev}
    \frac{d\rho_+(\eta;\mu)}{d\ln \mu} = -\biggl[ \Gamma_{\rm cusp} \ln \frac{\mu}{\hat{\eta}} + \gamma_+\biggr]\rho_+(\eta;\mu)\,,
\end{equation}
where $\hat{\eta} = e^{-2\gamma_E}\,\eta$ and the anomalous dimensions are given in \eqref{eq:betacuspdef} and below 
\eqref{eq:aVdef}. This equation is solved by
\begin{equation}
    \rho_+(\eta;\mu) = e^{V(\mu,\mu_s)} \biggl(\frac{\mu_s}{\hat{\eta}}\biggr)^{a(\mu,\mu_s)} \rho_+(\eta;\mu_s)\,.
\end{equation}
The zeroth cut-off moment can be conveniently converted to an integral over $\eta$~\cite{Feldmann:2014ika}
\begin{equation}
\label{eq:M0rho}
    M_0(\LamUV,\mu) = \int_0^\infty d\eta \,\frac{\LamUV}{\eta} J_2\bigg(2\sqrt{\frac{\LamUV}{\eta}}\,\bigg)\rho_+(\eta;\mu)\,.
\end{equation}
The fixed-order result $M_0^{\rm OPE}(\LamUV,\mu)$ can be reproduced by using the perturbative-partonic expression for $\rho_+(\eta;\mu)$~\cite{Feldmann:2014ika} (in analogy with the derivation of $M_0^{\rm OPE}$ in momentum space~\cite{Lee:2005gza})
\begin{align}
\label{eq:rhopert}
    \rho_+(\eta;\mu)_{\rm pert} &= \frac{C_0(\eta,\mu)}{\omega_0} J_2\bigg(2\sqrt{\frac{2\omega_0}{\eta}}\,\bigg) \,,\nonumber\\
    C_0(\eta,\mu) &= 1 + \frac{\as(\mu)C_F}{4\pi}\biggl(-2\ln^2 \frac{\mu}{\hat{\eta}} +2\ln \frac{\mu}{\hat{\eta}} - \frac{\pi^2}{12} -2 \biggr) + \mathcal{O}(\as^2)\,, 
\end{align}
where $\omega_0 \sim \LamQCD$ drops out after expanding 
the cut-off moment in $x=\omega_0/\LamUV$.

In order to implement the ``improved running'' in dual space the initial scale of the LCDA evolution is set to
\begin{equation}
\label{eq:FLWic}
    \mu_{s\eta} = \sqrt{\mu_s^2 + \hat{\eta}^2}\,,
\end{equation}
leading to the evolved
\begin{equation}
\label{eq:rhoFLW}
    \rho_+^{\rm FLW}(\eta; \mu) = e^{V(\mu,\mu_{s\eta})} \biggl(\frac{\mu_{s\eta}}{\hat{\eta}}\biggr)^{a(\mu,\mu_{s\eta})} \rho_+(\eta;\mu_{s\eta})\,,
\end{equation}
with initial condition
\begin{eqnarray}
    \rho_+(\eta,\mu_{s\eta}) &=& \frac{1}{\eta} e^{-\frac{\omega_0}{\eta}}\biggl(1+\frac{\as(\mu_{s\eta})C_F}{4\pi} \biggl(\frac{1}{2}-\frac{\pi^2}{12} \biggr) \biggr) 
\nonumber\\
&&\hspace*{-2.25cm}-\,\frac{\as(\mu_{s\eta}) C_F}{4\pi \eta} \biggl(2 \ln^2 \frac{\mu_{s\eta}}{\hat{\eta}} -2 \ln \frac{\mu_{s\eta}}{\hat{\eta}} -2\sqrt{e} \frac{\mu_{s\eta}}{\eta} \,_3F_4\Bigl(1,1,1;2,2,2,3;-\sqrt{e}\frac{\mu_{s\eta}}{\eta} \Bigr)+\frac{5}{2} \biggr)\,.\qquad
\end{eqnarray}
The important difference to standard evolution is that when $\hat{\eta}\gg \mu_s \sim \LamQCD$, the evolution does not start at the low scale but at $\hat{\eta}$.
The initial condition $\rho_+(\eta;\mu_{s\eta})$ is obtained from the transformation~\eqref{eq:gotodual} applied to our model~\eqref{eq:phimus} evaluated at $\mu_s = \mu_{s\eta}$.
From the evolved $\rho_+^{\rm FLW}(\eta;\mu)$ we compute numerically the zeroth moment through~\eqref{eq:M0rho}
\begin{equation}
    N_{\int}^{\rm FLW}(\LamUV,\mu) \equiv \int_0^\infty d\eta \,\frac{\LamUV}{\eta} J_2\bigg(2\sqrt{\frac{\LamUV}{\eta}}\,\bigg)\rho_+^{\rm FLW}(\eta;\mu)\,.
\end{equation}

\addtocontents{toc}{\protect\enlargethispage{\baselineskip}} 
\subsection{Numerical Comparisons of $M_0$}
\label{sec:appM0num}

Since in the following we want to compare the model-independent predictions with the model-dependent numerical calculations $N_{\int}(\LamUV,\mu)$, $N_{\int}^{\rm FLW}(\LamUV,\mu)$ of $M_0$, we need to keep in mind that the predictions are only valid in the limit $\LamUV \gg \LamQCD$, and therefore the comparison is affected by power corrections in $x=\omega_0/\LamUV$. 

The model~\eqref{eq:phimus} is constructed in order to have the correct asymptotic behaviour, and to match the fixed order $M_0^{\rm OPE}$ at the low scale $\mu_s$ up to corrections of order $\mathcal{O}(e^{-\LamUV/\omega_0})$, which are completely negligible for $\LamUV \geq 3~\rm{GeV}$.
However after evolution, the power corrections are turned into linear as can be seen from~\eqref{eq:Nint}. 
It suffices to expand~\eqref{eq:Nint} to linear order to 
obtain 
\begin{equation}
\label{eq:M0powerdiff}
N_{\int}(\LamUV,\mu) - M_0^{\rm{LL}-\mu_s}
\approx e^{V + 2 \gamma_E a} \,\frac{a \Gamma (a+3)}{\Gamma(1-a)} \left(\frac{\mu_s}{\LamUV}\right)^{\!a}
\,\frac{\omega_0}{\LamUV}+\mathcal{O}(x^2)\,,
\end{equation}
which evaluates to $-0.06$ for $\LamUV=2~\si{GeV}$ and $\mu=4.8~\rm{GeV}$. 
(For the $\bar{B}$ meson $\LamUV= \delta m_B = 2.38~\si{GeV}$ is used in Section~\ref{sec:numerics}.) We note that the 
power correction is negative and depends on $\mu$ only 
mildly through $a$.

\begin{figure}
\centering
\includegraphics[width=0.49\textwidth]{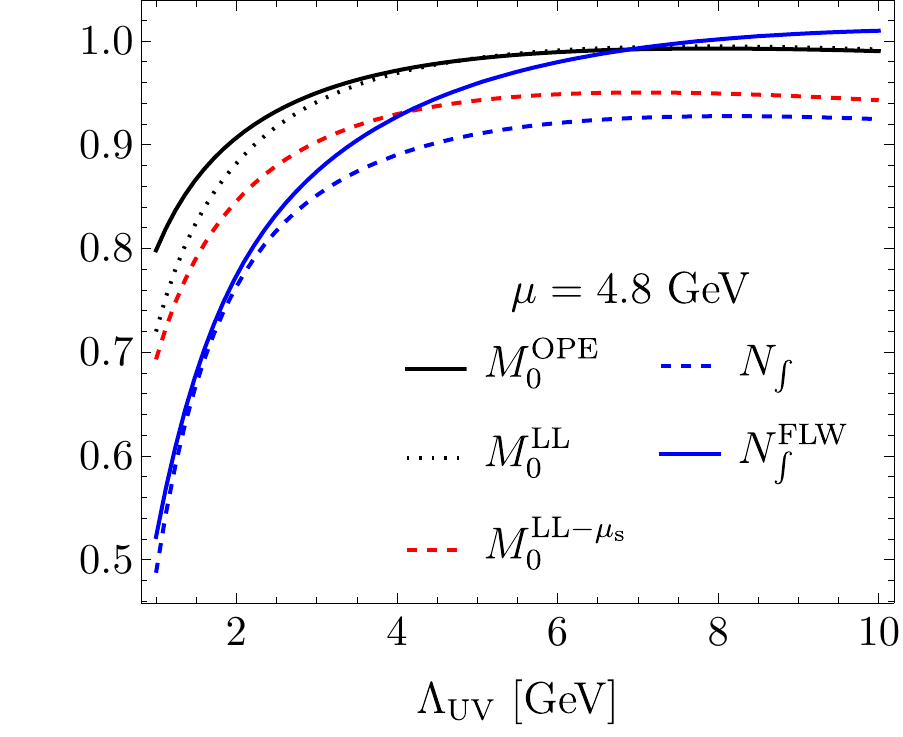}$\;$
\includegraphics[width=0.49\textwidth]{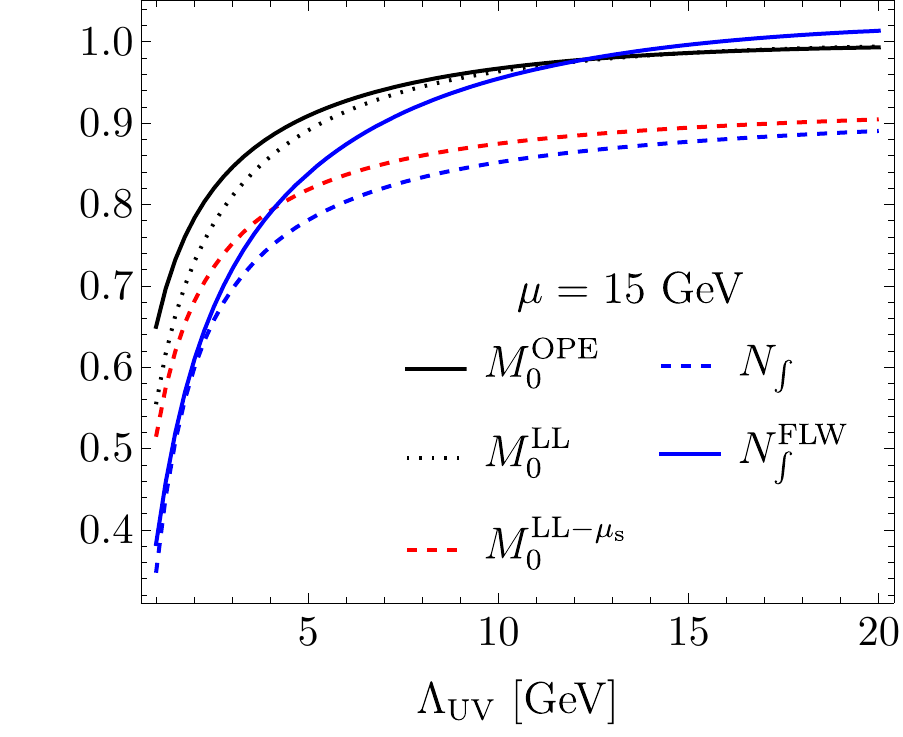}
\caption{\small Comparison of the five different determinations of $M_0$ at two values of the renormalization scale, $\mu_b$ and $\mu_{15}$.}
\label{fig:M0s}
\end{figure}

We can now compare the five different evaluation of $M_0$ 
that have been defined. We wish to demonstrate that with 
the improved evolution~\cite{Feldmann:2014ika}, the 
numerical integration of the evolved HQET LCDA agrees well 
with the OPE prediction. In Figure~\ref{fig:M0s}, 1) $M_0^{\rm OPE}$ (solid black) is the fixed-order OPE result, which serves as the reference. 
2) $M_0^{\rm LL}$ (dotted black) has the logarithms $\ln \mu/\LamUV$ correctly resummed to LL, improving the convergence of the perturbative series of $M_0^{\rm OPE}$. 
3) $M_0^{\rm{LL}-\mu_s}$ (dashed red) is the model-independent prediction (free from power corrections) obtained from integrating the LCDA evolved from $\mu_s$.
4) $N_{\int}$ (dashed blue) is the integration of the model $\varphi_+(\omega;\mu)$ evolved from $\mu_s$ including the power corrections, while 
5) $N_{\int}^{\rm FLW}$ (solid blue) is the integration of $\varphi_+(\omega;\mu)$ (in dual space) evolved from $\mu_{s\eta}$, called ``improved evolution''.

\begin{figure}
    \centering
\hskip-2cm\includegraphics[width=0.49\textwidth]{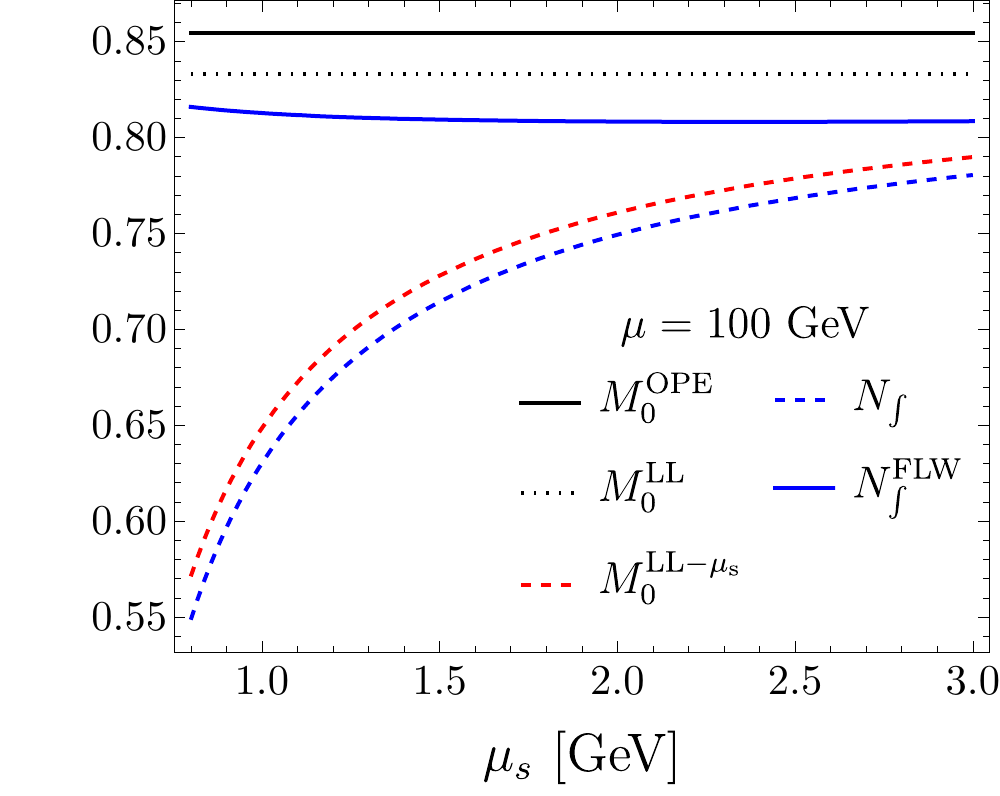}
\caption{\small As left panel of Figure~\ref{fig:M0intro} with $N_{\int}$ (dashed blue) and $N_{\int}^{\rm FLW}$ (blue) added. The scale $\mu$ has been set to $100~\rm{GeV}$ in order to reduce the effect of power corrections ($\LamUV = \Lambda_\mu = 14.1~\rm{GeV}$). }
\label{fig:M0musdep}
\end{figure}

From the above we already know that the dashed blue  
and dashed red curves differ only by the power 
correction \eqref{eq:M0powerdiff}.
At the same time, the dashed red curve differs from the solid black one by higher-order $\ln \mu_s/\LamUV$ terms. 
Due to the improved evolution, the solid blue curve 
should sum these corrections and be closer to the dotted 
black curve. These features are indeed visible in 
Figure~\ref{fig:M0s} for the two values of $\mu\sim m_Q$. 
In particular, the improved $N_{\int}^{\rm FLW}$ is much closer to the OPE result for $\mu_s\ll \LamUV\lesssim \mu$ than the problematic $N_{\int}$ used in the main text.
Furthermore it can be seen from Figure~\ref{fig:M0musdep} that the $\mu_s$-dependence is nearly absent after improved 
evolution, solving the main deficiency of the standard 
running.

However, for the $\LamUV$ values of interest for the $\bar{B}$ and $D$ meson (2.38 GeV and 1.22 GeV, respectively), the difference between $N_{\int}$ and $N_{\int}^{\rm FLW}$ is small (see Figure~\ref{fig:M0s}, left panel for the case of 
$\bar{B}$), proving that the $-10$\% deviation in the normalization of the QCD LCDA mentioned in the main text can be attributed only to the power corrections, and not the 
higher-order $\ln \mu_s/\LamUV$ terms.

\addtocontents{toc}{\protect\enlargethispage{\baselineskip}} 
\subsection{Large Meson-Mass Limit}
\label{sec:appLMMlimit}

\begin{figure}
\centering
\hskip2.4cm\includegraphics[width=0.8\textwidth]{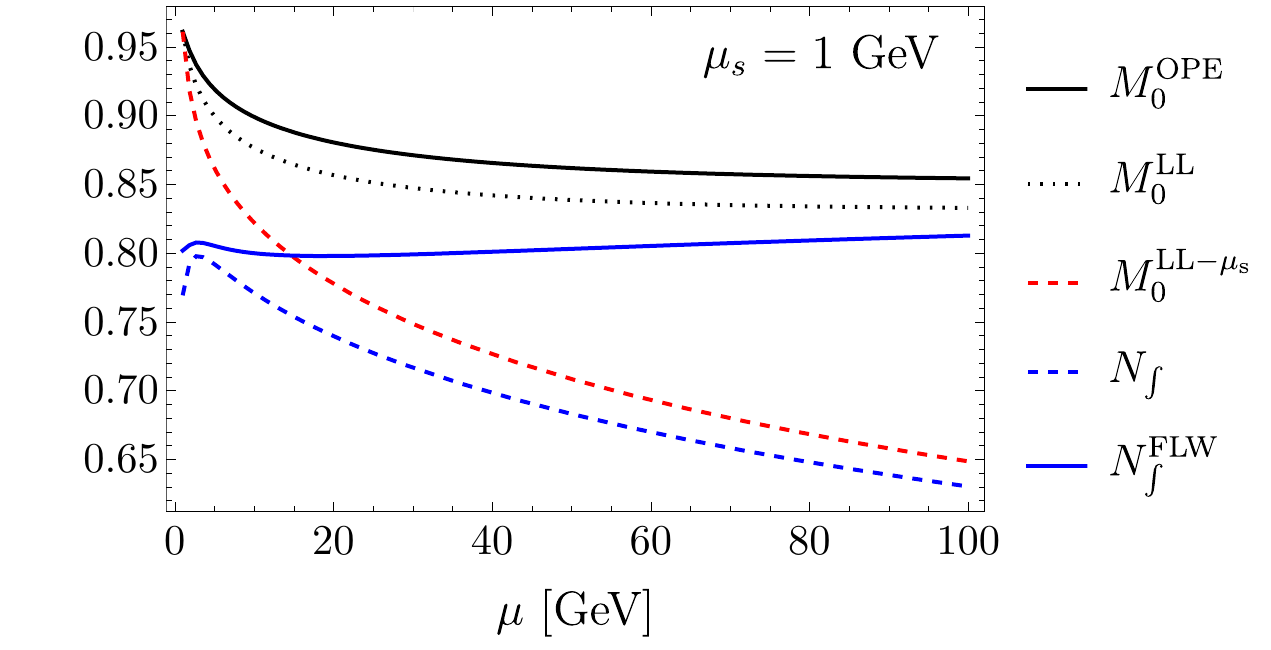}\\[0.5cm]

\includegraphics[width=0.6\textwidth]{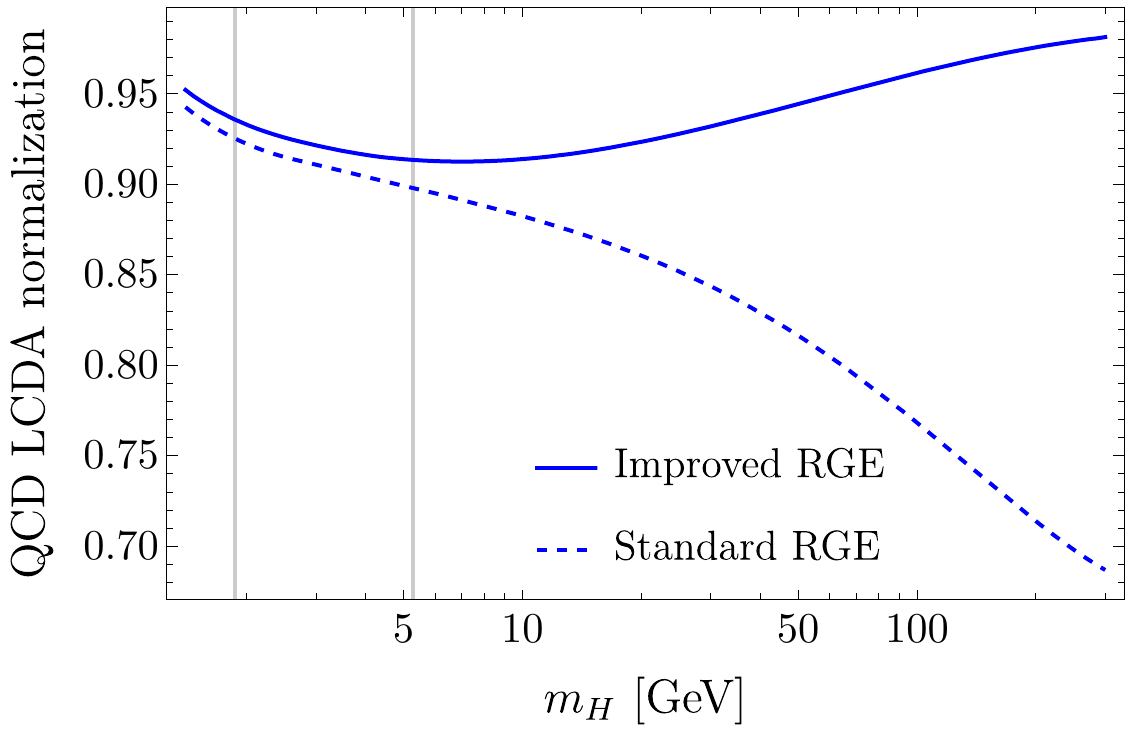}
\caption{\small Upper panel: $\mu$-dependence of the five 
evaluations of the cut-off moment discussed in the text. 
Lower panel: QCD LCDA normalization as a function of the meson mass.}
    \label{fig:M0mudep}
\end{figure}

We now address the heavy-quark limit of the cut-off 
moment evaluations and QCD LCDA normalization.  
The HQET LCDA is of course independent on the heavy mass, however when fixing the matching scale $\mu=m_H$, the large-mass limit corresponds to the evolution to very high scales.
The upper panel of Figure~\ref{fig:M0mudep} shows the five evaluations of $M_0$ for large scales, setting $\LamUV=\Lambda_\mu$. 
We observe that the improved resummation tends to $M_0^{\rm LL}$ for values of $\LamUV$ large enough while standard 
resummation (dashed red and blue) deviates more and more 
for increasing values of $\mu$ due to the 
unresummed $\ln \mu_s/\LamUV$ terms. This demonstrates 
that it is mandatory to use the FLW prescription  
\eqref{eq:FLWic} for the initial scale to ensure a 
consistent heavy-quark limit. 

Finally, we turn to the quantity of importance, the normalization of the QCD LCDA constructed from the matching to the universal HQET LCDA. We set $\mu=m_H$ and the peak--tail separation parameter $\delta$ to $\Lambda_\mu/m_H$, so that for 
high meson masses it will tend to the constant 0.1. The lower panel of Figure~\ref{fig:M0mudep} demonstrates that for increasing meson mass the normalization correctly tends to 1 
when the improved RGE is employed (solid blue), which is an important conceptual check of our result, while the standard 
RGE fails (dashed blue). The grey vertical lines mark the 
$D$ and $\bar{B}$ meson mass. For these, the difference is 
small, justifying the neglect of improvement for the 
numerical results in the main text.

\newpage
\bibliographystyle{JHEP} 
\bibliography{refs.bib}

\end{document}